\PassOptionsToPackage{table}{xcolor}
\documentclass[a4paper,11pt]{article}
\pdfoutput=1 % if you are submitting a pdf latex (i.e. if you have
\usepackage{jheppub} % for details on the use of the package, please
\usepackage[T1]{fontenc} % if needed
\usepackage{fontawesome}
\usepackage{pdflscape}
\usepackage[dvipsnames]{xcolor}
\usepackage[caption=false]{subfig}
\usepackage{epsfig}
\usepackage{makecell}
\usepackage{multirow}
\usepackage{colortbl}
\usepackage[T1]{fontenc} % if needed
\usepackage{tikz}
\tikzset{every picture/.style={remember picture}}
\usepackage{hepnames}
\usepackage{adjustbox}
\usepackage{rotating}
\usepackage{multirow}
\usepackage{bbold}
\usepackage{slashed}
\usepackage{bigints}
\usepackage[normalem]{ulem}
\usepackage{amsmath}
\usepackage{textcomp}
\usepackage{float}
\usepackage{graphicx}
\usepackage{orcidlink}
\makeatletter
\gdef\@fpheader{}
\makeatother
\usepackage[toc,page]{appendix}
\usepackage[compat=1.1.0]{tikz-feynman} 
\usepackage{silence}
\usepackage{booktabs}
\tikzfeynmanset{warn luatex=false}
\newcommand{\stkout}[1]{\ifmmode\text{\sout{\ensuremath{#1}}}\else\sout{#1}\fi}
\definecolor{calpolypomonagreen}{rgb}{0.12, 0.3, 0.17}
\newcommand{\bea}{\begin{eqnarray}}
	\newcommand{\eea}{\end{eqnarray}}
\newcommand{\beq}{\begin{equation}}
	\newcommand{\eeq}{\end{equation}}
\usepackage{array} 
\usepackage[most]{tcolorbox}
\usepackage{xcolor}
\usepackage{booktabs}   % for \toprule, \midrule, \bottomrule

\tcbset{highlight math style={
		colback=gray!20, colframe=gray!50!black,
		boxrule=0.5pt, sharp corners}}
\usepackage[utf8]{inputenc}
\usepackage[T1]{fontenc} 
\DeclareUnicodeCharacter{00B1}{\ensuremath{\pm}}
\DeclareUnicodeCharacter{2213}{\ensuremath{\mp}}
\DeclareUnicodeCharacter{2013}{--}     % en dash
\DeclareUnicodeCharacter{2014}{---}    % em dash
\DeclareUnicodeCharacter{2212}{-}      % minus sign
\DeclareUnicodeCharacter{00B1}{\ensuremath{\pm}}

\usepackage{upgreek}

\title{\textcolor{black}{Anomalous Dimensions, Matching, and Phenomenology of Dirac Fermionic Dark Matter Effective Interactions}}

\author[a]{Subhajit Kala\,\orcidlink{0000-0001-5915-0212}}
\author[b]{Lipika Kolay\,,\orcidlink{0000-0003-0715-6668}}
\author[a]{Soumitra Nandi\,\orcidlink{0000-0001-6567-0302}}
\affiliation[a]{Department of Physics, Indian Institute of Technology Guwahati, North Guwahati, Assam-781039, India.}
\affiliation[b]{Department of Physics, Indian Institute of Technology Gandhinagar, Palaj, Gujarat 382355, India}

\emailAdd{s.kala@iitg.ac.in}
\emailAdd{lipika.kolay@iitgn.ac.in}
\emailAdd{soumitra.nandi@iitg.ac.in}
\abstract{We explore a fermionic dark matter (DM) extension of the Standard Model Effective Field Theory (SMEFT) and establish a complete renormalisation-group framework for its phenomenological investigation. We derive the anomalous-dimension matrix of all relevant dimension-five and dimension-six operators involving Standard Model (SM) and DM fields, enabling the consistent evolution of the associated Wilson coefficients (WCs) across energy scales. By combining renormalisation-group running with matching at the relevant thresholds, we construct a robust bridge between high-scale new physics and experimental observables. We then perform a comprehensive phenomenological analysis, evaluating the contributions of these operators to observables spanning a wide range of energies and deriving constraints on the WCs from current data. We obtain stringent and complementary bounds on the DM effective field theory (DMEFT) WCs from electroweak precision observables, flavour processes, lepton-flavour-violating decays, top-quark flavour-changing neutral-current decays, and invisible meson decays. Interpreted in terms of the effective scale of new physics, the resulting limits demonstrate that current precision measurements probe energy scales ranging from the TeV regime to several tens or even hundreds of TeV, highlighting the remarkable sensitivity of indirect searches to dark-sector interactions.}


\begin{document}
\maketitle
\flushbottom
%%%%%%%%%%%%%%%%%%%%%%%%%%%%%
%%%%%%%%%%%%%%%%%%%%%%%%%%%%%
%--------------------------------------
\section{Introduction}
%--------------------------------------
Despite its remarkable success in describing fundamental particles and their interactions, the Standard Model (SM) of particle physics is widely regarded as incomplete. One of its most significant shortcomings is the lack of a viable dark matter (DM) candidate, despite overwhelming astrophysical and cosmological evidence indicating that approximately 85\% of the Universe's matter content is non-luminous and non-baryonic. The nature of DM, therefore, remains one of the most compelling open questions in contemporary particle physics \cite{Taoso:2007qk,Peter:2012rz,Lin:2019uvt,CaboAlmeida:2025har}. For a recent review, see ~\cite{Cirelli:2024ssz} and the references therein.

The search for DM is pursued through a broad experimental programme, including direct-detection experiments, indirect searches for DM annihilation or decay products \cite{Ibarra:2013cra,Gaskins:2016cha,Hooper:2018kfv,PerezdelosHeros:2020qyt}, and collider searches at high-energy facilities such as the Large Hadron Collider (LHC) \cite{Gaitskell:2004gd,Goodman:2010ku,MarrodanUndagoitia:2015veg,Kahlhoefer:2017dnp,CMS:2024tkt,CMS:2024zqs,CMS:2025zxs}. On the theoretical side, numerous ultraviolet (UV)-complete scenarios have been proposed, ranging from supersymmetric models \cite{Jungman:1995df,Bertone:2004pz,Steffen:2008qp,Allahverdi:2009sb} and extended gauge sectors to theories with additional scalar or fermionic degrees of freedom \cite{Patt:2006fw,He:2007tt,Barger:2008jx,Petriello:2008pu,Arcadi:2024mli}. Complementary approaches based on effective field theories (EFTs) \cite{Cao:2009uw,Brod:2021xbb} and simplified models with explicit mediators have also been developed \cite{He:2009yd,DeSimone:2016fbz}, providing a systematic and phenomenologically versatile framework for interpreting experimental results in a largely model-independent manner.

More generally, the Standard Model Effective Field Theory (SMEFT) has emerged as a powerful, systematic framework for probing physics beyond the SM \cite{Grzadkowski:2010es,Alonso:2013hga,Jenkins:2013zja,Jenkins:2013wua,Jenkins:2017jig,Jenkins:2017dyc,Brivio:2017vri,Isidori:2023pyp}. If the underlying new degrees of freedom are sufficiently heavy to evade direct production, their effects can still appear indirectly through deviations from SM predictions. SMEFT captures such effects through higher-dimensional operators built solely from SM fields and respecting SM gauge symmetries. The framework, therefore, provides a model-independent description of heavy new physics and enables precision measurements from flavour observables, electroweak precision tests, Higgs physics, electric dipole moments, and collider processes to be interpreted within a common theoretical framework \cite{Kala:2025srq, Kala:2026xzo}.

However, by construction, SMEFT contains only the particle content of the SM and introduces no additional propagating degrees of freedom. Consequently, it is unable to accommodate DM as an explicit degree of freedom. Given the compelling evidence for the existence of DM, it is therefore natural and well-motivated to extend the SMEFT framework by incorporating DM as an additional dynamical degree of freedom. Such an extension retains the model-independent virtues of SMEFT while simultaneously allowing for a systematic description of DM interactions with the visible sector.

Within this extended framework, the dark matter field is treated on equal footing with the SM fields, and one can construct the complete set of gauge-invariant higher-dimensional operators involving both SM and DM degrees of freedom \cite{Criado:2021trs, Aebischer:2022wnl}, known as dark matter Effective Field Theory (DMEFT). The corresponding Wilson coefficients (WCs) encode the effects of the underlying UV dynamics and provide a model-independent parametrisation of DM-SM interactions. This approach establishes a unified framework for studying dark matter phenomenology across a wide range of energy scales. Moreover, by combining matching calculations with renormalisation-group evolutions (RGEs), constraints obtained from low-energy observables, collider measurements, and DM searches can be consistently connected to the underlying high-scale theory. In this way, the effective field theory description provides a powerful bridge between experimental data and the fundamental origin of the DM.

A few studies have been carried out on subsets of the dimension-6 operator basis. Studies considering DM masses in the sub-TeV range have primarily focused on flavour-conserving operators, exploring DM and collider phenomenology \cite{Bai:2010hh, Cheung:2010zf, Zheng:2010js, Fox:2011pm, Haisch:2012kf, Abdallah:2014hon, Beniwal:2022rde, Demetriou:2025ewa, Fuyuto:2024oii}. In contrast, some studies have considered flavour-violating operators in the sub-GeV DM mass regime, mainly in the context of invisible decays, where the DM constitutes the invisible particle escaping detection \cite{Kumar:2024ivx, Buras:2024ewl, Mescia:2026xju, Liu:2025lbw, Barducci:2026teq}. By comparison, the phenomenology of dimension-5 operators with Dirac fermions as DM remains largely unexplored. Moreover, the effects of RGE evolution and operator mixing in connecting DMEFT interactions to low-energy observables have not been systematically investigated. This motivates a phenomenological study of Dirac fermionic DM that incorporates both dimension-5 and dimension-6 operators, with flavour-conserving and flavour-violating interactions treated on equal footing. In particular, we aim to obtain individual bounds on the WCs through a global analysis of a broad class of observables. In this work, we investigate their impact on a wide range of low-energy observables, including rare semileptonic $B$-meson decays, neutral-meson mixing, lepton-flavour-violating (LFV) observables, electroweak precision observables (EWPOs), and collider observables. We use these observables to derive constraints on the DMEFT WCs and assess their sensitivity to DM-SM interactions, thereby addressing an aspect that has not been systematically explored in the DMEFT framework.

The rest of the paper is organized as follows. Sec.~\ref{sec:DMEFT} discusses the dark matter effective field theory and the motivation for the choice of the operator bases. Sec.~\ref{sec:methodology} describes the methodology for the scale evolution of the WCs across different scales, while Sec.~\ref{sec:phenomenology} presents the phenomenological implications of the multi-scale evolution of the WCs. Sec.~\ref {sec:observables} discusses the bounds on the WCs obtained from different flavour and electroweak observables for sub-TeV mass DM, while Sec .~\ref {sec:invisible} explores the possibility of having DM as an invisible particle in the low-energy processes. Finally, we conclude our discussion in Sec.~\ref {sec:summary}.

%--------------------------------------
\section{Dark Matter Effective Field Theory}\label{sec:DMEFT}
%--------------------------------------
We consider an effective field theory obtained by extending the SM with a Dirac fermionic DM field, $\chi$. The DM field transforms as a singlet under the SM gauge group $\mathrm{SU}(3)_C \times \mathrm{SU}(2)_L \times \mathrm{U}(1)_Y$, with the representation $(\mathbf{1}, \mathbf{1}, 0)$, and is therefore electrically neutral. Consequently, the covariant derivative acting on $\chi$ reduces to a simple partial derivative. At dimension four, the renormalizable Lagrangian contains only the kinetic and mass terms of $\chi$; no tree-level interactions between the DM and SM fields are present. Interactions between the dark and visible sectors arise only at higher mass dimensions through non-renormalizable operators suppressed by the new physics scale $\Lambda$.

\begin{equation}\label{eq:total_Lagrangian}
    \mathcal{L} = \mathcal{L}_{\rm SM}^{(4)} + i\bar{\chi}\slashed{\partial}\chi - m_{\chi}\bar{\chi}\chi + \mathcal{L}_{\rm DM}^{(d>4)}\,,
\end{equation}
with
\begin{align}
    \mathcal{L}_{\rm SM}^{(4)} =& -\frac{1}{4}G^A_{\mu\nu}G^{A\;\mu\nu} - \frac{1}{4}W^I_{\mu\nu}W^{I\; \mu\nu} - \frac{1}{4}B_{\mu\nu}B^{\mu\nu} + \left(D_{\mu}H\right)^{\dagger}\left(D^{\mu}H\right) + m_H^2 H^{\dagger}H \nonumber\\
    & -\frac{1}{2}\lambda \left(H^{\dagger} H\right)^2 + i\left(\bar{l}_L \slashed{D} l_L + \bar{e}_R \slashed{D} e_R + \bar{q}_L \slashed{D} q_L + \bar{u}_R \slashed{D} u_R + \bar{d}_R\slashed{D} d_R \right) \\
    & -\left([\Gamma_e]_{pr} \bar{l}_{L,p} e_{R,r} H + [\Gamma_u]_{pr} \bar{q}_{L,p} u_{R,r} \tilde{H} + [\Gamma_d]_{pr} \bar{q}_{L,p} d_{R,r} H + h.c. \right) + \mathcal{L}_{\rm ghost} + \mathcal{L}_{\rm G.F.} \,, \nonumber
\end{align}
where the Yukawa couplings $[\Gamma_{u,d,e}]_{pr}$ are matrices in generation space with flavour indices $p$ and $r$, and $\tilde{H}^j=\epsilon_{jk}(H^k)^*$. Throughout this work, we evaluate the DMEFT framework strictly in the unbroken electroweak phase without considering Spontaneous Symmetry Breaking (SSB). This is theoretically justified because, in dimensional regularization combined with a mass-independent subtraction scheme like $\overline{\text{MS}}$, the UV divergences are inherently insensitive to dimensionful parameters such as the Higgs vacuum expectation value (VEV). The covariant derivative is given by:
\begin{equation}
    D_{\mu} = \partial_{\mu} + ig_1 B_{\mu}Y + ig_2 W_{\mu}^I \tau^I + i g_s G^A_{\mu}T^A \,, 
\end{equation}
where the weak hypercharge $Y$ is given by $Y_{l_L}=-1/2$, $Y_{e_R}=-1$, $Y_{q_L}=1/6$, $Y_{u_R}=2/3$, $Y_{d_R}=-1/3$, and $Y_H=1/2$. In the fundamental representation, the generators for $\mathrm{SU}(2)_L$ read $\tau^I=\sigma^I/2$ with $\sigma^I\;(I=1,2,3)$ being the Pauli matrices, and for $\mathrm{SU}(3)_C$ read $T^A=\lambda^A/2$ with $\lambda^{A}\; (A=1,...,8)$ being the Gell-Mann matrices. The field strength tensors are given by:
\begin{align}
    G^A_{\mu\nu} &= \partial_{\mu}G_{\nu}^A - \partial_{\nu}G_{\mu}^A - g_s f^{ABC} G_{\mu}^B G_{\nu}^C \,, \nonumber\\
    W^I_{\mu\nu} &= \partial_{\mu}W_{\nu}^I - \partial_{\nu}W_{\mu}^I - g_2 \epsilon^{IJK} W_{\mu}^J W_{\nu}^K \,,  \\
    B_{\mu\nu} &= \partial_{\mu}B_{\nu} - \partial_{\nu}B_{\mu} \,. \nonumber
\end{align}
In eq.~\eqref{eq:total_Lagrangian}, $m_\chi$ denotes the mass of the Dirac DM particle, and $\mathcal{L}_{\rm DM}^{(d>4)}$ represents the effective Lagrangian containing a tower of higher-dimensional operators constructed from the SM degrees of freedom and the DM field. A discrete $\mathbb{Z}_2$ symmetry is imposed, under which $\chi$ is odd while all SM fields are even, thereby preventing the decay of $\chi$ and ensuring its stability as a viable DM candidate. The complete list of gauge-invariant operators up to mass dimension six built from the DM and SM fields is discussed in the following paragraph. 
%--------------------------------------
\paragraph{\textbf{\underline{Operator basis} :}}
%--------------------------------------
The higher-dimensional effective Lagrangian with DM, can be written as
%-----------------------
\begin{equation}
    \mathcal{L}_{\rm DM}^{(d>4)} = \sum_{i,d} \frac{C_i^{(d)}}{\Lambda^{d-4}} \mathcal{O}_i^{(d)} \,, 
\end{equation}
%-----------------------
where $\mathcal{O}_i^{(d)}$ denote the higher-dimensional operators and $C_i^{(d)}$ are the corresponding dimensionless WCs. For clarity, we use the shorthand notation $\mathcal{C}_i$ for the WCs throughout the analysis, where for dimension-five operators $\mathcal{C}_i\equiv C_i/\Lambda$, while for dimension-six operators $\mathcal{C}_i\equiv C_i/\Lambda^2$. The complete operator basis of the DMEFT is constructed following Ref.~\cite{Brod:2017bsw}.

% %--------------------------
% \begin{subequations}\label{eq:dim_5}
% \begin{align}
%     \mathcal{O}^{(5,1)}_{B\chi} &= \left(\bar{\chi} \sigma_{\mu \nu} \chi\right)B^{\mu\nu}\,, &\quad \mathcal{O}^{(5,2)}_{B\chi} &= \left(\bar{\chi} i\sigma_{\mu \nu}\gamma_5 \chi\right)B^{\mu\nu}\,,\\
%     \mathcal{O}^{(5,1)}_{H\chi} &= (\bar{\chi}\chi)(H^{\dagger} H)\,, &\quad \mathcal{O}^{(5,2)}_{H\chi} &= (\bar{\chi} i\gamma^5\chi)(H^{\dagger} H)\,.
% \end{align}
% \end{subequations}
% %---------------------------------
\begin{table}[hbt!]
    \centering
    \footnotesize
    \renewcommand{\arraystretch}{1.5}
    \begin{tabular}{llcll}
        \toprule
        \textbf{Operator} & \textbf{Structure} && \textbf{Operator} & \textbf{Structure} \\
        \midrule
        \multicolumn{5}{c}{\textbf{Dimension-5}} \\
        \midrule
        $\mathcal{O}^{(5,1)}_{B\chi}$  & $\left(\bar{\chi} \sigma_{\mu \nu} \chi\right)B^{\mu\nu}$ && $\mathcal{O}^{(5,2)}_{B\chi}$  & $\left(\bar{\chi} i\sigma_{\mu \nu}\gamma_5 \chi\right)B^{\mu\nu}$ \\
        $\mathcal{O}^{(5,1)}_{H\chi}$  & $(\bar{\chi}\chi)(H^{\dagger} H)$ && $\mathcal{O}^{(5,2)}_{H\chi}$  & $(\bar{\chi} i\gamma^5\chi)(H^{\dagger} H)$ \\
        \midrule
        \multicolumn{5}{c}{\textbf{Dimension-6}} \\
        \midrule
        $\mathcal{O}^{(6,1)}_{\chi q}$ & $\left(\bar{\chi} \gamma_{\mu}\chi\right)\left(\bar{q}_s \gamma^{\mu}q_t\right)$ && $\mathcal{O}^{(6,2)}_{\chi q}$ & $\left(\bar{\chi} \gamma_{\mu}\gamma_5\chi\right)\left(\bar{q}_s \gamma^{\mu}q_t\right)$ \\
        $\mathcal{O}^{(6,1)}_{\chi \ell}$ & $\left(\bar{\chi} \gamma_{\mu}\chi\right)\left(\bar{\ell}_s \gamma^{\mu}\ell_t\right)$ && $\mathcal{O}^{(6,2)}_{\chi \ell}$ & $\left(\bar{\chi} \gamma_{\mu}\gamma_5\chi\right)\left(\bar{\ell}_s \gamma^{\mu}\ell_t\right)$ \\
        $\mathcal{O}^{(6,1)}_{\chi u}$ & $\left(\bar{\chi} \gamma_{\mu}\chi\right)\left(\bar{u}_s \gamma^{\mu} u_t\right)$ && $\mathcal{O}^{(6,2)}_{\chi u}$ & $\left(\bar{\chi} \gamma_{\mu}\gamma_5\chi\right)\left(\bar{u}_s \gamma^{\mu} u_t\right)$ \\
        $\mathcal{O}^{(6,1)}_{\chi d}$ & $\left(\bar{\chi} \gamma_{\mu}\chi\right)\left(\bar{d}_s \gamma^{\mu} d_t\right)$ && $\mathcal{O}^{(6,2)}_{\chi d}$ & $\left(\bar{\chi} \gamma_{\mu}\gamma_5\chi\right)\left(\bar{d}_s \gamma^{\mu} d_t\right)$ \\
        $\mathcal{O}^{(6,1)}_{\chi e}$ & $\left(\bar{\chi} \gamma_{\mu}\chi\right)\left(\bar{e}_s \gamma^{\mu} e_t\right)$ && $\mathcal{O}^{(6,2)}_{\chi e}$ & $\left(\bar{\chi} \gamma_{\mu}\gamma_5\chi\right)\left(\bar{e}_s \gamma^{\mu} e_t\right)$ \\
        $\mathcal{O}^{(6,1)}_{\chi DH}$ & $(\bar{\chi}\gamma_{\mu} \chi)\left(H^{\dagger}i\overleftrightarrow{D}^{\mu}H\right)$ && $\mathcal{O}^{(6,2)}_{\chi DH}$ & $(\bar{\chi} \gamma_{\mu}\gamma^5\chi)\left(H^{\dagger}i\overleftrightarrow{D}^{\mu}H\right)$ \\
        \bottomrule
    \end{tabular}
    \caption{The complete basis of dimension-5 and dimension-6 DMEFT operators coupling the singlet dark fermion $\chi$ to the SM. Fermion generation indices are denoted by $s$ and $t$.}
    \label{tab:dmeft_operators}
\end{table}
%-----------------------------
The set of operators relevant to our work is listed in Table~\ref{tab:dmeft_operators}. The first row summarises the set of dimension-five operators. In total, four independent operators arise at this order. Among them, the dipole operators $\mathcal{O}^{(5,1)}_{B\chi}$ and $\mathcal{O}^{(5,2)}_{B\chi}$ are generated at the loop level, while the scalar and pseudoscalar Higgs portals $\mathcal{O}^{(5,1)}_{H\chi}$ and $\mathcal{O}^{(5,2)}_{H\chi}$ can be generated already at tree level in a perturbative UV-complete theory. We absorb the loop factor $\frac{g_1}{8\pi^2}$  for the loop-generated dipole operators directly into the definition of the respective WCs to streamline the notation.

The second row of Table~\ref{tab:dmeft_operators}, summarises the set of dimension-6 operators involving both the SM and DM fields, where $s$ and $t$ denote the SM fermion generation indices, and the Hermitian derivative is defined as $H^{\dagger}i\overleftrightarrow{D}^{\mu}H \equiv H^\dagger i D^\mu H - i (D^\mu H)^\dagger H$. This list does not include operators composed purely of DM fields. While such pure-DM dimension-6 operators do arise, we omit them from the present analysis as they are phenomenologically less relevant. For simplicity, we restrict our framework to a single-component DM scenario $(\chi)$; the extension to a multicomponent singlet DM model follows straightforwardly by considering different generations of DM states $(\chi_i)$.
%--------------------------------------
\section{RGEs of the WCs}
\label{sec:methodology}
%--------------------------------------
One of the important tasks is to compute the RGEs that describe the evolution of the WCs with the energy scale, running from the high new-physics scale $\Lambda$ down to either the electroweak scale or the dark matter mass scale $m_{\chi}$, in case the DM mass $m_{\chi} << \Lambda$. 
% In this work, we restrict our calculation specifically to the mixed DM-SM effective operators up to dimension-six, omitting purely DM-DM operators from the analysis as they remain decoupled from the primary visible-sector observables at this order.

We derive the one-loop anomalous dimension matrices (ADMs) associated with the operator bases introduced above and determine the corresponding operator-mixing patterns. To this end, we compute the UV divergences of the relevant one-particle-irreducible (1PI) one-loop Feynman diagrams. The calculation is performed in the unbroken gauge of the SM, which suffices because anomalous dimensions are insensitive to electroweak symmetry breaking and thus to the Higgs vacuum expectation value. Furthermore, the SM-DM effective operators considered in this work do not mix with SMEFT operators at $\mathcal{O}(1/\Lambda^{2})$. Although such mixing may arise at higher orders in the EFT expansion, its impact is expected to be phenomenologically subleading at the current level of precision.

To maintain explicit gauge invariance under the SM gauge group throughout the intermediate stages of the calculation, we employ the Background Field Method (BFM) \cite{Abbott:1981ke}. In this approach, all gauge fields are split into classical background components and quantum fluctuating components. This choice significantly simplifies the tracking of counterterms and ensures that the relations between gauge couplings and wave-function renormalization constants are tightly constrained by background gauge symmetry. For the explicit formulation of the one-loop Feynman rules for the non-Abelian gauge fields within the BFM framework, we follow the conventions and derivations outlined in Ref.~\cite{Abbott:1980hw}.

% We regulate ultraviolet (UV) divergences using dimensional regularization in
% $d=4-2\epsilon$ dimensions and adopt the $\overline{\rm MS}$ renormalization
% scheme. The bare WCs $\mathcal{C}_i^{(0)}$ are related to the
% renormalized, scale-dependent coefficients $\mathcal{C}_i(\mu)$ through the
% operator renormalization matrix $Z_{ij}$,
% %
% \begin{equation}
% 	\mathcal{C}_i^{(0)}
% 	=
% 	\mu^{2\epsilon} Z_{ij}\,\mathcal{C}_j(\mu)\,.
% \end{equation}
% %
% At one loop, the renormalization constants admit the expansion
% %
% \begin{equation}
% 	Z_{ij}
% 	=
% 	\delta_{ij}
% 	+
% 	\frac{1}{16\pi^2\epsilon}\,\gamma_{ij}
% 	+
% 	\mathcal{O}\!\left(\frac{1}{\epsilon^2}\right),
% \end{equation}
% %
% where $\gamma_{ij}$ denotes the anomalous dimension matrix (ADM). Requiring the
% bare WCs to be independent of the arbitrary renormalization
% scale $\mu$ leads to the standard renormalization-group equations (RGEs),
% %
% \begin{equation}
% 	\mu\frac{d}{d\mu}\mathcal{C}_i(\mu)
% 	=
% 	\frac{1}{16\pi^2}\gamma^{T}_{ij}\,
% 	\mathcal{C}_j(\mu)\,.
% \end{equation}

The anomalous dimensions of the higher-dimensional operators are extracted from the UV-divergent $1/\epsilon$ poles of the corresponding one-loop Green functions. Following the conventions of Ref.~\cite{Buras:1998raa}, the bare amputated Green function with an insertion of the operator basis $\vec{\mathcal{O}}^{(0)}$ is related to the renormalized matrix element $\langle \vec{\mathcal{O}} \rangle$ through
\begin{align}
	\langle \vec{\mathcal{O}} \rangle^{(0)}
	=
	\left(
	\prod_{\psi}
	\mathcal{Z}_{\rm WR}^{\psi}
	\right)^{-1/2}
	\mathcal{Z}_{jk}
	\langle \vec{\mathcal{O}} \rangle
	\equiv
	\mathcal{Z}^{(0)}_{jk}
	\langle \vec{\mathcal{O}} \rangle ,
	\label{eq:ADM_rel}
\end{align}
where $\mathcal{Z}_{\rm WR}^{\psi}$ denotes the wave-function renormalization
constant of the external field $\psi$, and $\mathcal{Z}_{jk}$ is the operator
renormalization matrix. The corresponding coupling renormalization matrix is
given by the inverse transpose of the operator renormalization matrix,
$\mathcal{Z}^{(c)}_{ij}=(\mathcal{Z}_{ji})^{-1}$.
In the $\overline{\rm MS}$ scheme, the renormalization constants take the form
\begin{align}
	(\mathcal{Z})_{ij}
	&=
	\delta_{ij}
	+
	\frac{\alpha_n}{4\pi}
	\frac{1}{\epsilon}
	a^{n}_{ij},
	\nonumber\\
	(\mathcal{Z}_{\rm WR})_{\psi}
	&=	1 + \frac{\alpha_n}{4\pi} \frac{1}{\epsilon} b^{n}_{\psi} -
	\frac{1}{16\pi^2} \frac{1}{\epsilon} b^{y}_{\psi},\\
	(\mathcal{Z}^{(0)})_{ij}
	&=
	\delta_{ij}
	+
	\frac{\alpha_n}{4\pi}
	\frac{1}{\epsilon}
	c^{n}_{ij},
	\nonumber
	\label{eq:renormalisation_mat}
\end{align}
where $\alpha_n=g_n^2/(4\pi)$ corresponds to the gauge couplings of
$\mathrm{U}(1)_Y$ ($n=1$), $\mathrm{SU}(2)_L$ ($n=2$), and
$\mathrm{SU}(3)_C$ ($n=3$), while the superscript $y$ denotes contributions
arising from Yukawa interactions.

The scale evolution of the WCs is governed by the equation
\begin{align}
	\dot{\mathcal{C}}_i(\mu) \equiv 16\pi^2\, \mu\frac{d}{d\mu}\mathcal{C}_i(\mu) = \gamma^T_{ij}\, \mathcal{C}_j(\mu),
\end{align}\label{eq:rgewc}
where $\gamma^{T}$ is the ADM in the Wilson-coefficient basis. The
Anomalous dimensions are obtained directly from the UV counterterms according to
\begin{equation}
	\frac{\gamma}{16\pi^2}
	=
	- (\mathcal{Z})^{-1} \mu 
	\frac{d\mathcal{Z}}{d\mu}
	%=
	%\beta(g,\epsilon)
	%(\mathcal{Z})^{-1}
	%\frac{d\mathcal{Z}}{dg(\mu)}
	%\nonumber\\	
	=-\frac{g_n^2}{16\pi^2}
	\,2a^{n}_{ij} \,. 
\end{equation}
%
%where the $D$-dimensional beta function is defined by
%
%\begin{equation}
%	\beta(g,\epsilon)
%	=
%	-\epsilon g
%	-
%	g\,Z_g^{-1}\,
%	\mu \frac{dZ_g}{d\mu}.
%\end{equation}
%
Combining eq.~\eqref{eq:ADM_rel} with the renormalization constants above, the
ADM can be expressed in terms of the divergent wave-function and operator
counterterms as
\begin{align}
	\gamma_{ij}=-2g_n^2\left(\sum_{\psi}\frac{1}{2}b_{\psi}^{n}\delta_{ij}
	+ a_{ij}^{n}\right)+b_{\psi}^{y}\delta_{ij}.
	\label{eq:ADM}
\end{align}

% A detailed derivation of extracting this ADM can be found in the Appendix~\ref{Append:List_of_diagrams}.

To manage the extensive algebraic complexity, the calculation was performed using an automated computational pipeline. We cross-verified our independent operator basis using \texttt{BasisGen} \cite{Criado:2019ugp}. The model and its BFM-compatible Feynman rules were implemented via \texttt{FeynRules} \cite{Alloul:2013bka}, while 1PI diagram topologies were generated using \texttt{FeynArts} \cite{Hahn:2000kx}. Finally, we have calculated the ultraviolet $1/\epsilon$ poles using an in-house code and cross-verified against \texttt{Package-X}\cite{Patel:2016fam} and \texttt{FeynCalc}\cite{Shtabovenko:2023idz}.

The algebraic structure of the calculated anomalous dimension matrix is inherently similar to that of standard SMEFT under a proper mapping and replacement of fields. Consequently, wherever applicable, we cross-verified our results against the established SMEFT ADMs Refs.~\cite{Jenkins:2013zja, Jenkins:2013wua, Alonso:2013hga}. A notable exception arises for the dimension-five dipole-type operators, such as $\mathcal{O}_{B\chi}^{(5)}$, which do not map directly to dimension-six SMEFT structures. In that case, we successfully cross-verified our analytical results against the dimension-five dipole operators of the LEFT framework, finding perfect agreement.
%--------------------------------------
\paragraph{\underline{Anomalous Dimension Matrix} :} \label{sec:ADM}
%--------------------------------------
Here, we present the RGEs of the WCs following the eq.~\eqref{eq:rgewc} after calculating the one-loop ADMs of the DMEFT operator basis. The renormalisation group evolution of the WCs is driven by operator mixing induced via quantum corrections. Generically, the matrix elements $\gamma_{ij}$ are functions of the fundamental parameters of the SM. Specifically, they depend on the gauge couplings $\alpha_n \equiv g_n^2/(4\pi)$ (where $n = 1, 2, 3$ correspond to the $\mathrm{U}(1)_Y$, $\mathrm{SU}(2)_L$, and $\mathrm{SU}(3)_C$ gauge groups, respectively), the generation-space Yukawa matrices $\Gamma_{u,d,e}$, the Higgs quartic self-coupling $\lambda$, and the wave-function renormalization constants $(b_{\psi})$ associated with the external fields of the respective operators. Their explicit expressions can be found in Appendix~\ref{Append:List_of_diagrams}. Furthermore, the complete set of one-loop topologies correcting the DMEFT operators that arise during these calculations is detailed in Appendix~\ref{Append:List_of_diagrams}. 

Following eq.~\eqref{eq:rgewc}, the RGEs of the dimension-five operators are as given below 
%--------------------------------------
%\subsection{Dimension $5$}
%\label{subsec:dim5_adm}
%--------------------------------------
\begin{subequations}
% \begin{align}
%     \dot{\mathcal{C}}_{B\chi}^{(5,1)}&=
% \end{align}
\begin{align}
    \dot{\mathcal{C}}_{H\chi}^{(5,1)} = & \; \left[  12\lambda + 2\gamma_H^{(\Gamma)} \right] \mathcal{C}_{H\chi }^{(5,1)} 
     - 2 m_{\chi} \left[ 3 |\mathcal{C}^{(5,1)}_{H\chi}|^2- |\mathcal{C}^{(5,2)}_{H\chi}|^2\right]+24 m_{\chi}\left[ 3 |\mathcal{C}^{(5,1)}_{B\chi}|^2- |\mathcal{C}^{(5,2)}_{B\chi}|^2\right] \,. 
\end{align}
\end{subequations}
Here, $\gamma_H^{(\Gamma)}=\mathrm{Tr}\left(N_c\Gamma_u\Gamma_u^{\dagger} + N_c\Gamma_d\Gamma_d^{\dagger} + \Gamma_e\Gamma_e^{\dagger}\right)$. The operator $\mathcal{O}_{B\chi}$ receives no corrections up to $\mathcal{O}(1/\Lambda^2)$.
For the dimension-six operators the RGEs of the respective WCs are given by the equations as given below:   
%--------------------------------------
%\subsection{Dimension $6$}
%\label{subsec:dim6_adm}
%--------------------------------------
\begin{subequations}
%     \begin{align}
%     \dot{\mathcal{C}}_{\substack{\chi q \\ prst}}^{(6,1)} = & \; g_1^2 \delta_{st} \left[ \frac{8}{3} N_c Y_q^2 \mathcal{C}_{\substack{\chi q \\ prst}}^{(6,1)} + \frac{8}{3} Y_q Y_\ell \mathcal{C}_{\substack{\chi \ell \\ prst}}^{(6,1)} + \frac{4}{3} N_c Y_q Y_u \mathcal{C}_{\substack{\chi u \\ prst}}^{(6,1)} + \frac{4}{3} N_c Y_q Y_d \mathcal{C}_{\substack{\chi d \\ prst}}^{(6,1)} + \frac{4}{3} Y_q Y_e \mathcal{C}_{\substack{\chi e \\ prst}}^{(6,1)} \right] \nonumber \\
%     & + b_{sv}^{(q)} \mathcal{C}_{\substack{\chi q \\ prvt}}^{(6,1)} + \mathcal{C}_{\substack{\chi q \\ prsv}}^{(6,1)} b_{vt}^{(q)} - [\Gamma_u]_{sw} [\Gamma_u^\dagger]_{vt} \mathcal{C}_{\substack{\chi u \\ prwv}}^{(6,1)} - [\Gamma_d]_{sw} [\Gamma_d^\dagger]_{vt} \mathcal{C}_{\substack{\chi d \\ prwv}}^{(6,1)} \nonumber \\
%     & + \left[ \frac{4}{3} g_1^2 Y_H Y_q + \left( \Gamma_u \Gamma_u^\dagger - \Gamma_d \Gamma_d^\dagger \right)_{st} \right] \mathcal{C}_{\substack{\chi DH \\ pr}}^{(6,1)} + 12 g_1^2 Y_q^2 \delta_{st} \mathcal{C}_{\substack{B\chi \\ pv}}^{(5,1)} \mathcal{C}_{\substack{B\chi \\ vr}}^{(5,1)*}
% \end{align}
%
\begin{align}
    \dot{\mathcal{C}}_{\chi \underset{st}{q}}^{(6,1)} = & \; g_1^2 \delta_{st} \left[ \frac{8}{3} N_c Y_q^2 \mathcal{C}_{\chi \underset{st}{q}}^{(6,1)} + \frac{8}{3} Y_q Y_\ell \mathcal{C}_{\chi \underset{st}{\ell}}^{(6,1)} + \frac{4}{3} N_c Y_q Y_u \mathcal{C}_{\chi \underset{st}{u}}^{(6,1)} + \frac{4}{3} N_c Y_q Y_d \mathcal{C}_{\chi \underset{st}{d}}^{(6,1)} + \frac{4}{3} Y_q Y_e \mathcal{C}_{\chi \underset{st}{e}}^{(6,1)} \right] \nonumber \\
    & + b_{sv}^{(q)} \mathcal{C}_{\chi \underset{vt}{q}}^{(6,1)} + \mathcal{C}_{\chi \underset{sv}{q}}^{(6,1)} b_{vt}^{(q)} - [\Gamma_u]_{sw} [\Gamma_u^\dagger]_{vt} \mathcal{C}_{\chi \underset{wv}{u}}^{(6,1)} - [\Gamma_d]_{sw} [\Gamma_d^\dagger]_{vt} \mathcal{C}_{\chi \underset{wv}{d}}^{(6,1)} \nonumber \\
    & + \left[ \frac{4}{3} g_1^2 Y_H Y_q + \left( \Gamma_u \Gamma_u^\dagger - \Gamma_d \Gamma_d^\dagger \right)_{st} \right] \mathcal{C}_{\chi DH}^{(6,1)} -12 g_1^2 Y_q^2 \delta_{st} \Big[|\mathcal{C}_{B\chi}^{(5,1)}|^2 + |\mathcal{C}_{B\chi}^{(5,2)}|^2\Big] \,,
\end{align}
% 
%    \begin{align}
%     \dot{\mathcal{C}}_{\substack{\chi \ell \\ prst}}^{(6,1)} = & \; g_1^2 \delta_{st} \left[ \frac{8}{3} N_c Y_q Y_\ell \mathcal{C}_{\substack{\chi q \\ prst}}^{(6,1)} + \frac{8}{3} Y_\ell^2 \mathcal{C}_{\substack{\chi \ell \\ prst}}^{(6,1)} + \frac{4}{3} N_c Y_u Y_\ell \mathcal{C}_{\substack{\chi u \\ prst}}^{(6,1)} + \frac{4}{3} N_c Y_d Y_\ell \mathcal{C}_{\substack{\chi d \\ prst}}^{(6,1)} + \frac{4}{3} Y_e Y_\ell \mathcal{C}_{\substack{\chi e \\ prst}}^{(6,1)} \right] \nonumber \\
%     & + b_{sv}^{(\ell)} \mathcal{C}_{\substack{\chi \ell \\ prvt}}^{(6,1)} + \mathcal{C}_{\substack{\chi \ell \\ prsv}}^{(6,1)} b_{vt}^{(\ell)} - [\Gamma_e]_{sw} [\Gamma_e^\dagger]_{vt} \mathcal{C}_{\substack{\chi e \\ prwv}}^{(6,1)} \nonumber \\
%     & + \left[ \frac{4}{3} g_1^2 Y_H Y_\ell - [\Gamma_e \Gamma_e^\dagger]_{st} \right] \mathcal{C}_{\substack{\chi DH \\ pr}}^{(6,1)} + 12 g_1^2 Y_\ell^2 \delta_{st} \mathcal{C}_{\substack{B\chi \\ pv}}^{(5,1)} \mathcal{C}_{\substack{B\chi \\ vr}}^{(5,1)*}
% \end{align}
%
\begin{align}
    \dot{\mathcal{C}}_{\chi \underset{st}{\ell}}^{(6,1)} = & \; g_1^2 \delta_{st} \left[ \frac{8}{3} N_c Y_q Y_\ell \mathcal{C}_{\chi \underset{st}{q}}^{(6,1)} + \frac{8}{3} Y_\ell^2 \mathcal{C}_{\chi \underset{st}{\ell}}^{(6,1)} + \frac{4}{3} N_c Y_u Y_\ell \mathcal{C}_{\chi \underset{st}{u}}^{(6,1)} + \frac{4}{3} N_c Y_d Y_\ell \mathcal{C}_{\chi \underset{st}{d}}^{(6,1)} + \frac{4}{3} Y_e Y_\ell \mathcal{C}_{\chi \underset{st}{e}}^{(6,1)} \right] \nonumber \\
    & + b_{sv}^{(\ell)} \mathcal{C}_{\chi \underset{vt}{\ell}}^{(6,1)} + \mathcal{C}_{\chi \underset{sv}{\ell}}^{(6,1)} b_{vt}^{(\ell)} - [\Gamma_e]_{sw} [\Gamma_e^\dagger]_{vt} \mathcal{C}_{\chi \underset{wv}{e}}^{(6,1)} \nonumber \\
    & + \left[ \frac{4}{3} g_1^2 Y_H Y_\ell - [\Gamma_e \Gamma_e^\dagger]_{st} \right] \mathcal{C}_{\chi DH}^{(6,1)}  -12 g_1^2 Y_\ell^2 \delta_{st} \Big[|\mathcal{C}_{B\chi}^{(5,1)}|^2 + |\mathcal{C}_{B\chi}^{(5,2)}|^2\Big] \,,
\end{align}
% 
%     \begin{align}
%     \dot{\mathcal{C}}_{\substack{\chi u \\ prst}}^{(6,1)} = & \; g_1^2 \delta_{st} \left[ \frac{8}{3} N_c Y_q Y_u \mathcal{C}_{\substack{\chi q \\ prst}}^{(6,1)} + \frac{8}{3} Y_u Y_\ell \mathcal{C}_{\substack{\chi \ell \\ prst}}^{(6,1)} + \frac{4}{3} N_c Y_u^2 \mathcal{C}_{\substack{\chi u \\ prst}}^{(6,1)} + \frac{4}{3} N_c Y_u Y_d \mathcal{C}_{\substack{\chi d \\ prst}}^{(6,1)} + \frac{4}{3} Y_u Y_e \mathcal{C}_{\substack{\chi e \\ prst}}^{(6,1)} \right] \nonumber \\
%     & + b_{sv}^{(u)} \mathcal{C}_{\substack{\chi u \\ prvt}}^{(6,1)} + \mathcal{C}_{\substack{\chi u \\ prsv}}^{(6,1)} b_{vt}^{(u)} - 2 [\Gamma_u^\dagger]_{sw} [\Gamma_u]_{vt} \mathcal{C}_{\substack{\chi q \\ prwv}}^{(6,1)} \nonumber \\
%     & + \left[ \frac{4}{3} g_1^2 Y_H Y_u - 2 [\Gamma_u^\dagger \Gamma_u]_{st} \right] \mathcal{C}_{\substack{\chi DH \\ pr}}^{(6,1)} + 12 g_1^2 Y_u^2 \delta_{st} \mathcal{C}_{\substack{B\chi \\ pv}}^{(5,1)} \mathcal{C}_{\substack{B\chi \\ vr}}^{(5,1)*}
% \end{align}
% 
\begin{align}
    \dot{\mathcal{C}}_{\chi \underset{st}{u}}^{(6,1)} = & \; g_1^2 \delta_{st} \left[ \frac{8}{3} N_c Y_q Y_u \mathcal{C}_{\chi \underset{st}{q}}^{(6,1)} + \frac{8}{3} Y_u Y_\ell \mathcal{C}_{\chi \underset{st}{\ell}}^{(6,1)} + \frac{4}{3} N_c Y_u^2 \mathcal{C}_{\chi \underset{st}{u}}^{(6,1)} + \frac{4}{3} N_c Y_u Y_d \mathcal{C}_{\chi \underset{st}{d}}^{(6,1)} + \frac{4}{3} Y_u Y_e \mathcal{C}_{\chi \underset{st}{e}}^{(6,1)} \right] \nonumber \\
    & + b_{sv}^{(u)} \mathcal{C}_{\chi \underset{vt}{u}}^{(6,1)} + \mathcal{C}_{\chi \underset{sv}{u}}^{(6,1)} b_{vt}^{(u)} - 2 [\Gamma_u^\dagger]_{sw} [\Gamma_u]_{vt} \mathcal{C}_{\chi \underset{wv}{q}}^{(6,1)} \nonumber \\
    & + \left[ \frac{4}{3} g_1^2 Y_H Y_u - 2 [\Gamma_u^\dagger \Gamma_u]_{st} \right] \mathcal{C}_{\chi DH}^{(6,1)} - 12 g_1^2 Y_u^2 \delta_{st} \Big[|\mathcal{C}_{B\chi}^{(5,1)}|^2 + |\mathcal{C}_{B\chi}^{(5,2)}|^2\Big] \,, 
\end{align}
% 
%     \begin{align}
%     \dot{\mathcal{C}}_{\substack{\chi d \\ prst}}^{(6,1)} = & \; g_1^2 \delta_{st} \left[ \frac{8}{3} N_c Y_q Y_d \mathcal{C}_{\substack{\chi q \\ prst}}^{(6,1)} + \frac{8}{3} Y_d Y_\ell \mathcal{C}_{\substack{\chi \ell \\ prst}}^{(6,1)} + \frac{4}{3} N_c Y_u Y_d \mathcal{C}_{\substack{\chi u \\ prst}}^{(6,1)} + \frac{4}{3} N_c Y_d^2 \mathcal{C}_{\substack{\chi d \\ prst}}^{(6,1)} + \frac{4}{3} Y_d Y_e \mathcal{C}_{\substack{\chi e \\ prst}}^{(6,1)} \right] \nonumber \\
%     & + b_{sv}^{(d)} \mathcal{C}_{\substack{\chi d \\ prvt}}^{(6,1)} + \mathcal{C}_{\substack{\chi d \\ prsv}}^{(6,1)} b_{vt}^{(d)} - 2 [\Gamma_d^\dagger]_{sw} [\Gamma_d]_{vt} \mathcal{C}_{\substack{\chi q \\ prwv}}^{(6,1)} \nonumber \\
%     & + \left[ \frac{4}{3} g_1^2 Y_H Y_d + 2 [\Gamma_d^\dagger \Gamma_d]_{st} \right] \mathcal{C}_{\substack{\chi DH \\ pr}}^{(6,1)} + 12 g_1^2 Y_d^2 \delta_{st} \mathcal{C}_{\substack{B\chi \\ pv}}^{(5,1)} \mathcal{C}_{\substack{B\chi \\ vr}}^{(5,1)*}
% \end{align}
% 
\begin{align}
    \dot{\mathcal{C}}_{\chi \underset{st}{d}}^{(6,1)} = & \; g_1^2 \delta_{st} \left[ \frac{8}{3} N_c Y_q Y_d \mathcal{C}_{\chi \underset{st}{q}}^{(6,1)} + \frac{8}{3} Y_d Y_\ell \mathcal{C}_{\chi \underset{st}{\ell}}^{(6,1)} + \frac{4}{3} N_c Y_u Y_d \mathcal{C}_{\chi \underset{st}{u}}^{(6,1)} + \frac{4}{3} N_c Y_d^2 \mathcal{C}_{\chi \underset{st}{d}}^{(6,1)} + \frac{4}{3} Y_d Y_e \mathcal{C}_{\chi \underset{st}{e}}^{(6,1)} \right] \nonumber \\
    & + b_{sv}^{(d)} \mathcal{C}_{\chi \underset{vt}{d}}^{(6,1)} + \mathcal{C}_{\chi \underset{sv}{d}}^{(6,1)} b_{vt}^{(d)} - 2 [\Gamma_d^\dagger]_{sw} [\Gamma_d]_{vt} \mathcal{C}_{\chi \underset{wv}{q}}^{(6,1)} \nonumber \\
    & + \left[ \frac{4}{3} g_1^2 Y_H Y_d + 2 [\Gamma_d^\dagger \Gamma_d]_{st} \right] \mathcal{C}_{\chi DH}^{(6,1)} - 12 g_1^2 Y_d^2 \delta_{st} \Big[|\mathcal{C}_{B\chi}^{(5,1)}|^2 + |\mathcal{C}_{B\chi}^{(5,2)}|^2\Big] \,,
\end{align}
% 
%      \begin{align}
%   \dot{\mathcal{C}}_{\substack{\chi e \\ prst}}^{(6,1)} = & \; g_1^2 \delta_{st} \left[ \frac{8}{3} N_c Y_q Y_e \mathcal{C}_{\substack{\chi q \\ prst}}^{(6,1)} + \frac{8}{3} Y_e Y_\ell \mathcal{C}_{\substack{\chi \ell \\ prst}}^{(6,1)} + \frac{4}{3} N_c Y_u Y_e \mathcal{C}_{\substack{\chi u \\ prst}}^{(6,1)} + \frac{4}{3} N_c Y_d Y_e \mathcal{C}_{\substack{\chi d \\ prst}}^{(6,1)} + \frac{4}{3} Y_e^2 \mathcal{C}_{\substack{\chi e \\ prst}}^{(6,1)} \right] \nonumber \\
%     & + b_{sv}^{(e)} \mathcal{C}_{\substack{\chi e \\ prvt}}^{(6,1)} + \mathcal{C}_{\substack{\chi e \\ prsv}}^{(6,1)} b_{vt}^{(e)} - 2 [\Gamma_e^\dagger]_{sw} [\Gamma_e]_{vt} \mathcal{C}_{\substack{\chi \ell \\ prwv}}^{(6,1)} \nonumber \\
%     & + \left[ \frac{4}{3} g_1^2 Y_H Y_e + 2 [\Gamma_e^\dagger \Gamma_e]_{st} \right] \mathcal{C}_{\substack{\chi DH \\ pr}}^{(6,1)} + 12 g_1^2 Y_e^2 \delta_{st} \mathcal{C}_{\substack{B\chi \\ pv}}^{(5,1)} \mathcal{C}_{\substack{B\chi \\ vr}}^{(5,1)*}
% \end{align}
% 
\begin{align}
    \dot{\mathcal{C}}_{\chi \underset{st}{e}}^{(6,1)} = & \; g_1^2 \delta_{st} \left[ \frac{8}{3} N_c Y_q Y_e \mathcal{C}_{\chi \underset{st}{q}}^{(6,1)} + \frac{8}{3} Y_e Y_\ell \mathcal{C}_{\chi \underset{st}{\ell}}^{(6,1)} + \frac{4}{3} N_c Y_u Y_e \mathcal{C}_{\chi \underset{st}{u}}^{(6,1)} + \frac{4}{3} N_c Y_d Y_e \mathcal{C}_{\chi \underset{st}{d}}^{(6,1)} + \frac{4}{3} Y_e^2 \mathcal{C}_{\chi \underset{st}{e}}^{(6,1)} \right] \nonumber \\
    & + b_{sv}^{(e)} \mathcal{C}_{\chi \underset{vt}{e}}^{(6,1)} + \mathcal{C}_{\chi \underset{sv}{e}}^{(6,1)} b_{vt}^{(e)} - 2 [\Gamma_e^\dagger]_{sw} [\Gamma_e]_{vt} \mathcal{C}_{\chi \underset{wv}{\ell}}^{(6,1)} \nonumber \\
    & + \left[ \frac{4}{3} g_1^2 Y_H Y_e + 2 [\Gamma_e^\dagger \Gamma_e]_{st} \right] \mathcal{C}_{\chi DH}^{(6,1)} - 12 g_1^2 Y_e^2 \delta_{st} \Big[|\mathcal{C}_{B\chi}^{(5,1)}|^2 + |\mathcal{C}_{B\chi}^{(5,2)}|^2\Big] \,, 
\end{align}
% 
%    \begin{align}
%     \dot{\mathcal{C}}_{\substack{\chi DH \\ pr}}^{(6,1)} = & \; \frac{8}{3} g_1^2 Y_H \left[ N_c Y_q \mathcal{C}_{\substack{\chi q \\ prss}}^{(6,1)} + Y_\ell \mathcal{C}_{\substack{\chi \ell \\ prss}}^{(6,1)} \right] + \frac{4}{3} g_1^2 Y_H \left[ N_c Y_u \mathcal{C}_{\substack{\chi u \\ prss}}^{(6,1)} + N_c Y_d \mathcal{C}_{\substack{\chi d \\ prss}}^{(6,1)} + Y_e \mathcal{C}_{\substack{\chi e \\ prss}}^{(6,1)} \right] \nonumber \\
%     & - 2 N_c \mathcal{C}_{\substack{\chi q \\ prst}}^{(6,1)} \left[ \Gamma_d \Gamma_d^\dagger - \Gamma_u \Gamma_u^\dagger \right]_{ts} - 2 \mathcal{C}_{\substack{\chi \ell \\ prst}}^{(6,1)} \left[ \Gamma_e \Gamma_e^\dagger \right]_{ts} \nonumber \\
%     & - 2 N_c \mathcal{C}_{\substack{\chi u \\ prst}}^{(6,1)} \left[ \Gamma_u^\dagger \Gamma_u \right]_{st} + 2 N_c \mathcal{C}_{\substack{\chi d \\ prst}}^{(6,1)} \left[ \Gamma_d^\dagger \Gamma_d \right]_{st} + 2 \mathcal{C}_{\substack{\chi e \\ prst}}^{(6,1)} \left[ \Gamma_e^\dagger \Gamma_e \right]_{st}  \\
%     & + \left[ \frac{4}{3} g_1^2 Y_H^2 + 2 \gamma_H^{(\Gamma)} \right] \mathcal{C}_{\substack{\chi DH \\ pr}}^{(6,1)} \nonumber
% \end{align}
% 
\begin{align}
    \dot{\mathcal{C}}_{\chi DH}^{(6,1)} = & \; \frac{8}{3} g_1^2 Y_H \left[ N_c Y_q \mathcal{C}_{\chi \underset{ss}{q}}^{(6,1)} + Y_\ell \mathcal{C}_{\chi \underset{ss}{\ell}}^{(6,1)} \right] + \frac{4}{3} g_1^2 Y_H \left[ N_c Y_u \mathcal{C}_{\chi \underset{ss}{u}}^{(6,1)} + N_c Y_d \mathcal{C}_{\chi \underset{ss}{d}}^{(6,1)} + Y_e \mathcal{C}_{\chi \underset{ss}{e}}^{(6,1)} \right] \nonumber \\
    & - 2 N_c \mathcal{C}_{\chi \underset{st}{q}}^{(6,1)} \left[ \Gamma_d \Gamma_d^\dagger - \Gamma_u \Gamma_u^\dagger \right]_{ts} - 2 \mathcal{C}_{\chi \underset{st}{\ell}}^{(6,1)} \left[ \Gamma_e \Gamma_e^\dagger \right]_{ts} \nonumber \\
    & - 2 N_c \mathcal{C}_{\chi \underset{st}{u}}^{(6,1)} \left[ \Gamma_u^\dagger \Gamma_u \right]_{st} + 2 N_c \mathcal{C}_{\chi \underset{st}{d}}^{(6,1)} \left[ \Gamma_d^\dagger \Gamma_d \right]_{st} + 2 \mathcal{C}_{\chi \underset{st}{e}}^{(6,1)} \left[ \Gamma_e^\dagger \Gamma_e \right]_{st}  \\
    & + \left[ \frac{4}{3} g_1^2 Y_H^2 + 2 \gamma_H^{(\Gamma)} \right] \mathcal{C}_{\chi DH}^{(6,1)}  \,. \nonumber
\end{align}
 \end{subequations}

It is important to emphasize that the ADMs associated with the dimension-six axial-vector operators, $\mathcal{C}^{(6,2)}_{\chi\psi}$, are identical to those of the corresponding vector operators. This follows from the fact that the DM fermion $\chi$ is a gauge singlet and does not possess any renormalisable (dimension-four) interactions with the SM fields. Consequently, at $\mathcal{O}(1/\Lambda^2)$ there are no loop-induced wavefunction renormalisation effects involving the external $\chi$ legs, since such contributions necessarily require insertions of higher-dimensional operators and therefore arise only at higher orders in the EFT expansion. As a result, the DM bilinear acts merely as a spectator at this order, and the renormalisation of the operators is entirely governed by the SM fermion current. Therefore, the vector and axial-vector operators exhibit exactly the same one-loop Feynman topologies and, consequently, the same anomalous dimensions. Here, we present the ADMs for the case in which the same generation index is assigned to $\chi$. The same anomalous dimensions can also be applied to multicomponent dark matter, since the loop corrections affect only the SM part of the operators.

For precisely the same reason, the DMEFT operators do not mix with pure SMEFT operators at $\mathcal{O}(1/\Lambda^2)$. Any such mixing would require the generation of purely SM operators through loop diagrams containing insertions of DMEFT operators. However, since the dark matter field $\chi$ has no renormalisable interactions with the SM sector, closing the $\chi$ lines necessarily involves additional insertions of higher-dimensional operators, leading to further suppression by powers of $1/\Lambda$. Hence, DMEFT-SMEFT mixing can only arise at a higher order in the EFT expansion than the one considered here. Such effects are beyond the accuracy and scope of the present work, and therefore our analysis is consistently restricted to operator mixing at $\mathcal{O}(1/\Lambda^2)$, where no DMEFT-SMEFT mixing occurs.

%------------------------------------
\section{Framework for Multiscale Evolution}\label{sec:phenomenology}
%--------------------------------------
In this section, we discuss the phenomenological relevance of the DMEFT operators and the procedure for constraining their corresponding WCs using currently available experimental data. To this end, it is necessary to establish a connection between the DMEFT operator coefficients and the experimentally measurable observables. The relevant observables include not only low-energy measurements, such as flavour and precision observables, but also observables at the EW scale, including EWPO, Higgs production and decay rates, and other precision collider observables.

A systematic analysis requires determining the matching relations between the DMEFT operators and the effective operators relevant for the observables of interest. Since the electroweak-scale observables are conveniently described within the SMEFT, whereas those below the electroweak scale are more naturally formulated in the Low-Energy Effective Field Theory (LEFT), the analysis proceeds through a sequence of matching and RGE evolution steps.

Our phenomenological procedure follows a rigorous top-down approach. Specifically, we first compute the one-loop matching of the DMEFT operators of interest onto the SMEFT operators that contribute at tree level to the low-energy observables, EWPOs, and Higgs observables. This matching is performed at the mass threshold of the heavy particle, where the DM is integrated out. The resulting SMEFT WCs are then evolved down to the electroweak scale $\mu_{\rm EW}$ using the RGEs by utilising the one-loop anomalous dimension matrices of SMEFT, thereby accounting for operator mixing and resumming potentially large logarithmic corrections of the form $\ln(m_{\chi}/\mu_{\rm EW})$.
 
An alternative but formally equivalent procedure can also be adopted. In this approach, the DMEFT WCs are first evolved directly from the scale $\Lambda$ down to $\mu_{\rm EW}$ using the DMEFT RGEs as discussed in the previous section. The matching onto the relevant SMEFT operators is then performed at the electroweak scale, after which the resulting SMEFT coefficients are used to evaluate the observables. Since matching and running commute up to higher-order effects beyond the accuracy of the calculation, both approaches yield identical results when implemented consistently at the same perturbative order. Throughout this work, however, we adopt the former strategy, namely matching at the high scale followed by SMEFT running, as it provides a more transparent understanding of the SMEFT operators generated by the DMEFT interactions and allows for a direct comparison with existing SMEFT analyses and global-fit results.
 
At the scale $\mu_{\rm EW}$, the SMEFT coefficients are matched onto the corresponding WCs of the LEFT. The SMEFT-LEFT matching relations have been extensively studied and are available in the literature. Subsequently, the LEFT WCs are evolved from $\mu_{\rm EW}$ down to the characteristic low-energy scale $\mu_b$ using the known LEFT RGEs. This procedure enables us to express the WCs relevant for both electroweak-scale and low-energy observables entirely in terms of the DMEFT WCs defined at the scale $\Lambda$. Consequently, experimental constraints on a wide range of observables can be translated into bounds on the DMEFT parameter space in a consistent and systematic manner.
% 
%-----------------------------------
\begin{figure}[t]
    \centering
    \begin{tikzpicture}[
        node distance=1.5cm and 2cm,
        % Define styles for the boxes
        scalebox/.style={rectangle, draw=black!50, fill=gray!10, thick, minimum width=2.5cm, minimum height=1cm, align=center, font=\bfseries\small},
        processbox/.style={rectangle, draw=blue!70, fill=blue!5, thick, rounded corners, minimum width=4cm, minimum height=1cm, align=center, font=\small},
        decisionbox/.style={rectangle, draw=orange!80, fill=orange!10, thick, rounded corners, minimum width=4cm, minimum height=1cm, align=center, font=\bfseries\small},
        resultbox/.style={rectangle, draw=red!70, fill=red!5, thick, rounded corners, minimum width=5cm, minimum height=1cm, align=center, font=\bfseries\small},
        arrow/.style={-{Stealth[scale=1.2]}, thick, draw=blue!70},
        textlabel/.style={font=\scriptsize\itshape, align=center, color=black!80}
        ]

        % --- Nodes ---
        
        % Scales (Left Column)
        \node[scalebox] (scaleL) at (-5.5, 0) {High Scale \\ $\mu = \Lambda$};
        \node[scalebox] (scaleEW) at (-5.5, -3) {Weak Scale \\ $\mu = m_Z$};
        
        % Shifted Hadronic Scale further down (from -8 to -9.5)
        \node[scalebox] (scaleHad) at (-5.5, -9.5) {Hadronic Scale \\ $\mu \sim 2$ GeV};

        % Top Process
        \node[processbox, minimum width=5cm] (dmeft) at (2, 0) {New Physics / DMEFT Basis \\ $\mathcal{C}_i(\Lambda)$};

        % EW Scale Decision Box
        \node[decisionbox, minimum width=5cm] (ewmatch) at (2, -3) {EWSB Matching \\ (Integrate out $W, Z, t, H$)};

        % NEW: EWPO and Higgs Observables Box (placed to the right of the matching box)
        \node[resultbox, minimum width=3.5cm] (ewpo) at (8, -3) {EWPO \& Higgs \\ Observables};

        % Branching Scenarios (Heavy vs Light)
        \node[processbox, minimum width=4.5cm] (heavy) at (-1.5, -5.5) {Heavy DM ($M_{\text{DM}} \gtrsim m_Z$) \\ Integrate out DM \\ \textbf{Match to LEFT}};
        \node[processbox, minimum width=4.5cm] (light) at (5.5, -5.5) {Light DM ($M_{\text{DM}} \lesssim m_Z$) \\ Keep DM active \\ \textbf{Match to DLEFT} \\ \scriptsize (Accommodates inv. channels)};

        % Low Energy & Pheno (Shifted further down to match the scale)
        \node[processbox, minimum width=7cm] (hadronic) at (2, -9.5) {Non-perturbative QCD \\ (Hadronic Form Factors, Lattice Inputs)};
        \node[resultbox] (pheno) at (2, -12) {Low Energy Observables \\ (e.g., $B \to K^* \mu^+\mu^-$, Inv. Decays, Direct Detection)};

        % --- Arrows and Labels ---
        
        % High to EW
        \draw[arrow] (dmeft) -- node[right, textlabel] {RGE Evolution \\ (1-loop ADMs)} (ewmatch);
        
        % To EWPO and Higgs
        \draw[arrow] (ewmatch) -- node[above, textlabel] {} (ewpo);

        % The Split
        \draw[arrow] (ewmatch) -- node[above left=0.1cm, textlabel] {DM is Heavy} (heavy);
        \draw[arrow] (ewmatch) -- node[above right=0.1cm, textlabel] {DM is Light} (light);
        
        % Converging to Hadronic
        \draw[arrow] (heavy) -- node[below left=0.1cm, textlabel] {QED/QCD RGE \\ $\mathcal{C}_i \to \mathcal{L}_{\text{LEFT}}$} (hadronic);
        \draw[arrow] (light) -- node[below right=0.1cm, textlabel] {QED/QCD RGE \\ $\mathcal{C}_i \to \mathcal{L}_{\text{DLEFT}}$} (hadronic);
        
        % To Phenomenology
        \draw[arrow] (hadronic) -- node[right, textlabel] {Compute Amplitudes / Rates} (pheno);

        % --- Dashed Scale Lines ---
        \draw[dashed, draw=gray] (scaleL) -- (dmeft);
        \draw[dashed, draw=gray] (scaleEW) -- (ewmatch);
        \draw[dashed, draw=gray] (scaleHad) -- (hadronic);

    \end{tikzpicture}
    \caption{Schematic flowchart illustrating the phenomenological roadmap connecting the high-scale DMEFT basis to low-energy observables across diverse energy regimes.}
    \label{fig:RGE_flowchart}
\end{figure}
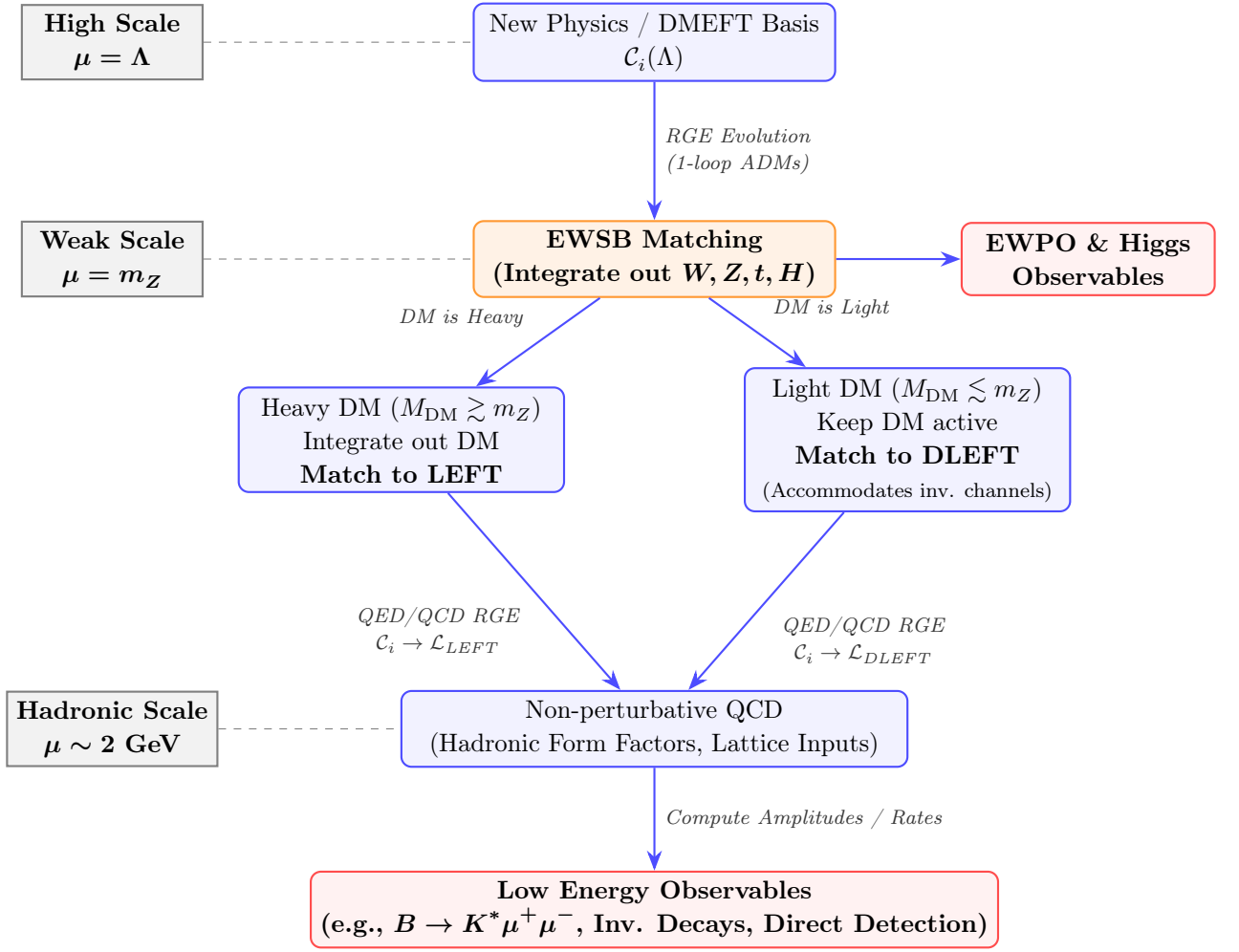
%-----------------------------------
To summarise, fig.~\ref{fig:RGE_flowchart} provides a schematic overview of the EFT framework used in our analysis. It illustrates the sequence of matching and RG-evolution steps that connect the DMEFT WCs defined at the high scale $\Lambda$ to the electroweak and low-energy observables evaluated at their respective characteristic scales.

%{\color{red}At the electroweak scale, we perform the matching associated with electroweak symmetry breaking (EWSB), integrating out the heavy SM degrees of freedom, namely the top quark, the Higgs boson, and the $W^\pm$ and $Z$ gauge bosons. The resulting low-energy effective interactions can then be probed directly through electroweak precision observables and high-energy Higgs measurements.}

Below the electroweak scale, the structure of the effective theory depends on the mass of the DM particle relative to the electroweak threshold. For heavy dark matter, $m_{\rm \chi}\gtrsim m_Z$, the DM field is not part of the low-energy spectrum and must be integrated out at an appropriate matching scale. To account for the associated threshold effects, we consider two representative benchmark scenarios. In the \textit{first benchmark}, $m_\chi=250~{\rm GeV}$, the DM mass lies close to the electroweak scale. In this case, a single-step matching procedure at $\mu_{\rm EW}$ is sufficient, where the DM particle is integrated out, yielding a direct matching of the DMEFT onto the LEFT.
In the \textit{second benchmark}, $m_\chi=800~{\rm GeV}$, the DM mass is significantly above the electroweak scale, necessitating a two-step matching procedure. The DMEFT WCs are defined at the cutoff scale $\Lambda=1~{\rm TeV}$ and evolved to the DM threshold $\mu_\chi=m_\chi$. Since $\Lambda$ and $\mu_\chi$ are relatively close, the DMEFT running in this interval is numerically small. At $\mu_\chi$, the DM field is integrated out, generating threshold corrections that match the DMEFT onto the SMEFT. The resulting SMEFT WCs are then evolved down to the electroweak scale using the SMEFT RGEs. Finally, at $\mu_{\rm EW}$, the heavy SM fields are integrated out, and the theory is matched onto the LEFT, whose WCs are subsequently evolved to the relevant low-energy scale using the known LEFT RGEs.
    
In the case of light DM ($m_{\mathrm{\chi}} \lesssim m_Z$), the dark matter particle remains an active degree of freedom below the electroweak scale. Consequently, the matching is performed onto the Dark Low-Energy Effective Field Theory (DLEFT), which retains the DM field in the low-energy description. This framework enables the study of phenomenological signatures involving missing energy, such as invisible decay channels. A more detailed discussion of these effects is presented in Sec.~\ref{sec:invisible}.

Finally, the resulting low-energy effective theory, either $\mathcal{L}_{\mathrm{LEFT}}$ or $\mathcal{L}_{\mathrm{DLEFT}}$, is evolved down to the characteristic hadronic scale, $\mu \sim 2~\mathrm{GeV}$, through QCD and QED renormalisation-group running. For the evolution of the SMEFT WCs from the new-physics scale $\Lambda$ to the electroweak scale, we employ the complete SMEFT anomalous dimension matrices presented in refs.~\cite{Jenkins:2013zja,Jenkins:2013wua,Alonso:2013hga}. Below the electroweak scale, the running of the LEFT WCs is performed using the anomalous dimensions and renormalisation-group equations derived in refs.~\cite{Aebischer:2017gaw,Jenkins:2017jig}. Wherever applicable, the running is carried out using the \texttt{DsixTools} package~\cite{Fuentes-Martin:2020zaz}, which implements the SMEFT and LEFT renormalisation-group equations using the aforementioned references.

%-----------------------------------

%--------------------------------------
\section{Observables and Phenomenological Constraints}
\label{sec:observables}
%--------------------------------------
Having established the RGEs and the framework of multiscale matching across the relevant energies, we now turn to the phenomenological implications of the DMEFT framework. To constrain the parameter space and assess the impact of operator mixing, we consider a broad set of observables sensitive to different sectors of the theory. For clarity, these observables are organised according to the characteristic energy scale of the process and the effective degrees of freedom governing the corresponding low-energy dynamics. 
%--------------------------------------
\subsection{Electroweak Precision Observables }
\label{subsec:EWPOs}
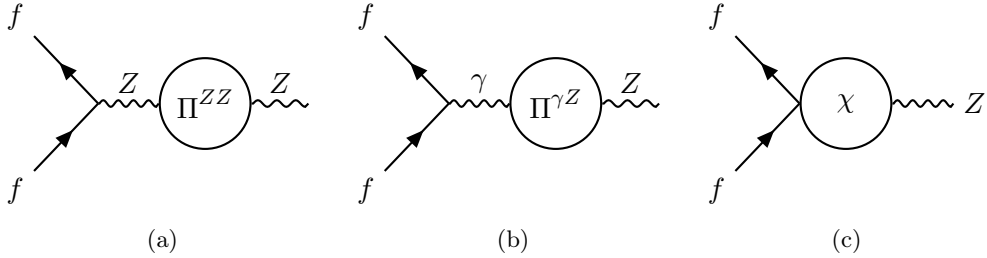
\begin{figure}[h!]
	\centering
	%-------------------
	\subfloat[]{
		\begin{tikzpicture}
			\begin{feynman}
				
				\vertex (a1);
				\vertex [above left=1.2cm of a1] (a2) {\(f\)};
				\vertex [below left=1.2cm of a1] (a3) {\(f\)};
				\vertex [thick, draw, circle,minimum size=12mm, right=0.8 cm of a1] (a4) {\(\Pi^{ZZ}\)};
				\vertex [right=1.5cm of a4] (a5) {};
				
				\diagram*{
					(a3) -- [fermion, thick] (a1) -- [fermion, thick] (a2),
					(a1) -- [boson, thick, edge label=\(Z\)] (a4) -- [boson, thick, edge label=\(Z\)] (a5),
				};
				
			\end{feynman}
	\end{tikzpicture} \label{fig:Feyn_EWPO1}}
	%----------------------
	\subfloat[]{
		\begin{tikzpicture}
			\begin{feynman}
				
				\vertex (a1);
				\vertex [above left=1.2cm of a1] (a2) {\(f\)};
				\vertex [below left=1.2cm of a1] (a3) {\(f\)};
				\vertex [thick, draw, circle,minimum size=12mm, right=0.8 cm of a1] (a4) {\(\Pi^{\gamma Z}\)};
				\vertex [right=1.5cm of a4] (a5) {};
				
				\diagram*{
					(a3) -- [fermion, thick] (a1) -- [fermion, thick] (a2),
					(a1) -- [boson, thick, edge label=\(\gamma\)] (a4) -- [boson, thick, edge label=\(Z\)] (a5),
				};
				
			\end{feynman}
		\end{tikzpicture} \label{fig:Feyn_EWPO2}
	}
	%----------------------
	\subfloat[]{
		\begin{tikzpicture}
			\begin{feynman}
				
				\vertex[thick, draw, circle,minimum size=12mm] (a1);
				\vertex [above left=1.2cm of a1] (a2) {\(f\)};
				\vertex [below left=1.2cm of a1] (a3) {\(f\)};
				\vertex [thick, draw, circle,minimum size=12mm, right=0.0 cm of a1] (a4) {\(\chi\)};
				\vertex [right=1.7cm of a4] (a5) {\(Z\)};
				
				\diagram*{
					(a3) -- [fermion, thick] (a1) -- [fermion, thick] (a2),
					(a4) -- [boson, thick] (a5),
				};
				
			\end{feynman}
		\end{tikzpicture} \label{fig:Feyn_EWPO3}
	}
	%----------------------
	\caption{Modified $Z$ boson couplings under the one-loop corrections.}
	\label{fig:Z_boson_correction}
	
\end{figure}

%--------------------------------------
%%%%%%%%%%%%%%%%%%%%%%%%%%%%%%%%%%%%%%%%%%%%%%%%%%%%%%%%%%%%%%%%%%%%
\begin{table}[t]
	\centering
	\footnotesize
	\renewcommand{\arraystretch}{1.25}
	\setlength{\tabcolsep}{8pt}
	\begin{tabular}{c c c c}
		\toprule
		\textbf{Observables} & \textbf{SM values} & \textbf{Experimental values} & \textbf{Definition} \\
		\midrule
		$\Gamma_Z$ (GeV) & $2.4946 \pm 0.0006$ & $2.4955 \pm 0.0023$ & $\sum_f \Gamma(Z\to f \bar{f})$\\
		$\sigma_{\rm had}$ (nb) & $41.4875 \pm 0.0069$ & $41.541\pm 0.037$ & $\frac{12 \pi}{m_Z^2}\frac{\Gamma(Z\to e^+ e^-)\Gamma(Z\to q \bar{q})}{\Gamma_Z^2}$ \\
		\midrule
		$A_e$ & $ 0.14684 \pm 0.00048 $& $ 0.1516 \pm 0.0021 $ & $\frac{\Gamma(Z\to e_L^+ e_L^-)-\Gamma(Z\to e_R^+ e_R^-)}{\Gamma(Z\to e^+ e^-)}$  \\
		$A_{\mu}$ & $0.14684 \pm 0.00048$ & $0.142 \pm 0.015$ &$\frac{\Gamma(Z\to \mu_L^+ \mu_L^-)-\Gamma(Z\to \mu_R^+ \mu_R^-)}{\Gamma(Z\to \mu^+ \mu^-)}$ \\
		$A_{\tau}$ & $0.14684 \pm 0.00048$ & $0.143 \pm 0.004$ & $\frac{\Gamma(Z\to \tau_L^+ \tau_L^-)-\Gamma(Z\to \tau_R^+ \tau_R^-)}{\Gamma(Z\to \tau^+ \tau^-)}$ \\
		$A_b$ & $0.93472 \pm 0.00004$ & $0.923 \pm 0.020$ &$\frac{\Gamma(Z\to b_L\bar{b}_L)-\Gamma(Z\to b_R\bar{b}_R)}{\Gamma(Z\to b\bar{b})}$\\
		\midrule
		$A^{\rm FB}_{e}$ & $(1.606 \pm 0.006)\%$ & $(1.45 \pm 0.25)\%$ & $\frac{3}{4}A_e^2$\\
		$A^{\rm FB}_{\mu}$ & $(1.617 \pm 0.011)\%$ & $(1.69 \pm 0.13)\%$& $\frac{3}{4}A_e A_{\mu}$\\
		$A^{\rm FB}_{\tau}$ &$(1.617 \pm 0.011)\%$ & $(1.88 \pm 0.17)\%$& $\frac{3}{4}A_e A_{\tau}$\\
		$A^{\rm FB}_b$ & $(10.295 \pm 0.034) \%$& $(9.92 \pm 0.16)\%$ &$\frac{3}{4}A_e A_b$\\
		\midrule
		$R_{e}$ & $20.736 \pm 0.010$ & $20.804 \pm 0.050$ & $\frac{\Gamma(Z \to q \bar{q})}{\Gamma(Z\to e^+ e^-)}$ \\
		$R_{\mu}$ & $20.736 \pm 0.010$ & $20.784 \pm 0.034$ & $\frac{\Gamma(Z \to q \bar{q})}{\Gamma(Z\to \mu^+ \mu^-)}$\\
		$R_{\tau}$ & $20.781 \pm 0.010$ & $20.764 \pm 0.045$ & $\frac{\Gamma(Z \to q \bar{q})}{\Gamma(Z\to \tau^+ \tau^-)}$\\
		$R_b$ & $0.21588 \pm 0.00010$ & $ 0.21629 \pm 0.00066 $ & $\frac{\Gamma(Z\to b\bar{b})}{\sum_q \Gamma(Z \to q \bar{q})}$\\
		\bottomrule
	\end{tabular}
	\caption{Updated SM predictions and experimental measurements of the $Z$-Pole Observables. All values are taken from Refs.~\cite{ALEPH:2005ab, Janot:2019oyi, dEnterria:2020cgt, SLD:2000jop, ParticleDataGroup:2024cfk, Reina:2025suh}.} \label{tab:Z_pole}
\end{table}
%%%%%%%%%%%%%%%%%%%%%%%%%%%%%%%%%%%%%%%%%%%%%%%%%%%%%%%%%%%%%%%%%%%%%%%%%%
\noindent To constrain the parameter space using EWPOs, we first determine the contributions of the relevant DMEFT operators to the $Z$-pole observables. The general effective Lagrangian parameterizing the $Zf\bar{f}$ interactions can be written as 
\begin{align}\label{eq:Z_eff_EWPO} 
	\mathcal{L}^{\rm EFT}_Z=-g_Z Z_{\mu}\left[\left(\left(T_f^3-Q_f s_{\rm eff}^2\right)\delta_{ij}+\delta g_L^{Zf}\right)\bar{f}_{iL} \gamma^{\mu} f_{jL}+\left(-Q_f s_{\rm eff}^2 \delta_{ij} +\delta g_R^{Zf} \right)\bar{f}_{iR} \gamma^{\mu} f_{jR}\right].
 \end{align}
 Here, $g_Z=\sqrt{g_1^2+g_2^2}$ denotes the neutral-current gauge coupling, while the tree-level SM couplings are given by $g_L^{\rm SM}=T_f^3-Q_f s_W^2$ and $g_R^{\rm SM}=-Q_f s_W^2$, with $s_W^2$ representing the weak mixing angle. New physics effects are encoded in the shifts $\delta g_L^{Zf}$, $\delta g_R^{Zf}$, and the effective weak mixing angle $s_{\rm eff}^2$. In the DMEFT framework considered here, these quantities receive no tree-level contributions and are generated only at the one-loop level through corrections to the effective $Zf\bar{f}$ interaction, as illustrated by the Feynman topologies in fig.~\ref{fig:Z_boson_correction}. The gauge-boson self-energy corrections shown in figs.~\ref{fig:Feyn_EWPO1} and \ref{fig:Feyn_EWPO2} modify the electroweak input parameters and induce universal shifts in $s_{\rm eff}^2$, whereas the vertex corrections depicted in fig.~\ref{fig:Feyn_EWPO3} generate flavour-dependent contributions to $\delta g_L^{Zf}$ and $\delta g_R^{Zf}$.

The radiatively corrected effective weak mixing angle, $s_{\rm eff}^2$, can be parameterized in terms of the transverse parts of the gauge-boson self-energies ($\Pi^{VV'}$) as:
\begin{equation}\label{eq:seff}
	s^2_{\rm eff} = s_W^2 + \frac{s_W^2 c_W^2}{c_W^2-s_W^2}\Pi_R - s_W c_W \frac{\Pi^{\gamma Z}(M_Z^2)}{M_Z^2} \,, 
\end{equation}
with
\begin{equation}
	\Pi_R = - \frac{\Pi^{\gamma\gamma}(M_Z^2)}{M_Z^2} + \frac{\Pi^{ZZ}(M_Z^2)}{M_Z^2} - \frac{\Pi^{WW}(0)}{M_W^2}.
\end{equation}
Here, $s_W$ and $c_W$ denote the sine and cosine of the tree-level weak mixing angle, respectively.  In our framework, the self energy corrections are associated the operators $\mathcal{O}_{\chi B}$ and $\mathcal{O}_{\chi DH}$, respectively, while the contributions in $\delta g_{L}^{Zf}$ and $\delta g_{R}^{Zf}$ arise from the simultaneous insertions of the dimension-5 and dimension-6 DMEFT operators introduced above.

These contributions can be tightly constrained by the precision measurements of the $Z$-pole observables, particularly through the partial-width ratios $R_f$ and the asymmetry observables $A_f$ and $A_f^{\rm FB}$, which are highly sensitive to deviations in the effective $Zf\bar{f}$ couplings. The full set of $Z$-pole observables employed in this analysis is summarised in Table~\ref{tab:Z_pole}, together with their theoretical definitions, SM predictions, and current experimental measurements. For the oblique parameters $S, T,$ and $U$, we have used the fitted values from the GFitter collaboration \cite{Baak:2014ora}.

\begin{align}
    S &= 0.03 \pm 0.10, \quad T = 0.05 \pm 0.12, \quad U = -0.03 \pm 0.09, \quad 
    \rho = \begin{pmatrix} 
    1 & +0.92 & -0.67 \\ 
    +0.92 & 1 & -0.88 \\ 
    -0.67 & -0.88 & 1 
    \end{pmatrix} \,.
\end{align}

%%%%%%%%%%%%%%%%%%%%%%%%%%%%%%%%%%%%%%%%%%%%%%%%%%%%

%-----------------------------------
\begin{table}[t]
	\centering
	\footnotesize
	\renewcommand{\arraystretch}{1.5}
	\begin{tabular}{c c c}
		\toprule
		\textbf{Scenario} & \textbf{$m_\chi = 250$ GeV} & \textbf{$m_\chi = 800$ GeV} \\
		\midrule
		$\mathcal{C}_{B\chi}^{(5,1)}$ & $(-1.98 \pm 1.62)\times 10^{-4}$ & $(1.61\pm 0.67) \times 10^{-4}$\\
		$\mathcal{C}_{B\chi}^{(5,2)}$ & $(1.79 \pm 1.49) \times 10^{-4}$ &$(1.08 \pm 0.38) \times 10^{-4}$ \\
		$\mathcal{C}_{\chi D H}^{(6,1)}$ & $(1.51 \pm 0.69) \times 10^{-3}$ & $(-2.09\pm 1.45) \times 10^{-3}$\\
		\midrule
		$\mathcal{C}_{B\chi}^{(5,1)},\mathcal{C}_{B\chi}^{(5,2)} $ & $(-0.83 \pm 0.32) \times 10^{-3},~(1.39 \pm 0.63)\times 10^{-3}$ & $(-2.11 \pm 0.49)\times 10^{-4},~(0.0 \pm 4.65)\times 10^{-3}$\\
		
		$\mathcal{C}_{B\chi}^{(5,1)},\mathcal{C}_{\chi D H}^{(6,1)} $ & $(2.30 \pm 1.17)\times 10^{-4},~(0.88\pm0.73)\times 10^{-3}$ & $(2.15 \pm 0.62)\times 10^{-4},~(0.11 \pm 1.10)\times 10^{-3}$\\
		
		$\mathcal{C}_{B\chi}^{(5,2)},\mathcal{C}_{\chi D H}^{(6,1)} $ & $(0.0 \pm 2.59)\times 10^{-4},~(1.28 \pm 0.82)\times 10^{-3}$& $(-1.29 \pm 0.32)\times 10^{-4},~ (-0.80 \pm 3.97)\times 10^{-3}$\\
		\bottomrule
	\end{tabular}
	\caption{Constraints on the DMEFT WCs from EWPOs for two benchmark values
of the DM mass. The top row shows the results when the same operator
contributes to both vertices, while the bottom row presents the bounds
when two different operators contribute to the two vertices.}
	\label{tab:EWPO_fit}
\end{table}
%--------------------------------

\begin{table}[t]
	\centering
	\footnotesize
	\renewcommand{\arraystretch}{1.5}
	\begin{tabular}{c c c c}
		\toprule
		\multicolumn{4}{c}{\textbf{$m_\chi = 800$ GeV}} \\
		\midrule
		\textbf{Scenario} & \textbf{Values} & \textbf{Scenario} & \textbf{Values} \\
		\midrule
		
		$\left[\mathcal{C}_{B \chi}^{(5,1)}, \mathcal{C}_{\chi \underset{11}{\ell}}^{(6,1)}\right] \times 10^4$ & $(2.33 \pm 0.94)\,,(-0.35 \pm 0.15)$ & $\left[\mathcal{C}_{B \chi }^{(5,1)}, \mathcal{C}_{\chi \underset{33}{\ell}}^{(6,1)}\right] \times 10^4$ & $(-1.74 \pm 1.02)\,,(0.12 \pm 0.16)$ \\
		
		$\left[\mathcal{C}_{B \chi}^{(5,1)}, \mathcal{C}_{\chi \underset{11}{e}}^{(6,1)}\right]\times 10^4$ & $(1.97 \pm 1.13)\,,(-5.73 \pm 4.37)$ & $\left[\mathcal{C}_{B \chi}^{(5,1)}, \mathcal{C}_{\chi \underset{33}{e}}^{(6,1)}\right] \times 10^4$ & $(1.38 \pm 1.29)\,,(0.32 \pm 0.39)$ \\
		
		$\left[\mathcal{C}_{\chi D H}^{(6,1)}, \mathcal{C}_{\chi \underset{11}{\ell}}^{(6,1)}\right] \times 10^3$ & $(0.0\pm 0.65)\,,(-0.11\pm 20.3)$ & $\left[\mathcal{C}_{\chi D H}^{(6,1)}, \mathcal{C}_{\chi \underset{33}{\ell}}^{(6,1)}\right]\times 10^3$ & $(0.0\pm 0.25)\,,(0.28\pm 46.8)$ \\
		
		$\left[\mathcal{C}_{\chi D H}^{(6,1)}, \mathcal{C}_{\chi \underset{11}{e}}^{(6,1)}\right]\times 10^3$ & $(0.01 \pm 0.48) \,,(-0.05 \pm 4.86)$ & $\left[\mathcal{C}_{\chi D H}^{(6,1)}, \mathcal{C}_{\chi \underset{33}{e}}^{(6,1)}\right] \times 10^3$ & $(0.0 \pm 1.31)\,,(-0.10 \pm 17.7)$ \\
		
		\cmidrule(lr){1-2} \cmidrule(lr){3-4}
		
		$\left[\mathcal{C}_{B \chi}^{(5,1)}, \mathcal{C}_{\chi \underset{22}{\ell}}^{(6,1)}\right]\times 10^4$ & $(1.5 \pm 1.18)\,,(-0.17 \pm 0.20)$ & $\left[\mathcal{C}_{B \chi}^{(5,1)}, \mathcal{C}_{\chi \underset{33}{q}}^{(6,1)}\right] \times 10^4$ & $(1.22 \pm 1.37)\,,(0.37 \pm 0.45)$ \\
		
		$\left[\mathcal{C}_{B \chi}^{(5,1)}, \mathcal{C}_{\chi \underset{22}{e}}^{(6,1)}\right]\times 10^4$ & $(1.47 \pm 1.20)\,, (0.21 \pm 0.25)$ & $\left[\mathcal{C}_{B \chi}^{(5,1)}, \mathcal{C}_{\chi \underset{33}{d}}^{(6,1)}\right] \times 10^4$ & $(1.61 \pm 1.02)\,,(1.36 \pm 1.05)$ \\
		
		$\left[\mathcal{C}_{\chi D H}^{(6,1)}, \mathcal{C}_{\chi \underset{22}{\ell}}^{(6,1)}\right] \times 10^3$ & $(-0.01\pm 1.21)\,,(0.08 \pm 17.1)$ & $\left[\mathcal{C}_{\chi D H}^{(6,1)}, \mathcal{C}_{\chi \underset{33}{q}}^{(6,1)}\right]\times 10^3$ & $(0.0\pm 3.81)\,,(0.13\pm 76.6)$ \\
		
		$\left[\mathcal{C}_{\chi D H}^{(6,1)}, \mathcal{C}_{\chi \underset{22}{e}}^{(6,1)}\right]\times 10^3$ & $(0.0 \pm0.93)\,,(-0.12\pm27.2)$ & $\left[\mathcal{C}_{\chi D H}^{(6,1)}, \mathcal{C}_{\chi \underset{33}{d}}^{(6,1)}\right] \times 10^3$ & $(0.01 \pm 0.32)\,,(-0.43 \pm 13.3)$ \\
		
		\bottomrule
	\end{tabular}
	\caption{Constraints on the WCs of the DMEFT operators from EWPOs, considering one four-fermion operator at the vertex, with \(\mu^{\rm ren}=1~\mathrm{TeV}\).}
	\label{tab:EWPOs}
\end{table}
%%%%%%%%%%%%%%%%%%%%%%%%%%%%%%%%%%%%%%%%%%%%%%%%%%%%%%%%%%%%

Furthermore, the operators $\mathcal{O}_{\chi B}$ and $\mathcal{O}_{\chi DH}$ contribute to the electroweak oblique parameters $S$, $T$, and $U$ through their effects on the gauge-boson vacuum-polarisation functions. These operators also induce corrections to the $W$-boson mass through the radiative parameter $\Delta r$, defined by 
\begin{align} \label{eq:deltar_exp}
	 M_W^2 \left( 1 - \frac{M_W^2}{M_Z^2} \right) = \frac{\pi \alpha_{em}}{\sqrt{2} G_F} \frac{1}{1 - \Delta r}. 
 \end{align} 
The quantity $\Delta r$ is sensitive to the oblique parameters and can be expressed as 
\begin{align} \label{eq:deltar_theo} 
	\Delta r=\Delta r_{\rm SM}+\frac{\alpha_{em}}{s_W^2}\left( \frac{S}{2}-c_W^2 T-\frac{c_W^2-s_W^2}{4s_W^2}U\right). 
\end{align} 

The explicit definitions of the oblique parameters in terms of the gauge-boson vacuum-polarisation amplitudes are provided in eq.~\eqref{eq:oblique_param} of Appendix~\ref{Append:EWPO_match}. To consistently incorporate these effects, we compute the transverse vacuum-polarisation functions $\Pi^{VV'}$ and express them in terms of the relevant DMEFT WCs. The corresponding analytical expressions and matching relations are also presented in Appendix~\ref{Append:EWPO_match}. For completeness, the experimental determinations of $\Delta r$ employed in our analysis are summarised in Table~\ref{tab:W_mass_deltar} of the same appendix. In summary, the numerical analysis incorporates the constraints arising from the $\Delta r$, $S$, $T$, and $U$ parameters, together with the full set of $Z$-pole observables discussed above.

We present our results from the EWPOs in two separate tables, namely Tables~\ref{tab:EWPO_fit} and~\ref{tab:EWPOs}. The results shown in the top row of Table~\ref{tab:EWPO_fit} correspond to scenarios in which only one operator is considered. The row below shows the results when two different operators are considered together. In this case, only the Feynman diagrams shown in figs.~\ref{fig:Feyn_EWPO1} and~\ref{fig:Feyn_EWPO2} contribute to the process. The relevant operators affect both the oblique parameters
($\Delta r$, $S$, $T$, and $U$) and the effective $Zf\bar f$ couplings,
thereby modifying the full set of EWPOs included in our fit. We obtain bounds on the WCs $\mathcal{C}_{B\chi}^{(5,1(2))}$ and $\mathcal{C}_{\chi DH}^{(6,1)}$, respectively. The results indicate that the allowed magnitudes of these coefficients are typically constrained to be at the level of $\lesssim 10^{-4}$ in both the single-operator and two-operator scenarios. We also observe that the constraint on $\mathcal{C}_{\chi DH}^{(6,1)}$ is somewhat weaker than that on $\mathcal{C}_{B\chi}^{(5,1(2))}$. This behaviour originates from the different operator normalisations and electroweak symmetry-breaking structures of $\mathcal{O}_{\chi DH}^{(6,1)}$ and $\mathcal{O}_{B\chi}^{(5,1(2))}$. In particular, the contributions associated with $\mathcal{O}_{\chi DH}^{(6,1)}$ involve explicit factors of the Higgs vacuum expectation value, leading to a different overall scaling of the induced self-energy and vertex corrections relative to those generated by $\mathcal{O}_{B\chi}^{(5,1(2))}$. Consequently, the EWPO observables exhibit different sensitivities to the corresponding WCs, resulting in the observed hierarchy in the derived bounds. The relevant analytical expressions illustrating this behaviour are provided in the Appendix.

In contrast, the results presented in Table~\ref{tab:EWPOs} correspond to fits in which one four-fermion operator is taken into account. Here, all three diagrams of fig.~\ref{fig:Z_boson_correction} will contribute. In such scenarios, additional contributions arise in the effective $Zf\bar{f}$ couplings through the shifts $\delta g_{L}^{Zf}$ and $\delta g_{R}^{Zf}$. The relevant two-operator combinations involving one dimension-5 and one dimension-6 operator are listed in the table. These fits allow us to constrain the WCs $\mathcal{C}_{\chi DH}^{(6,1)}$, $\mathcal{C}_{B\chi}^{(5,1(2))}$, and $\mathcal{C}_{\chi \underset{ii}{\ell}}^{(6,1)}$. As can be seen from Table~\ref{tab:EWPOs}, the coefficients $\mathcal{C}_{B\chi}^{(5,1(2))}$ are constrained at the level of $\lesssim 10^{-4}$, while $\mathcal{C}_{\overset{\chi \ell}{pq ii}}^{(6,1)}$ can be constrained to $\lesssim 10^{-5}$ when considered in combination with $\mathcal{C}_{B\chi}^{(5,1(2))}$. On the other hand, when combined with $\mathcal{C}_{\chi DH}^{(6,1)}$, relatively weaker limits are obtained, allowing values of $\mathcal{O}(10^{-2})$ for $\mathcal{C}_{\chi \underset{ii}{\ell}}^{(6,1)}$. This hierarchy follows the same pattern observed in Table~\ref{tab:EWPO_fit} and can likewise be attributed to the different normalisations and electroweak symmetry-breaking structures of the operators entering the fit. As a result, the EWPO observables exhibit a reduced sensitivity to operator combinations involving $\mathcal{O}_{\chi DH}^{(6,1)}$, leading to comparatively weaker bounds on the associated WCs.

We have extracted the results reported in Table~\ref{tab:EWPO_fit} for two benchmark DM masses, $m_\chi = 800$~GeV and $250$~GeV, in order to assess the dependence of the EWPO constraints on the dark matter mass. As expected, the benchmark point with $m_\chi = 800$~GeV yields relatively more precise constraints on the WCs, since the dark matter mass enters explicitly into the matching relations, as discussed in Appendix~\ref{Append:EWPO_match}. This demonstrates the non-trivial interplay between the matching scale and the resulting electroweak precision constraints. 

%It is important to note that for $m_{\chi}=250~\mathrm{GeV}$, we identify the renormalisation scale with the electroweak scale, whereas for $m_{\chi}=800~\mathrm{GeV}$, we set the high-energy renormalisation scale to $\mu=1~\mathrm{TeV}$. In both cases, this choice of scales prevents the appearance of large logarithmic corrections, perfectly maintaining the perturbativity of the theory. 

%It is evident from Table~\ref{tab:EWPO_fit} that increasing the DM mass results in a more constrained parameter space. This is because the vacuum polarisation amplitudes, $\Pi^{VV'}$, are directly dependent on the DM mass, as discussed in detail in Appendix~\ref{Append:EWPO_match}.
%------------------------------------

%---------------------------------
% \begin{table}[h]
% \footnotesize
% \centering
% \begin{tabular}{l c c c c}
% \toprule
% \makecell{\textbf{$\mu_{\rm ren}$}\\ (GeV)} &
% \makecell{\textbf{$m_{\chi}$}\\ (GeV)} &
% \makecell{\textbf{$\mathcal{C}_{B\chi}^{5,1}\times 10^4$}\\ $(\mathrm{GeV}^{-1})$} &
% \makecell{\textbf{$\mathcal{C}_{B\chi}^{5,2}\times 10^4$}\\ $(\mathrm{GeV}^{-1})$} &
% \makecell{\textbf{$\mathcal{C}_{\chi DH}^{6,1}\times 10^3$}\\ $(\mathrm{GeV}^{-2})$}
% \\
% \midrule
%  % & $1$ & $(0.0\pm 0.22)$&  $(0.0\pm 0.23)$ & $(0.0\pm 0.37)$\\
%  $\mu_{\rm EW}$ & $250$ & $(0.0\pm 57.2)$ & $(1.79 \pm 1.49)$ & $(0.0\pm 3.06)$ \\
% $\mu_{\Lambda}$ & $800$ & $(0.87\pm 2.04)$ & $(0.89 \pm 0.74)$ & $(0.0\pm 2.47)$\\
% \bottomrule
% \end{tabular}
% \caption{Constraints on NP operators from EWPOs.}
% \label{tab:EWPO_fit}
% \end{table}

\subsection{Low energy FCNC processes}\label{subsec:Meson_decay}
%--------------------------------------
In this subsection, we discuss the constraints on the DMEFT WCs arising from low-energy flavour-changing neutral current (FCNC) observables. Owing to the absence of tree-level FCNCs in the SM, such processes are highly suppressed and therefore provide a particularly sensitive probe of new physics contributions. In the present analysis, we focus on two classes of observables involving $B$-mesons, to which the considered DMEFT operators contribute at the one-loop level: the rare semileptonic decays, $b\to s(d)\ell^+\ell^-$ and neutral meson mixing observables. These processes probe complementary aspects of the underlying effective interactions and are sensitive to different combinations of WCs. The resulting constraints therefore play an important role in assessing the viability of the DMEFT parameter space.
 
\paragraph{\underline {\bf Semileptonic $B$ Meson Decays} :}

%--------------------------------------
%-------------------------------------
\begin{figure}[t!]
	\centering
	%----------------------
	\subfloat[]{
		\begin{tikzpicture}
			\begin{feynman}
				
				\vertex[thick, draw, circle,minimum size=12mm] (a1);
				\vertex [above left=1.2cm of a1] (a2) {\(d_i\)};
				\vertex [below left=1.2cm of a1] (a3) {\(b\)};
				\vertex [thick, draw, circle,minimum size=12mm, right=0.0 cm of a1] (a4) {\(\chi\)};
				\vertex [right=1.3cm of a4] (a5) ;
				\vertex[above right=1.2cm of a5] (a6) {\(\ell\)};
				\vertex[below right=1.2cm of a5] (a7) {\(\ell\)};
				
				\diagram*{
					(a3) -- [fermion, thick] (a1) -- [fermion, thick] (a2),
					(a4) -- [boson, thick,  edge label=\(\gamma/Z\)] (a5),
					(a7) --[fermion, thick](a5) --[fermion, thick](a6),
				};
			\end{feynman}
		\end{tikzpicture} \label{fig:Feyn_b2dill_1}
	}
	%----------------------
	\subfloat[]{
		\begin{tikzpicture}
			\begin{feynman}
				
				\vertex[thick, draw, circle,minimum size=12mm] (a1);
				\vertex [above left=1.2cm of a1] (a2) {\(d_i\)};
				\vertex [below left=1.2cm of a1] (a3) {\(b\)};
				\vertex [thick, draw, circle,minimum size=12mm, right=0.0 cm of a1] (a4) {\(\chi\)};
				\vertex [right=0.6cm of a4] (a5) ;
				\vertex[above right=1.2cm of a5] (a6) {\(\ell\)};
				\vertex[below right=1.2cm of a5] (a7) {\(\ell\)};
				
				\diagram*{
					(a3) -- [fermion, thick] (a1) -- [fermion, thick] (a2),
					% (a4) -- [boson, thick,  edge label=\(Z\)] (a5),
					(a7) --[fermion, thick](a5) --[fermion, thick](a6),
				};
			\end{feynman}
		\end{tikzpicture} \label{fig:Feyn_b2dill_2}
	}
	%----------------------
	\caption{Feynman diagrams contributing to semileptonic FCNC decays of $B$ meson in the presence of DMEFT operators.}
	\label{fig:b2dill}
	
\end{figure}
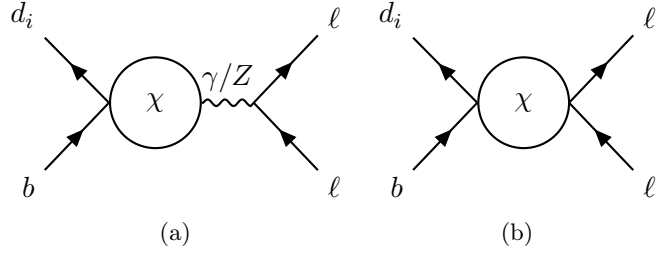
%-------------------------------------

The DMEFT operators considered in this work contribute to semileptonic decays of $B$-mesons, $b\to s(d)\ell^+\ell^-$, through the loop-induced Feynman topologies shown in fig.~\ref{fig:b2dill}. As illustrated in the figure, the relevant amplitudes arise from double insertions of DMEFT operators. This feature originates from the fact that the underlying FCNC transitions are generated through loop diagrams containing the DM particle propagating in the internal lines, thereby requiring the simultaneous insertion of two effective operators. Consequently, the resulting contributions depend on products of WCs rather than on individual coefficients. The various operator combinations relevant for our analysis, together with the corresponding WC products entering the flavour observables, are summarised in Table~\ref{tab:Rare_B_decays}.

The most general low-energy weak effective theory (WET) Hamiltonian describing the $b \to d_j \ell^+\ell^-$ transition is given by
%------------------------
\begin{equation} \label{eq:Heff_b2di}
	\mathcal{H}^{\text{eff}}_{b\to d_j\ell\ell} = - \frac{4\,G_F}{\sqrt{2}} V_{tb}V_{td_j}^\ast
	\left[ \sum_{i=1}^{6} C_i (\mu) O_i(\mu) + \sum_{i=7,8,9,10,P,S} \left(C_i (\mu) O_i + C'_i (\mu) O'_i\right)\right] \,.
\end{equation}
%------------------------
In our framework as discussed above, new physics contributions exclusively modify the vector and axial-vector WCs, namely $C_9^{(\prime)}$ and $C_{10}^{(\prime)}$. The corresponding semileptonic operators are defined as:
\begin{subequations}
	\begin{align}
		O_{9} &= \frac{e^2}{16 \pi^2} (\bar{d}_j \gamma_{\mu} P_L b)(\bar{\ell} \gamma^\mu \ell) \,,&
		O_{9}^\prime &= \frac{e^2}{16 \pi^2} (\bar{d}_j \gamma_{\mu} P_R b)(\bar{\ell} \gamma^\mu \ell) \,, \\
		O_{10} &=\frac{e^2}{16 \pi^2} (\bar{d}_j \gamma_{\mu} P_L b)( \bar{\ell} \gamma^\mu \gamma_5 \ell) \,,&
		O_{10}^\prime &=\frac{e^2}{16 \pi^2} (\bar{d}_j \gamma_{\mu} P_R b)( \bar{\ell} \gamma^\mu \gamma_5 \ell) \,.
\end{align} \end{subequations}

For completeness, the explicit matching relations bridging the high-scale DMEFT couplings to the WET WCs are provided in Appendix~\ref{Append:Rare_B_decay}, where we also discussed the connection of these WCs with observables.
To constrain the relevant parameter space, we perform a global $\chi^2$ analysis incorporating a comprehensive set of low-energy $b\to s(d)\ell^+\ell^-$ observables. The dataset includes angular observables \cite{LHCb:2015svh,LHCb:2020lmf}, differential branching fractions \cite{LHCb:2016ykl,CMS:2024syx}, branching ratios of rare decay modes \cite{CMS:2022mgd,LHCb:2021vsc}, lepton-flavour-universality-violating (LFUV) ratios $R_K$ and $R_{K^*}$ \cite{LHCb:2022qnv}, as well as CP asymmetries \cite{LHCb:2014mit}. The resulting constraints are summarised in Table~\ref{tab:Rare_B_decays} in the Appendix~\ref{Append:Rare_B_decay}, where we also provide a detailed discussion of the underlying phenomenology and the resulting bounds. 

As discussed earlier, the DMEFT contributions to the rare semileptonic transitions arise through loop-induced amplitudes involving double insertions of effective operators. Consequently, these observables are sensitive to products of WCs rather than to individual couplings. In particular, the channels receive contributions from combinations such as $\mathcal{C}_{\chi \underset{i3}{q(d)}}^{(6,2)} \mathcal{C}_{\chi \underset{22}{\ell}}^{(6,2)}$ and $\mathcal{C}_{\chi \underset{i3}{q(d)}}^{(6,2)} \mathcal{C}_{\chi \underset{22}{e}}^{(6,2)}$ (with $i=1,2$) through the topology shown in fig.~\ref{fig:Feyn_b2dill_2}. The decays are sensitive to the combination $\mathcal{C}_{\chi \underset{i3}{q(d)}}^{(6,2)} \mathcal{C}_{B\chi}^{(5,1)}$, which arise through the diagram in fig.~\ref{fig:Feyn_b2dill_2}. On the other hand, neutral meson mixing observables probe a complementary set of interactions. Since the corresponding one-loop amplitudes involve identical operator insertions at both vertices, the mixing amplitudes are proportional to the squares of the individual WCs. As a result, meson mixing observables provide direct constraints on individual flavour-changing couplings, while the semileptonic decays constrain their products. At the end of this subsection, we combine the information from rare semileptonic decays and neutral meson mixing observables to derive bounds on the individual WCs and to identify the most strongly constrained regions of the DMEFT parameter space.

%---------------------------------
\paragraph{\underline {\bf Neutral Meson Mixing} :} \label{subsec:Meson_Mixing}

The most general low-energy effective Hamiltonian describing $\Delta F=2$ transitions is expressed as
\begin{align}\label{eq:mixing_delF_2}
	\mathcal{H}_{\rm eff}^{\Delta F=2} = \sum_{i=1}^5 C_i(\mu) Q_i + \sum_{i=1}^3 \tilde{C}_i(\mu) \tilde{Q}_i + \text{h.c.} \,,
\end{align}
where the corresponding operator basis is given by
\begin{align}\label{eq:meson_mixing_basis}
	Q_1 &= (\bar{q}_j^\alpha \gamma_\mu P_L q_i^\alpha)(\bar{q}_j^\beta \gamma^\mu P_L q_i^\beta)\,, & \tilde{Q}_1 &= (\bar{q}_j^\alpha \gamma_\mu P_R q_i^\alpha)(\bar{q}_j^\beta \gamma^\mu P_R q_i^\beta) \,,\nonumber\\
	Q_2 &= (\bar{q}_j^\alpha P_L q_i^\alpha)(\bar{q}_j^\beta P_L q_i^\beta)\,, & \tilde{Q}_2 &= (\bar{q}_j^\alpha P_R q_i^\alpha)(\bar{q}_j^\beta P_R q_i^\beta) \,,\nonumber\\
	Q_3 &= (\bar{q}_j^\alpha P_L q_i^\beta)(\bar{q}_j^\beta P_L q_i^\alpha)\,, & \tilde{Q}_3 &= (\bar{q}_j^\alpha P_R q_i^\beta)(\bar{q}_j^\beta P_R q_i^\alpha) \,,\\
	Q_4 &= (\bar{q}_j^\alpha P_L q_i^\alpha)(\bar{q}_j^\beta P_R q_i^\beta)\,, & \nonumber\\
	Q_5 &= (\bar{q}_j^\alpha P_L q_i^\beta)(\bar{q}_j^\beta P_R q_i^\alpha) \, . \nonumber
\end{align}
Here, $q_{i,j}$ denote the quark fields, while $\alpha$ and $\beta$ are color indices. Appropriate choices of the external quark flavours correspond to the $B^0-\bar{B}^0$, $B_s^0-\bar{B}_s^0$ and $K^0-\bar{K}^0$ neutral meson mixing systems.

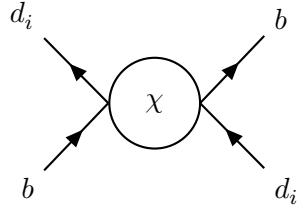
\begin{figure}[t!]
	\centering
	%----------------------
		\begin{tikzpicture}
			\begin{feynman}
				
				\vertex[thick, draw, circle,minimum size=12mm] (a1);
				\vertex [above left=1.2cm of a1] (a2) {\(d_i\)};
				\vertex [below left=1.2cm of a1] (a3) {\(b\)};
				\vertex [thick, draw, circle,minimum size=12mm, right=0.0 cm of a1] (a4) {\(\chi\)};
				\vertex [right=0.6cm of a4] (a5) ;
				\vertex[above right=1.2cm of a5] (a6) {\(b\)};
				\vertex[below right=1.2cm of a5] (a7) {\(d_i\)};
				
				\diagram*{
					(a3) -- [fermion, thick] (a1) -- [fermion, thick] (a2),
					% (a4) -- [boson, thick,  edge label=\(Z\)] (a5),
					(a7) --[fermion, thick](a5) --[fermion, thick](a6),
				};
			\end{feynman}
		\end{tikzpicture}
	%----------------------
	\caption{One-loop Feynman diagram contributing to $B^0$-$\bar{B}^0$ and $B_s^0-\bar{B}_s^0$ meson mixing in the presence of DMEFT operators.}
	\label{fig:mixing}	
\end{figure}

The DMEFT operators contribute to neutral meson mixing processes such as $B_q-\bar{B}_q$ mixing via the 1-loop level diagram shown in fig.~\ref{fig:mixing}. Therefore, the contribution to neutral meson mixing observables will originate from the double insertion of a relevant individual operator with the DM in the loop. In our analysis, these $\Delta F=2$ amplitudes are modified via the double insertion of a single underlying DMEFT operator (such as $\mathcal{O}_{\chi q}$ or $\mathcal{O}_{\chi d}$). The $B^0-\bar{B}^0$, $B_s^0-\bar{B}_s^0$ and $K^0-\bar{K}^0$ mixings are sensitive to $|\mathcal{C}_{\chi \underset{23}{q(d)}}^{(6,2)}|^2$, $|\mathcal{C}_{\chi \underset{13}{q(d)}}^{(6,2)}|^2$ and $|\mathcal{C}_{\chi \underset{23}{q(d)}}^{(6,2)}|^2$, respectively.
%----------------------------------
%-------------------------------------
\begin{table}[t!]
	\centering
	\footnotesize
	\renewcommand{\arraystretch}{1.25}
	\setlength{\tabcolsep}{8pt}
	\begin{tabular}{c c c c}
		\toprule
		\multirow{2}{*}{\textbf{Process} }& \multirow{2}{*}{\textbf{Observables}} & \textbf{SM values} & \textbf{Experimental values}  \\
		& & \textbf{$(\mathrm{ps^{-1}})$} & \textbf{$(\mathrm{ps^{-1}})$} \\
		\midrule
		$B^0 -\bar{B}^0$ & $\Delta M_d$ & $(0.533^{+0.022}_{-0.036})$ ~\cite{Saha:2003tq,FermilabLattice:2016ipl}& $(0.5062 \pm 0.0019)$~\cite{HFLAV:2022esi}\\
		$B_s^0 -\bar{B}_s^0$ &$\Delta M_s$ & $(18.4^{+0.7}_{-1.2})$~\cite{Saha:2003tq,FermilabLattice:2016ipl} & $(17.765 \pm 0.006)$~\cite{HFLAV:2022esi}\\
		$K^0-\bar{K}^0 $ &$\Delta M_K$ & $(5.80 \pm 2.37)\times 10^{-3}$~\cite{ParticleDataGroup:2024cfk} & $(5.289 \pm 0.001) \times 10^{-3}$~\cite{ParticleDataGroup:2024cfk}\\
		\bottomrule
	\end{tabular}
	\caption{SM predictions and experimental measurements of the meson mixing observable $\Delta M$ .}
	\label{tab:Meson_mixing_obs}
\end{table}

%--------------------------------
% 
%--------------------------------------
\begin{table}[t]
	\centering
	\footnotesize
	\renewcommand{\arraystretch}{1.5}
	\begin{tabular}{c c c}
		\toprule
		\textbf{Scenario} & \textbf{$m_\chi = 800$ GeV} & \textbf{$m_\chi = 250$ GeV} \\
		\midrule
		$|\mathcal{C}_{\chi \underset{23}{q(d)}}^{(6,2)}|$ & $(1.15 \pm 0.61) \times 10^{-8}$ & $(1.74 \pm 0.92 ) \times 10^{-8}$ \\
		
		$|\mathcal{C}_{\chi \underset{13}{q(d)}}^{(6,2)}|$ &  $(2.89 \pm 1.13) \times 10^{-9}$&$(4.35 \pm 1.70) \times 10^{-9}$ \\
		
		$|\mathcal{C}_{\chi \underset{12}{q(d)}}^{(6,2)}|$ & $(2.00 \pm 4.25 )\times 10^{-9}$ & $(3.01 \pm 6.40 ) \times 10^{-9}$\\

		$|\mathcal{C}_{\chi \underset{22}{\ell}}^{(6,2)}|$ & $(2.28\pm 2.06) \times 10^{-7}$& $(3.43 \pm 3.13) \times 10^{-7}$ \\
		
		$|\mathcal{C}_{\chi \underset{22}{e}}^{(6,2)}|$ & $(4.04 \pm 6.97)\times 10^{-7}$& $(0.61 \pm 1.05) \times 10^{-6}$ \\
		
		% \rowcolor{magenta!10}
		$|\mathcal{C}_{B \chi}^{(5,1)}|$ & $(7.91 \pm 6.33) \times 10^{-4}$ & $ (3.70 \pm 2.96) \times 10^{-4} $ \\
		
		\bottomrule
	\end{tabular}
	\caption{Constraints on DMEFT WCs from combined analysis of meson mixing and rare semileptonic decays.}
	\label{tab:meson_mixing}
\end{table}
%--------------------------------------
Since our theoretical framework restricts the new physics to vector and axial-vector interactions, this double-insertion mechanism naturally modifies only the $Q_1$ and $\tilde{Q}_1$ operators. The neutral meson mixing observable, mass difference $\Delta M$, is defined as:  
\begin{align}
\Delta M = 2\left|M_{12}\right| = \frac{\left| \mathcal{M}\right|}{m_{P}} \,,
\end{align}
with $m_P$ is the mass of the corresponding pseudoscalar meson. Assuming the NP contributions do not introduce new CP-violating phases, we can express the total mass difference as the linear sum of the SM and NP contributions:
\begin{align}\Delta M^{\text{tot}} = \Delta M^{\text{SM}} + \Delta M^{\text{NP}} = \Delta M^{\text{SM}}\left( 1+\Delta\right) .\end{align}
In order to quantify the NP contributions, we define the ratio $\Delta \equiv \Delta M^{\text{NP}}/\Delta M^{\text{SM}}$. We will evaluate this ratio and compare our estimates against current experimental measurements \cite{Kolay:2024wns, Kolay:2025jip, Bhattacharya:2025mlg, Kolay:2026wca}. The primary advantage of constructing this observable is that it partially factors out leading sources of uncertainty, such as bag factors and decay constants, thereby mitigating theoretical uncertainties associated with the bag parameters. 
$\Delta M_{\rm SM}$ is the contribution coming from the SM, can be expressed as: 
\begin{equation}
	\Delta M_{\rm SM}^{q} =  \frac{G_{F}^2}{6 \pi ^2}  M_W^2 m_{B_{q}} f_{B_{q}}^2 B_{B_{q}} \eta_{B_{q}} |V_{tq}^{*} V_{tb} |^2 S_{0}(x_t) \,,
\end{equation}
where $S_{0}(x_t)$ is the loop factor, known as the Inami-Lim function \cite{Inami:1980fz}, defined by: 
\begin{equation}\label{eq:mixing_SM}
	S_{0}(x_t) = x_{t} \left( \frac{1}{4} + \frac{9}{4} \frac{1}{1-x_{t}} - \frac{3}{2} \frac{1}{(1-x_{t})^2 }\right)-\frac{3}{2}\frac{x_t^3\log x_t}{(1-x_t)^3}\,,
\end{equation}
with $x_{t} = m_{t}^2/M_W^2$. $f_{B_{q}} $ is the decay constant defined as below: 
\begin{equation}
	\langle 0 | \bar{q}_{d} \gamma_{\mu} b | B (p) \rangle = i f_{B_{q}} p_{\mu}\,.
\end{equation}
$B_{B_{q}}$ is the bag factor and $\eta_{B_q}$ is the QCD correction at next-to-leading order \cite{Gay:2000utx}. In the presence of the DMEFT operators, the amplitude for the meson mixing can be written as:
\begin{align}
    \mathcal{M}^{\rm NP}_q = \left\langle B_q^0 \left| \left( \Delta C_1^{qb} Q_1^{qb} + \Delta \tilde{C}_1^{qb} \tilde{Q}_1^{qb}\right) \right| \bar{B}_q^0 \right\rangle \,,
\end{align}
where, $\Delta C_1^{qb}, \Delta \tilde{C}_1^{qb} $ are effective couplings corresponding to the operators, defined in eq.~\eqref{eq:meson_mixing_basis}. The matching relation for the DMEFT WCs can be found in Appendix~\ref{Append:Meson_mixing} for different meson-mixing processes. The SM predictions of the relevant mixing amplitudes and their respective measured values are shown in Table~\ref{tab:Meson_mixing_obs}.

Using the available data on the rare semileptonic decays $b\to s(d)\ell^+\ell^-$ together with the neutral meson mixing observables, we perform a global $\chi^2$ minimisation to constrain the relevant DMEFT WCs. The resulting bounds are summarised in Table~\ref{tab:meson_mixing}. Since the neutral meson mixing amplitudes depend on the squares of the individual flavour-changing couplings, while the rare semileptonic decays are primarily sensitive to products of WCs, the combined analysis allows us to disentangle the individual operator contributions and derive direct constraints on the corresponding WCs.

The bounds obtained on the dimension-5 coefficient $\mathcal{C}_{B\chi}^{(5,1)}$ are found to be comparable to those derived from the electroweak precision observables discussed in the previous section, thereby demonstrating the consistency and complementarity of the two approaches. We further observe that the neutral meson mixing observables provide significantly stronger constraints on the flavour-changing quark-DM operators than the rare semileptonic decay data alone. This can be seen by comparing the limits obtained on the quark-DM operators, which contribute directly to the meson mixing amplitudes, with those involving lepton-DM interactions, whose effects are probed predominantly through the semileptonic decays. Owing to the high precision of the meson mixing measurements and the quadratic dependence of the mixing amplitudes on the relevant WCs, these observables emerge as the dominant source of flavour constraints in a large region of the parameter space.

To illustrate the dependence of the constraints on the dark matter mass, we present the results for two benchmark points, namely $m_\chi = 250$~GeV and $m_\chi = 800$~GeV. We find that the resulting bounds are largely unchanged between the two scenarios, indicating a relatively mild sensitivity of the low-energy flavour observables to the dark matter mass within the considered parameter range. This behaviour suggests that the flavour constraints are primarily driven by the underlying operator structure and the precision of the experimental measurements, rather than by the specific choice of the benchmark dark matter mass.

% %--------------------------------------
 \subsection{Lepton Flavour Violating Observables}
%--------------------------------------
In this section, we investigate a wide array of lepton flavour violating (LFV) processes sensitive to the DMEFT WCs. These include various mesonic LFV decays, both leptonic ($P \to \ell_1 \ell_2$) and semileptonic ($P \to M \, \ell_1 \ell_2$) transitions, with $P$ and $M$ being different mesons, as well as $\tau$, $Z$, and top-quark LFV processes. Since, no evidence of LFV decay sare found, various experiments provide strict upper limit on such processes. Also, since LFV decays are forbidden in the SM, any definitive observation in these channels would constitute an unambiguous signature of NP. Here, we constrain the allowed parameter space for the LFV couplings by leveraging these experimental upper limits.
To provide a structured analysis, we will systematically examine each of these individual LFV sectors in the following subsections.

%-------------------------------------
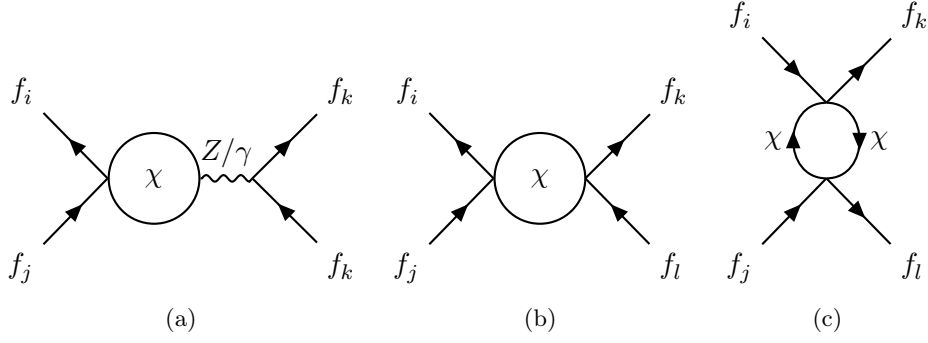
\begin{figure}[t]
\centering
%----------------------
\subfloat[]{
\begin{tikzpicture}
\begin{feynman}

\vertex[thick, draw, circle,minimum size=12mm] (a1);
\vertex [above left=1.2cm of a1] (a2) {\(f_i\)};
\vertex [below left=1.2cm of a1] (a3) {\(f_j\)};
\vertex [thick, draw, circle,minimum size=12mm, right=0.0 cm of a1] (a4) {\(\chi\)};
\vertex [right=1.3cm of a4] (a5) ;
\vertex[above right=1.2cm of a5] (a6) {\(f_k\)};
\vertex[below right=1.2cm of a5] (a7) {\(f_k\)};

\diagram*{
(a3) -- [fermion, thick] (a1) -- [fermion, thick] (a2),
(a4) -- [boson, thick,  edge label=\(Z/\gamma\)] (a5),
(a7) --[fermion, thick](a5) --[fermion, thick](a6),
};
\end{feynman}
\end{tikzpicture}
}
%----------------------
\subfloat[]{
\begin{tikzpicture}
\begin{feynman}

\vertex[thick, draw, circle,minimum size=12mm] (a1);
\vertex [above left=1.2cm of a1] (a2) {\(f_i\)};
\vertex [below left=1.2cm of a1] (a3) {\(f_j\)};
\vertex [thick, draw, circle,minimum size=12mm, right=0.0 cm of a1] (a4) {\(\chi\)};
\vertex [right=0.6cm of a4] (a5) ;
\vertex[above right=1.2cm of a5] (a6) {\(f_k\)};
\vertex[below right=1.2cm of a5] (a7) {\(f_l\)};

\diagram*{
(a3) -- [fermion, thick] (a1) -- [fermion, thick] (a2),
% (a4) -- [boson, thick,  edge label=\(Z\)] (a5),
(a7) --[fermion, thick](a5) --[fermion, thick](a6),
};
\end{feynman}
\end{tikzpicture}
}
%-----------------------
\subfloat[]{
\begin{tikzpicture}
\begin{feynman}

% Top contact interaction vertex
\vertex[] (vtop);
\vertex [above left=1.2cm of vtop] (fi) {\(f_i\)};
\vertex [above right=1.2cm of vtop] (fk) {\(f_k\)};

% Bottom contact interaction vertex
\vertex[ below=1.cm of vtop] (vbot) ;
\vertex [below left=1.2cm of vbot] (fj) {\(f_j\)};
\vertex [below right=1.2cm of vbot] (fl) {\(f_l\)};

\diagram*{
% Top external fermion lines
(fi) -- [fermion, thick] (vtop) -- [fermion, thick] (fk),

% Bottom external fermion lines
(fj) -- [fermion, thick] (vbot) -- [fermion, thick] (fl),

% The internal loop connecting the two contact interactions
    (vtop) -- [fermion, thick, half left, edge label=\(\chi\)] (vbot),
    (vbot) -- [fermion, thick, half left, edge label=\(\chi\)] (vtop),};
\end{feynman}
\end{tikzpicture}
}
%----------------------
\caption{Leading-order topologies contributing to the different LFV processes evaluated in this analysis.}
\label{fig:LFV_topologies}

\end{figure}
%-------------------------------------

% Unlike typical FCNC decays and meson mixing processes, these LFV transitions are mediated via $s$-channel dark matter topologies where the DM particles run inside the loop. In Appendix~\ref{}, we provide a comprehensive, sector-by-sector discussion of these LFV observables alongside their explicit matching relations onto the LEFT WCs. Because these processes are sensitive to the DM mass, we focus exclusively on the $m_\chi = 800\,\mathrm{GeV}$ benchmark scenario for brevity. The resulting bounds on the various products of couplings are summarised by sector in the tables below. We categorize these constraints into $B$ and $K$ meson leptonic and semileptonic decays (Tables~\ref{tab:B_LFV_leptonic} and \ref{tab:B_LFV_semileptonic}), charged LFV in purely leptonic $\tau$ decays (Table~\ref{tab:tau_cLFV}), hadronic $\tau$ LFV transitions (Table~\ref{tab:tau_LFV_Had}), $Z$ boson LFV (Table~\ref{tab:ZLFV}), and top-quark LFV (Table~\ref{tab:tLFV}).
%-----------------------------------
\subsubsection{Lepton Flavour Violating Meson Decays}
%----------------------------------
Rare meson decays involving both FCNC and LFV transitions provide excellent probes of physics beyond the SM. While FCNC processes are highly suppressed in the SM, occurring only at the loop level via the GIM mechanism, LFV decays are essentially forbidden, with only negligible contributions induced by neutrino masses. Consequently, the observation of such decays would provide a clear signature of NP. Moreover, the stringent experimental upper limits on their branching fractions make these channels particularly sensitive to constraining NP scenarios. In our framework, the LFV meson decays are generated at one loop by inserting two dimension-6 operators. The corresponding Feynman diagram is mediated via an s-channel dark matter topology, where the DM particle run inside the loop, as shown in fig.~\ref{fig:b2dill}. The most general effective Lagrangian describing these transitions can be written as:
\begin{equation}
    \mathcal{L}_{P\to \ell_{1} \ell_2} = \sum_{i,j=L,R} C_{V_{ij}} (\bar{q_{2}} \gamma_{\mu} P_{i}q_{1})(\bar{\ell_{2}} \gamma^{\mu} P_{j}\ell_{1}) + C_{S_{ij}} (\bar{q_{2}}  P_{i}q_{1})(\bar{\ell_{2}} P_{j}\ell_{1})\,.
\end{equation}
In our case, the tensor operator does not arise at the one-loop level; hence, we have not written the effective interactions for it. In the massless limit of the final state leptons, only the vector operators survive. The scalar contributions will be mass-suppressed by the final state leptons' mass. The branching ratio will be given by (only considering vector contributions): 
\begin{align}
\Gamma(P\to \ell_1\ell_2) & =
\frac{f_P^2\,\lambda^{1/2}(m_{P}^2,m_1^2,m_2^2)}{64\pi m_P^3}
\left[ -\left(C_{VaL}^2+C_{VaR}^2\right)\left(m_1^2-m_2^2\right)^2   \right.  \\ & \left. 
+ \left( -4C_{VaL}C_{VaR}\,m_1 m_2 +C_{VaL}^2\left(m_1^2+m_2^2\right) +C_{VaR}^2\left(m_1^2+m_2^2\right) \right)m_P^2 \right]. \nonumber
\end{align}
The couplings can be defined as: $C_{VaL} = (C_{V_{LL}} - C_{V_{RL}})$ and $C_{VaR} = (C_{V_{LR}} - C_{V_{RR}})$. The expressions obtained for the effective couplings $C_{ij}$ are given in the Appendix~\ref{Append:B_LFV}. As mentioned, with our choice of operators, these decays arise at one-loop level with two dimension-6 operator insertions. Hence, we will only obtain independent bounds on their multiplication. The bounds on the coupling combinations are given in Table~\ref{tab:B_LFV_leptonic}.

%---------------------------------
% Table B LFV Leptonic
\begin{table}[htbp]
\footnotesize
\centering
\renewcommand{\arraystretch}{1.5}
\label{tab:NPconstraints}
\begin{tabular}{l c c c}
\toprule
\textbf{Observable} &
\textbf{Constraint} &
\textbf{Scenario} & 
\textbf{Values} 
\\
\midrule
\multirow{2}{*}{$\mathcal{B}(B_s \to e^{\pm}\mu^{\mp})$} & \multirow{2}{*}{$<5.4\times 10^{-9}$ \cite{LHCb:2017hag}} 
& $\mathcal{C}_{\chi \underset{23}{q(d)}}^{(6,1)}\,\mathcal{C}_{\chi \underset{12}{\ell(e)}}^{(6,1)}$ & $< 4.47\times 10^{-7}$  \\
& & $\mathcal{C}_{\chi \underset{23}{q(d)}}^{(6,2)}\,\mathcal{C}_{\chi \underset{12}{\ell(e)}}^{(6,2)}$ & $< 3.35 \times 10^{-12}$ \\
\midrule
%---------------------------------
\multirow{2}{*}{$\mathcal{B}(B_s \to e^{\pm}\tau^{\mp})$} & \multirow{2}{*}{$<1.4\times 10^{-3}$ \cite{Belle:2023jwr}} 
& $\mathcal{C}_{\chi \underset{23}{q(d)}}^{(6,1)}\,\mathcal{C}_{\chi \underset{13}{\ell(e)}}^{(6,1)}$ & $<1.53 \times 10^{-5}$ \\
& & $\mathcal{C}_{\chi \underset{23}{q(d)}}^{(6,2)}\,\mathcal{C}_{\chi \underset{13}{\ell(e)}}^{(6,2)}$ & $<1.15 \times 10^{-10}$ \\
\midrule 
%-------------------------
\multirow{2}{*}{$\mathcal{B}(B_s \to \mu^{\pm}\tau^{\mp})$} & \multirow{2}{*}{$<4.2\times 10^{-5}$ \cite{LHCb:2019ujz}} 
& $\mathcal{C}_{\chi \underset{23}{q(d)}}^{(6,1)}\,\mathcal{C}_{\chi \underset{23}{\ell(e)}}^{(6,1)}$ & $<2.63 \times 10^{-6}$ \\
& & $\mathcal{C}_{\chi \underset{23}{q(d)}}^{(6,2)}\,\mathcal{C}_{\chi \underset{23}{\ell(e)}}^{(6,2)}$ & $<1.98 \times 10^{-11}$ \\
\midrule
%-------------------------
\multirow{2}{*}{$\mathcal{B}(B^0\to e^{\pm}\mu^{\mp})$} & \multirow{2}{*}{$<1.3\times 10^{-9}$ \cite{LHCb:2017hag}} 
& $\mathcal{C}_{\chi \underset{13}{q(d)}}^{(6,1)}\,\mathcal{C}_{\chi \underset{12}{\ell(e)}}^{(6,1)}$ & $<2.78 \times 10^{-7}$ \\
& & $\mathcal{C}_{\chi \underset{13}{q(d)}}^{(6,2)}\,\mathcal{C}_{\chi \underset{12}{\ell(e)}}^{(6,2)}$ & $<2.02 \times 10^{-12}$ \\ 
\midrule
%-------------------------
\multirow{2}{*}{$\mathcal{B}(B^0\to e^{\pm}\tau^{\mp})$} & \multirow{2}{*}{$<1.6\times 10^{-5}$ \cite{Belle:2021rod}} 
& $\mathcal{C}_{\chi \underset{13}{q(d)}}^{(6,1)}\,\mathcal{C}_{\chi \underset{13}{\ell(e)}}^{(6,1)}$ & $<2.07\times 10^{-6}$ \\
& & $\mathcal{C}_{\chi \underset{13}{q(d)}}^{(6,2)}\,\mathcal{C}_{\chi \underset{13}{\ell(e)}}^{(6,2)}$ & $<1.51\times 10^{-11}$ \\
\midrule
%-------------------------
\multirow{2}{*}{$\mathcal{B}(B^0\to \mu^{\pm}\tau^{\mp})$} & \multirow{2}{*}{$<1.4\times 10^{-5}$ \cite{LHCb:2019ujz}} 
& $\mathcal{C}_{\chi \underset{13}{q(d)}}^{(6,1)}\,\mathcal{C}_{\chi \underset{23}{\ell(e)}}^{(6,1)}$ & $<2.07 \times 10^{-6}$ \\
& & $\mathcal{C}_{\chi \underset{13}{q(d)}}^{(6,2)}\,\mathcal{C}_{\chi \underset{23}{\ell(e)}}^{(6,2)}$ & $<1.51 \times 10^{-11}$ \\
\midrule
%-------------------------
\multirow{2}{*}{$\mathcal{B}(K_{L} \to e^{\pm} \mu^{\mp})$} & \multirow{2}{*}{$< 4.7 \times 10^{-12}$ \cite{BNL:1998apv}} 
& $\mathrm{Im}\big[\mathcal{C}_{\chi \underset{12}{q(d)}}^{(6,1)}\,\mathcal{C}_{\chi \underset{12}{\ell(e)}}^{(6,1)}\big]$ & $<1.94 \times 10^{-8}$ \\
& & $\mathrm{Im}\big[\mathcal{C}_{\chi \underset{12}{q(d)}}^{(6,2)}\,\mathcal{C}_{\chi \underset{12}{\ell(e)}}^{(6,2)}\big]$ & $<6.41\times 10^{-15}$ \\
%------------------------
\bottomrule
\end{tabular}
\caption{Constraints on the products of DMEFT WCs derived from $B$ and $K$ meson LFV decays. The values in the table are presented for $m_{\chi} = 800 $ GeV and $\mu^{\rm ren} = 1$ TeV.}
\label{tab:B_LFV_leptonic}
\end{table}
%--------------------------------

%------------------------------
% B semileptonic LFV
\begin{table}[htbp!]
\footnotesize
\centering
\renewcommand{\arraystretch}{1.5}
\begin{tabular}{l c c c}
\toprule
\textbf{Observable} &
\textbf{Constraint} &
\textbf{Scenario} & 
\textbf{Values} 
\\
\midrule
$\mathcal{B}(B_s \to \phi \mu^\pm e^\mp)$ &  $<19.8\times 10^{-9}$ \cite{LHCb:2022lrd} & $\mathcal{C}_{\chi \underset{23}{q(d)}}^{(6,2)}\mathcal{C}_{\chi \underset{12}{\ell(e)}}^{(6,2)} $    & $<6.26 \times 10^{-14}$
 \\
\midrule
$\mathcal{B}(B^+ \to K^+ \mu^- e^+)$ & 
$<9.5\times 10^{-9}$ \cite{LHCb:2019bix} & $\mathcal{C}_{\chi \underset{23}{q(d)}}^{(6,2)}\mathcal{C}_{\chi \underset{12}{\ell(e)}}^{(6,2)} $  & $ < 5.70 \times 10^{-14}$
 \\
 % $\mathcal{B}(B^+ \to K^+ \mu^- e^+)$ & $<9.5\times 10^{-9}$ \cite{LHCb:2019bix} & $\mathcal{C}_{\chi \underset{23}{q(d)}}^{(6,2)}\mathcal{C}_{\chi \underset{12}{\ell(e)}}^{(6,2)} $  & $ < 1.29 \times 10^{-13}$ \\
\midrule
$\mathcal{B}(B^+ \to K^+ \mu^+ e^-)$ & $<8.8\times 10^{-9}$ \cite{LHCb:2019bix} & $\mathcal{C}_{\chi \underset{23}{q(d)}}^{(6,2)}\mathcal{C}_{\chi \underset{21}{\ell(e)}}^{(6,2)} $ & $< 5.48 \times 10^{-14}$ \\
% $\mathcal{B}(B^+ \to K^+ \mu^+ e^-)$ & $<8.8\times 10^{-9}$ \cite{LHCb:2019bix} & $\mathcal{C}_{\chi \underset{23}{q(d)}}^{(6,2)}\mathcal{C}_{\chi \underset{21}{\ell(e)}}^{(6,2)} $ & $< 1.24 \times 10^{-14}$ \\
\midrule
$\mathcal{B}(B^0 \to K^{*0} \mu^- e^+)$ & $<7.9\times 10^{-9}$ \cite{LHCb:2022lrd} & $\mathcal{C}_{\chi \underset{23}{q(d)}}^{(6,2)}\mathcal{C}_{\chi \underset{21}{\ell(e)}}^{(6,2)} $ & $< 3.69 \times 10^{-14} $
 \\
\midrule
$\mathcal{B}(B^0 \to K^{*0} \mu^+ e^-)$ & 
$<6.9\times 10^{-9}$ \cite{LHCb:2022lrd} & $\mathcal{C}_{\chi \underset{23}{q(d)}}^{(6,2)}\mathcal{C}_{\chi \underset{12}{\ell(e)}}^{(6,2)} $  & $< 3.45 \times 10^{-14}$
 \\
% $\mathcal{B}(B^+ \to \pi^{+} \mu^\pm e^\mp)$ & $<1.8\times 10^{-9}$ \cite{LHCb:2026luq} & & \\

% \midrule 
% \rowcolor{red!10}
% $\mathcal{B}(B^0 \to K^{*0} \mu^\pm e^\mp)$ & $<11.7\times 10^{-9}$ \cite{LHCb:2022lrd} & &  \\
\bottomrule
\end{tabular}
\caption{Constraints on the products of DMEFT WCs derived from semileptonic $B$ meson LFV decays. $m_{\chi} = 800$ GeV and $\mu^{\rm ren} = 1$ TeV.}
\label{tab:B_LFV_semileptonic}
\end{table}
%--------------------------------

%-----------------------------------
\subsubsection{Charged Lepton Flavour Violating decays}
%----------------------------------
We now consider lepton-flavour-violating decays of charged leptons. The relatively large mass of the $\tau$ provides a rich phenomenological testing ground, allowing us to probe both purely leptonic and semi-hadronic LFV decay modes. Conversely, while kinematically restricted, LFV transitions of the $\mu$ lepton are bounded by exceptionally stringent experimental limits, providing highly complementary constraints on our parameter space.

\paragraph{\textbf{Leptonic LFV:}}  For the purely leptonic transitions of the form $\ell_i\to \ell_j\ell_k\bar{\ell}_l$, the decay topologies can be classified into three distinct subclasses depending on the flavour and charge composition of the final-state leptons:
\begin{itemize}
    \item [(a)] \textbf{Three leptons of the same flavour:} $\mu^{\pm}\to e^{\pm}e^+e^-$, $\tau^{\pm}\to e^{\pm}e^+e^-$, and $\tau^{\pm}\to \mu^{\pm}\mu^+\mu^-$.
    \item [(b)] \textbf{Three distinguishable leptons:} $\tau^{\pm}\to e^{\pm}\mu^+\mu^-$ and $\tau^{\pm}\to\mu^{\pm} e^+ e^-$. 
    \item [(c)] \textbf{Charge-mismatched leptons:} Two leptons with the same flavour and charge, and one with a different flavour and opposite charge, i.e., $\tau^{\pm}\to e^{\mp}\mu^{\pm}\mu^{\pm}$ and $\tau^{\pm}\to \mu^{\mp}e^{\pm}e^{\pm}$. The amplitude for this channel is generated by the topology depicted in fig.~\ref{fig:LFV_topologies}(c).
\end{itemize}

The general amplitude for these purely leptonic decays can be parameterised in terms of effective four-fermion interactions. The matrix element is written as
\begin{align}
    \mathcal{M}_0 &= \sum_{\overset{X,Y=L,R}{A=S,V,T}} \mathcal{C}^A_{XY} \mathcal{O}^A_{XY} \,,
\end{align}
where the explicit chiral structures for the vector ($V$), scalar ($S$), and tensor ($T$) operators are defined as
\begin{align}
    \mathcal{O}^V_{XY} &= \left(\bar{u}(p_j)\gamma^{\mu}P_X\, u(p_i)\right)\left(\bar{u}(p_k)\gamma_{\mu}P_Y\,v(p_l)\right) \,, \\
    \mathcal{O}^S_{XY} &= \left(\bar{u}(p_j)P_X\, u(p_i)\right)\left(\bar{u}(p_k)P_Y\,v(p_l)\right) \,, \\
    \mathcal{O}^T_{XY} &= \delta_{XY}\left(\bar{u}(p_j)\sigma^{\mu\nu}P_X\, u(p_i)\right)\left(\bar{u}(p_k)\sigma_{\mu\nu}P_Y\,v(p_l)\right) \,.
\end{align}
Squaring this amplitude and integrating over the three-body phase space yields the general expression for the branching fraction:
\begin{align}
    \mathcal{B}(\ell_i\to\ell_j\ell_k\bar{\ell}_{l}) &= \frac{S_n\,m_i^5}{6144 \pi^3 \Gamma_i} \Bigg[ |\mathcal{C}^S_{LL}|^2 + |\mathcal{C}^S_{LR}|^2 + |\mathcal{C}^S_{RL}|^2 + |\mathcal{C}^S_{RR}|^2 \nonumber\\
    &\quad + 4 \left( |\mathcal{C}^V_{LL}|^2 + |\mathcal{C}^V_{LR}|^2 + |\mathcal{C}^V_{RL}|^2 + |\mathcal{C}^V_{RR}|^2 \right) \nonumber\\
    &\quad + 48 \left( |\mathcal{C}^T_{LL}|^2 + |\mathcal{C}^T_{RR}|^2 \right) + \mathcal{B}_{\gamma} \Bigg] \,.
\end{align}
Here, $S_n=\frac{1}{n!}$ is the symmetry factor accounting for $n$ identical particles in the final state, and $\mathcal{B}_{\gamma}$ represents the complementary three-body branching fraction mediated by a photon dipole transition. It is worth mentioning that topologies involving an intermediate $Z$ boson are also kinematically allowed for final states containing matching lepton flavours, namely the $\tau (\mu) \to e^-e^+e^-$, $\tau \to \mu^-\mu^+\mu^-$, $\tau \to e^-\mu^+\mu^-$ and $\tau \to \mu^-e^+e^-$ processes, and as shown in fig.~\ref{fig:LFV_topologies}(b). However, because the $Z$ boson is heavily suppressed at the tau mass scale ($q^2 \ll M_Z^2$), its propagator effectively collapses to a local interaction and is integrated out, mapping directly into the WCs of the vector-type four-fermion operators. However, these contributions are highly suppressed in our analysis, as they are proportional to the masses of the final-state leptons and are further suppressed by powers of $1/M_Z^2$.

Crucially, within our specific theoretical framework, the new physics contributions exclusively generate vector-type effective operators. Consequently, all scalar ($\mathcal{C}^S_{XY}$), tensor ($\mathcal{C}^T_{XY}$) WCs as well as the contribution due to photon dipole transition, $\mathcal{B}_{\gamma}$, strictly vanish. This drastically simplifies the decay rate, rendering these purely leptonic LFV observables directly sensitive only to the vector couplings $\mathcal{C}^V_{XY}$. For completeness, the explicit matching relations for the surviving vector couplings, $\mathcal{C}^V_{XY}$, in terms of the underlying DMEFT WCs are provided in Appendix~\ref{Append:lepton_cLFV} for each of the $\tau$ LFV decay topologies discussed in this section. We present the resulting bounds on the products of the DMEFT WCs in Table~\ref{tab:tau_cLFV}, assuming a dark fermion mass of $m_{\chi} = 800~\mathrm{GeV}$.
%------------------------------
% tau cLFV
\begin{table}[t]
\footnotesize
\centering
\renewcommand{\arraystretch}{1.5}
\begin{tabular}{l c c c}
\toprule
\textbf{Observable} &
\textbf{Constraint} &
\textbf{Scenario} & 
\textbf{Values} 
\\
\midrule
\multirow{2}{*}{$\mathcal{B}(\tau \to e e e)$} & 
 \multirow{2}{*}{$< 2.7 \times 10^{-8}$ \cite{Hayasaka:2010np}} & 
$\mathcal{C}_{\chi \underset{13}{\ell(e)}}^{(6,2)}\,\mathcal{C}_{\chi \underset{11}{\ell(e)}}^{(6,2)}$ & 
$< 2.23 \times 10^{-12}$ \\
& & $\mathcal{C}_{\chi \underset{13}{\ell(e)}}^{(6,1)} \mathcal{C}_{B\chi}^{(5,1)}$ & $< 6.49 \times 10^{-9}$\\
\midrule
\multirow{2}{*}{$\mathcal{B}(\tau \to \mu \mu \mu)$} & 
\multirow{2}{*}{$< 1.9 \times 10^{-8}$ \cite{LHCb:2026eod} }& 
$\mathcal{C}_{\chi \underset{23}{\ell(e)}}^{(6,2)}\,\mathcal{C}_{\chi \underset{22}{\ell(e)}}^{(6,2)}$ & 
$< 1.87 \times 10^{-12}$ \\
& & $\mathcal{C}_{\chi \underset{23}{\ell(e)}}^{(6,1)} \mathcal{C}_{B\chi}^{(5,1)}$ & $<5.45 \times 10^{-9}$\\
\midrule
\multirow{2}{*}{$\mathcal{B}(\mu \to e e e)$} & 
\multirow{2}{*}{$< 1.0 \times 10^{-12}$ \cite{SINDRUM:1987nra}} & 
$\mathcal{C}_{\chi \underset{12}{\ell(e)}}^{(6,2)}\,\mathcal{C}_{\chi \underset{11}{\ell(e)}}^{(6,2)}$ & 
$< 3.39 \times 10^{-16}$ \\
& & $\mathcal{C}_{\chi \underset{12}{\ell(e)}}^{(6,1)} \mathcal{C}_{B\chi}^{(5,1)}$ & $<2.71 \times 10^{-9}$ \\
\midrule
\multirow{2}{*}{$\mathcal{B}(\tau^\pm \to e^\pm \mu^+ \mu^-)$} & 
\multirow{2}{*}{$< 2.7 \times 10^{-8}$ \cite{Hayasaka:2010np}} & 
$\mathcal{C}_{\chi \underset{13}{\ell(e)}}^{(6,2)}\,\mathcal{C}_{\chi \underset{22}{\ell(e)}}^{(6,2)}$ & 
$< 4.46 \times 10^{-12}$ \\
& & $\mathcal{C}_{\chi \underset{13}{\ell(e)}}^{(6,1)} \mathcal{C}_{B\chi}^{(5,1)}$ & $<1.03\times 10^{-8}$\\
\midrule
\multirow{2}{*}{$\mathcal{B}(\tau^\pm \to \mu^\pm e^+ e^-)$} & 
\multirow{2}{*}{$< 1.8 \times 10^{-8}$ \cite{Hayasaka:2010np}} & 
$\mathcal{C}_{\chi \underset{23}{\ell(e)}}^{(6,2)}\,\mathcal{C}_{\chi \underset{11}{\ell(e)}}^{(6,2)}$ & 
$< 3.64 \times 10^{-12}$ \\
& & $\mathcal{C}_{\chi \underset{23}{\ell(e)}}^{(6,1)} \mathcal{C}_{B\chi}^{(5,1)}$ & $< 0.97 \times 10^{-8}$\\
\midrule
$\mathcal{B}(\tau^\pm \to e^\mp \mu^\pm \mu^\pm)$ & 
$< 1.7 \times 10^{-8}$ \cite{Hayasaka:2010np} & 
$\mathcal{C}_{\chi \underset{23}{\ell(e)}}^{(6,2)}\,\mathcal{C}_{\chi \underset{12}{\ell(e)}}^{(6,2)}$ & 
$< 3.54 \times 10^{-12}$ \\
\midrule
$\mathcal{B}(\tau^\pm \to \mu^\mp e^\pm e^\pm)$ & 
$< 1.5 \times 10^{-8}$ \cite{Hayasaka:2010np} & 
$\mathcal{C}_{\chi \underset{13}{\ell(e)}}^{(6,2)}\,\mathcal{C}_{\chi \underset{21}{\ell(e)}}^{(6,2)}$ & 
$< 3.33 \times 10^{-12}$ \\
\bottomrule
\end{tabular}
\caption{Constraints on the products of DMEFT WCs derived from charged Lepton Flavour Violating $\tau$  decays for $m_{\chi}=800~\text{GeV}$.}
\label{tab:tau_cLFV}
\end{table}
%---------------------------------

\paragraph{\textbf{Hadronic $\tau$ LFV:}}
%------------------------------
In addition to the purely leptonic channels, the heavy mass of the $\tau$ lepton kinematically allows for a variety of semi-hadronic LFV transitions. In this analysis, we systematically investigate two-body decays featuring a charged lepton and a single neutral meson, specifically evaluating both vector ($V = \rho^0, \phi$) and pseudoscalar ($P = \pi^0$) final states.

\begin{itemize}
    \item \textbf{$\tau \to \ell V (\rho,\phi)$:}
    In the WET, the effective Lagrangian describing these hadronic decays via vector operators can be written as~\cite{Aebischer:2018iyb}:
%---------------------------------
\begin{align}
    \mathcal{L}_{\text{eff}} \subset \sum_{\ell \in \{e,\mu\}, q \in \{u,d,s\}} \Big\{ 
    & C_{VLL}^{\tau\ell qq} \, (\bar{\ell}_L \gamma^\mu \tau_L)(\bar{q}_L \gamma_\mu q_L) \nonumber\\
    & + C_{VLR}^{\tau\ell qq} \, (\bar{\ell}_L \gamma^\mu \tau_L)(\bar{q}_R \gamma_\mu q_R) \\
    & + C_{VLR}^{qq\tau\ell} \, (\bar{\ell}_R \gamma^\mu \tau_R)(\bar{q}_L \gamma_\mu q_L) \nonumber\\
    & + C_{VRR}^{\tau\ell qq} \, (\bar{\ell}_R \gamma^\mu \tau_R)(\bar{q}_R \gamma_\mu q_R) 
\Big\} + \text{h.c.} \nonumber
\end{align}
%---------------------------------
In our analysis, the new physics strictly generates semi-leptonic vector-type interactions, meaning there are no contributions from dipole or tensor operators. Consequently, the general branching fraction simplifies to~\cite{Aebischer:2018iyb}:
\begin{align}
    \mathcal{B}(\tau\to \ell V) &= \tau_{\tau}\frac{\lambda^{1/2}(m_{\tau}^2,m_{\ell}^2,m_V^2)}{128\pi\,m_{\tau}^3}m_V^2 f_V^2 \nonumber\\ 
    &\quad \times \left\{ \left( \left| g_L^{\tau\ell V} \right|^2 + \left| g_R^{\tau\ell V} \right|^2 \right) \left( \frac{(m_\tau^2 - m_\ell^2)^2}{m_V^2} + m_\tau^2 + m_\ell^2 - 2\,m_V^2 \right) \right. \nonumber\\
    &\quad \left. - 12\,m_\tau m_\ell \, \text{Re} \left( g_R^{\tau\ell V} \left(g_L^{\tau\ell V}\right)^* \right) \right\} \,.
\end{align}
%--------------------------------
For the $V=\phi$ meson, the vacuum-to-meson matrix element of the quark vector current is defined as
\begin{align}
    \langle\phi|\bar{s}\gamma_{\mu}s|0\rangle &= m_{\phi}f_{\phi}\epsilon_{\mu}^* \,,
\end{align}
%--------------------------------
where $f_{\phi}$ is the $\phi$ decay constant and $m_{\phi}$ is the $\phi$ mass. The corresponding effective couplings are given by
%--------------------------
\begin{align}
    g_L^{\tau\ell\phi} &= \left(C_{V\,LL}^{\tau\ell ss}+C_{V\,LR}^{\tau\ell ss}\right)\,, \quad &
    g_R^{\tau\ell\phi} &= \left(C_{V\,RR}^{\tau\ell ss}+C_{V\,LR}^{ss\tau\ell}\right)\,.
\end{align}
%--------------------------
For the $V=\rho$ meson, the relevant matrix element is
%--------------------------
\begin{align}
    \left\langle\rho\left|\frac{\bar{u}\gamma_{\mu}u-\bar{d}\gamma_{\mu}d}{\sqrt{2}}\right|0\right\rangle &= m_{\rho} f_{\rho}\epsilon_{\mu}^* \,,
\end{align}
%--------------------------
and the corresponding effective couplings are given by
%--------------------------
\begin{align}
    g_L^{\tau\ell\rho} &= \left( \frac{C_{VLL}^{\tau\ell uu} - C_{VLL}^{\tau\ell dd}}{\sqrt{2}} + \frac{C_{VLR}^{\tau\ell uu} - C_{VLR}^{\tau\ell dd}}{\sqrt{2}} \right)\,, 
    g_R^{\tau\ell\rho} =  \left( \frac{C_{VRR}^{\tau\ell uu} - C_{VRR}^{\tau\ell dd}}{\sqrt{2}} + \frac{C_{VLR}^{uu\tau\ell} - C_{VLR}^{dd\tau\ell}}{\sqrt{2}} \right)\,.
\end{align}
%--------------------------

%---------------------------
    \item \textbf{$\tau \to \ell P(\pi^0)$:} 
    For the $\tau\to \ell \pi^0$ transition, the relevant hadronic matrix elements are given by
%--------------------------
\begin{equation}
\begin{aligned}
    \langle \pi^0 | \bar{u}\gamma_\mu\gamma_5 u | 0 \rangle &= \frac{i f_\pi p_{\pi\mu}}{\sqrt{2}} \,, & \langle \pi^0 | \bar{d}\gamma_\mu\gamma_5 d | 0 \rangle &= -\frac{i f_\pi p_{\pi\mu}}{\sqrt{2}} \,, \\
    \langle \pi^0 | \bar{u}\gamma_5 u | 0 \rangle &= \frac{i f_\pi m_\pi^2}{\sqrt{2}(m_u + m_d)} \,, & \langle \pi^0 | \bar{d}\gamma_5 d | 0 \rangle &= -\frac{i f_\pi m_\pi^2}{\sqrt{2}(m_u + m_d)} \,,
\end{aligned}
\end{equation}
where $f_{\pi}$ is the neutral pion decay constant. The corresponding effective couplings evaluate to
%--------------------------
\begin{align}
    g_L^{\tau\ell\pi^0} &= -m_{\ell}\left(\frac{C_{V\,LR}^{\tau\ell uu}-C_{V\,LL}^{\tau\ell uu}}{2}-\frac{C_{V\,LR}^{\tau\ell dd}-C_{V\,LL}^{\tau\ell dd}}{2}\right)\nonumber\\
    &\quad + m_{\tau}\left(\frac{C_{V\,RR}^{\tau\ell uu}-C_{V\,LR}^{uu\tau\ell}}{2}-\frac{C_{V\,RR}^{\tau\ell dd}-C_{V\,LR}^{dd\tau\ell}}{2}\right) \,, \\
    g_R^{\tau\ell\pi^0} &= +m_{\tau}\left(\frac{C_{V\,LR}^{\tau\ell uu}-C_{V\,LL}^{\tau\ell uu}}{2}-\frac{C_{V\,LR}^{\tau\ell dd}-C_{V\,LL}^{\tau\ell dd}}{2}\right)\nonumber\\
    &\quad -m_{\ell}\left(\frac{C_{V\,RR}^{\tau\ell uu}-C_{V\,LR}^{uu\tau\ell}}{2}-\frac{C_{V\,RR}^{\tau\ell dd}-C_{V\,LR}^{dd\tau\ell}}{2}\right) \,.
\end{align}
%--------------------------
The resulting branching fraction is expressed as
%---------------------------
\begin{align}
    \mathcal{B}(\tau\to \ell \pi^0) &= \tau_{\tau} \frac{\lambda^{1/2}(m_{\tau}^2,m_{\ell}^2,m_{\pi^0}^2)}{32\pi\,m_{\tau}^3}f_{\pi}^2 \nonumber\\
    &\quad \times \left\{ \frac{1}{2} \left( \left| g_L^{\tau\ell \pi^0} \right|^2 + \left| g_R^{\tau\ell \pi^0} \right|^2 \right) (m_\tau^2 + m_\ell^2 - m_{\pi^0}^2) \right.\nonumber\\
    &\quad \left. + 2\,m_\ell m_\tau \text{Re}\left( g_L^{\tau\ell \pi^0} \left(g_R^{\tau\ell \pi^0}\right)^* \right) \right\} \,.
\end{align}
%---------------------------
\end{itemize}

Unless otherwise specified, all numerical values for the masses, decay constants, and lifetimes utilised in this section are adopted from the PDG~\cite{ParticleDataGroup:2024cfk}.
The explicit matching relations with DMEFT WCs are given in the Appendix~\ref{Append:lepton_cLFV}. Table~\ref{tab:tau_LFV_Had} summarises the derived upper bounds on the products of the DMEFT WCs for a benchmark dark fermion mass of $m_{\chi} = 800~\mathrm{GeV}$.
%----------------------------
% tau Hadronic LFV
\begin{table}[t]
\footnotesize
\centering
\renewcommand{\arraystretch}{1.5}
\begin{tabular}{l c c c}
\toprule
\textbf{Observable} &
\textbf{Constraint} &
\textbf{Scenario} & 
\textbf{Values} 
\\
\midrule
$\mathcal{B}(\tau\to \mu~\phi)$ & $< 2.3 \times 10^{-8}$ \cite{Belle:2023ziz} & $\mathcal{C}_{\chi \underset{32}{\ell(e)}}^{(6,2)}~\mathcal{C}_{\chi \underset{22}{q(d)}}^{(6,2)}$ & $<1.12 \times 10^{-12}$\\
\midrule
$\mathcal{B}(\tau\to  e~\phi)$ & $< 2.0 \times 10^{-8}$ \cite{Belle:2023ziz} & $\mathcal{C}_{\chi \underset{31}{\ell(e)}}^{(6,2)}~\mathcal{C}_{\chi \underset{22}{q(d)}}^{(6,2)}$ & $<6.31 \times 10^{-13}$\\
\midrule
\multirow{2}{*}{$\mathcal{B}(\tau\to \mu~\rho)$} & \multirow{2}{*}{$< 1.7 \times 10^{-8}$ \cite{Belle:2023ziz}} & $\mathcal{C}_{\chi \underset{32}{\ell(e)}}^{(6,2)}~\mathcal{C}_{\chi \underset{11}{u}}^{(6,2)}$ & $<0.77 \times 10^{-12}$ \\
& &  $\mathcal{C}_{\chi \underset{32}{\ell(e)}}^{(6,2)}~\mathcal{C}_{\chi \underset{11}{d}}^{(6,2)}$ & $< 0.77 \times 10^{-12}$ \\
\midrule
\multirow{2}{*}{$\mathcal{B}(\tau\to  e~\rho)$} & \multirow{2}{*}{$< 2.2 \times 10^{-8}$ \cite{Belle:2023ziz} }& $\mathcal{C}_{\chi \underset{31}{\ell(e)}}^{(6,2)}~\mathcal{C}_{\chi \underset{11}{u}}^{(6,2)}$ & $< 0.88 \times 10^{-12}$\\
&  & $\mathcal{C}_{\chi \underset{31}{\ell(e)}}^{(6,2)}~\mathcal{C}_{\chi \underset{11}{d}}^{(6,2)}$ & $< 0.88 \times 10^{-12}$\\
\midrule
\multirow{2}{*}{$\mathcal{B}(\tau\to \mu~\pi^0)$} & \multirow{2}{*}{$< 1.1 \times 10^{-7}$ \cite{BaBar:2006jhm}} & $\mathcal{C}_{\chi \underset{32}{\ell(e)}}^{(6,2)}~\mathcal{C}_{\chi \underset{11}{u}}^{(6,2)}$ & $<3.20\times 10^{-12}$\\
&  & $\mathcal{C}_{\chi \underset{32}{\ell(e)}}^{(6,2)}~\mathcal{C}_{\chi \underset{11}{d}}^{(6,2)}$ & $<3.20\times 10^{-12}$\\
\midrule
\multirow{2}{*}{$\mathcal{B}(\tau\to  e~\pi^0)$} & \multirow{2}{*}{$< 8.0 \times 10^{-8}$ \cite{Belle:2007cio}} & $\mathcal{C}_{\chi \underset{31}{\ell(e)}}^{(6,2)}~\mathcal{C}_{\chi \underset{11}{u}}^{(6,2)}$ & $<2.73 \times 10^{-12}$ \\
&  & $\mathcal{C}_{\chi \underset{31}{\ell(e)}}^{(6,2)}~\mathcal{C}_{\chi \underset{11}{d}}^{(6,2)}$ & $<2.73 \times 10^{-12}$ \\
% \midrule
% $\mathcal{B}(\tau\to \mu~ K^0)$ & $< 2.9 \times 10^{-8}$ & & \\
% \midrule
% $\mathcal{B}(\tau\to  e~K^0)$ & $<$ & & \\
\bottomrule
\end{tabular}
\caption{Constraints on the products of DMEFT WCs derived from $\tau$ hadronic LFV decays for $m_{\chi}=800~\text{GeV}$.}
\label{tab:tau_LFV_Had}
\end{table}
%----------------------------

%-----------------------------------
\subsubsection{Lepton Flavour Violating $Z$- boson decay}
%----------------------------------
In Section~\ref{subsec:EWPOs}, we comprehensively accounted for the flavour-diagonal $Z \to \ell^+ \ell^-$ processes when evaluating the electroweak precision observables. We now extend this analysis to the LFV channels, $Z \to \ell_i^\pm \ell_j^\mp$. Since the underlying mechanism relies on the same loop insertions, the one-loop matching equations for these LFV couplings are fundamentally identical to those derived for the $Z \to f \bar{f}$ transitions. By taking the kinematic limit of vanishing final-state lepton masses, the LFV vertex corrections can be directly read off from our previously established flavour-conserving expressions in eq.~\eqref{eq:delta_gLR}. Utilizing these modified couplings, the general expression for the $Z$ LFV branching fraction is given by
%----------------------------------
\begin{align}
\mathcal{B}(Z\to \ell_i^\pm \ell_j^\mp) &= \frac{g_Z^2M_Z}{24\pi \Gamma_Z}\left( \left|\delta g_L^{Z\ell_i \ell_j}\right|^2+\left|\delta g_R^{Z \ell_i \ell_j}\right|^2\right) \,.
\end{align}
%---------------------------------
Finally, we leverage the experimental bounds derived from this sector to constrain the products of the effective couplings, specifically $\mathcal{C}_{B\chi}^{(5,1)} \mathcal{C}_{\chi \underset{ij}{\ell(e)}}$ and $\mathcal{C}_{\chi D H}^{(6,1(2))} \mathcal{C}_{\chi \underset{ij}{\ell(e)}}$. The resulting limits are summarized in Table~\ref{tab:ZLFV}.
%------------------------------
% Z LFV
\begin{table}[htbp]
\footnotesize
\centering
\renewcommand{\arraystretch}{1.5}
\begin{tabular}{l c c c}
\toprule
\textbf{Observable} &
\textbf{Constraint} &
\textbf{Scenario} & 
\textbf{Values} 
\\
\midrule
\multirow{3}{*}{$\mathcal{B}(Z \to \mu^\pm e^\mp)$} & \multirow{3}{*}{$< 2.62 \times 10^{-7}$ \cite{ATLAS:2022uhq}} 
& $\mathcal{C}_{B\chi}^{(5,1)}\,\mathcal{C}_{\chi \underset{21}{\ell(e)}}^{(6,1)}$ & $< 5.55 \times 10^{-9}$ \\
& & $\mathcal{C}_{\chi D H}^{(6,1)}\,\mathcal{C}_{\chi \underset{21}{\ell(e)}}^{(6,1)}$ & $< 1.04 \times 10^{-9}$ \\
& & $\mathcal{C}_{\chi D H}^{(6,2)}\,\mathcal{C}_{\chi \underset{21}{\ell(e)}}^{(6,2)}$ & $< 3.77 \times 10^{-12}$ \\
\midrule
%---------------------------------
\multirow{3}{*}{$\mathcal{B}(Z \to \tau^\pm e^\mp)$} & \multirow{3}{*}{$< 5.0 \times 10^{-6}$ \cite{ATLAS:2021bdj}} 
& $\mathcal{C}_{B\chi}^{(5,1)}\,\mathcal{C}_{\chi \underset{31}{\ell(e)}}^{(6,1)}$ & $< 2.42 \times 10^{-8}$ \\
& & $\mathcal{C}_{\chi D H}^{(6,1)}\,\mathcal{C}_{\chi \underset{31}{\ell(e)}}^{(6,1)}$ & $< 4.52 \times 10^{-9}$ \\
& & $\mathcal{C}_{\chi D H}^{(6,2)}\,\mathcal{C}_{\chi \underset{31}{\ell(e)}}^{(6,2)}$ & $< 2.83 \times 10^{-10}$ \\
\midrule
%---------------------------------
\multirow{3}{*}{$\mathcal{B}(Z \to \tau^\pm \mu^\mp)$} & \multirow{3}{*}{$< 6.5 \times 10^{-6}$ \cite{ATLAS:2021bdj}} 
& $\mathcal{C}_{B\chi}^{(5,1)}\,\mathcal{C}_{\chi \underset{32}{\ell(e)}}^{(6,1)}$ & $< 2.76 \times 10^{-8}$ \\
& & $\mathcal{C}_{\chi D H}^{(6,1)}\,\mathcal{C}_{\chi \underset{32}{\ell(e)}}^{(6,1)}$ & $< 5.16 \times 10^{-9}$ \\
& & $\mathcal{C}_{\chi D H}^{(6,2)}\,\mathcal{C}_{\chi \underset{32}{\ell(e)}}^{(6,2)}$ & $< 1.12 \times 10^{-11}$ \\
\bottomrule
\end{tabular}
\caption{Constraints on the products of DMEFT WCs derived from $Z$ boson Lepton Flavour Violating decays for $m_{\chi}=800~\text{GeV}$.}
\label{tab:ZLFV}
\end{table}
%--------------------------------
\subsubsection{Lepton Flavour Violating Higgs decay}
%-------------------------------
Unlike flavour-violating processes involving quarks or charged leptons, flavour-violating Higgs interactions are comparatively less constrained by current experimental data. As a result, the corresponding Higgs flavour-violating operators in the DMEFT parameter space are not expected to yield constraints as stringent as those obtained from the other flavour observables considered in our analysis. Nevertheless, Higgs flavour-violating decays provide a complementary probe of these interactions, particularly for operators that induce flavour-off-diagonal Higgs couplings after electroweak symmetry breaking. Unlike the flavour-diagonal Higgs couplings in the SM, which are proportional to the corresponding fermion Yukawa couplings and hence suppressed for light fermions, the flavour-off-diagonal couplings induced by DMEFT operators need not be subject to such Yukawa suppression. The branching fraction for a flavour-violating Higgs decay can be expressed as
%---------------------------------
\begin{align}
    \mathcal{B}(H\to f_i \bar{f}_j)=\frac{N_c M_H}{16\pi \Gamma_H}\left[\left|\left( \frac{m_{f_i}}{v} \delta_{ij} + \delta Y_{ij}^L \right)\right|^2+\left| \left( \frac{m_{f_i}}{v} \delta_{ij} + \delta Y_{ij}^R \right)\right|^2\right]
\end{align}
%---------------------------
Here $N_c=1$ for leptons. Finally, the $H$- LFV constraints on the scenarios involving the $\mathcal{C}_{H\chi}^{(5,2)} \mathcal{C}_{\chi \underset{ij}{\ell(e)}}^{(6,2)}$ operator combinations are evaluated, and the resulting bounds are summarized in Table~\ref{tab:HLFV}.

%------------------------------
% HLFV
\begin{table}[htbp]
\footnotesize
\centering
\renewcommand{\arraystretch}{1.5}
\begin{tabular}{l c c c}
\toprule
\textbf{Observable} &
\textbf{Constraint} &
\textbf{Scenario} & 
\textbf{Values} 
\\
\midrule
$\mathcal{B}(H\to \mu e)$ & $<4.4 \times 10^{-5}$ \cite{CMS:2023pte} & $\mathcal{C}^{(5,2)}_{H\chi} \mathcal{C}_{\chi \underset{21}{\ell(e)}}^{(6,2)}$ & $<7.15 \times 10^{-9}$ \\
$\mathcal{B}(H\to \tau e)$ & $<2.0 \times 10^{-3}$ \cite{ATLAS:2023mvd} &$\mathcal{C}^{(5,2)}_{H\chi} \mathcal{C}_{\chi \underset{31}{\ell(e)}}^{(6,2)}$ & $<2.48 \times 10^{-9}$ \\
$\mathcal{B}(H\to \tau \mu)$ & $<1.5 \times 10^{-3}$ \cite{CMS:2021rsq} & $\mathcal{C}^{(5,2)}_{H\chi} \mathcal{C}_{\chi \underset{32}{\ell(e)}}^{(6,2)}$ & $<2.87 \times 10^{-9}$  \\
\bottomrule
\end{tabular}
\caption{Constraints on the products of DMEFT WCs derived from $H$ boson Lepton Flavour Violating decays for $m_{\chi}=800~\text{GeV}$.}
\label{tab:HLFV}
\end{table}
%-----------------------------------
\subsubsection{Lepton Flavour Violating top quark decay}
%----------------------------------
We have already investigated LFV processes in the down-quark sector, focusing on rare FCNC decays of $B$- and $K$-mesons. Complementary information can be obtained from the up-quark sector, where LFV interactions involving the top quark provide an equally important probe of new physics. Due to its large mass, the top quark is particularly sensitive to physics beyond the SM. In a model-independent effective field theory framework, the branching fraction for the three-body decay $t\to q\,\ell_i^\pm\ell_j^\mp$ ($q=u,c$) can be expressed in terms of the general scalar, vector, and tensor four-fermion operators as
%--------------------

\begin{align}
   & \mathcal{B}(t\to q_j f_k\bar{f}_l)=\frac{S_n\,m_t^5}{6144 \pi^3 \Gamma_t}\left(|\mathcal{C}^S_{LL}|^2+|\mathcal{C}^S_{LR}|^2+|\mathcal{C}^S_{RL}|^2+|\mathcal{C}^S_{RR}|^2\right.\nonumber\\
&\left.+4(|\mathcal{C}^V_{LL}|^2+|\mathcal{C}^V_{LR}|^2+|\mathcal{C}^V_{RL}|^2+|\mathcal{C}^V_{RR}|^2)+48(|\mathcal{C}^T_{LL}|^2+|\mathcal{C}^T_{RR}|^2)\right)
\end{align}
%------------------------------------
Crucially, our theoretical framework exclusively generates vector and axial-vector currents. Consequently, all scalar and tensor WCs identically vanish, restricting the non-zero contributions entirely to the vector (axial)-type operators. The leading-order topologies for these decays are analogous to those presented in fig.~\ref{fig:LFV_topologies}. However, in contrast to the purely leptonic LFV transitions discussed earlier, these top-quark FCNC processes do not receive tree-level or loop-induced contributions from intermediate $Z$-boson exchange, as final state particles are not identical. The matching relations of the couplings $C^V_{XY}$ is detailed in the Appendix~\ref{Append:top_LFV}.

%------------------------------
% top LFV
\begin{table}[htbp]
\footnotesize
\centering
\renewcommand{\arraystretch}{1.5}
\begin{tabular}{l c c c}
\toprule
\textbf{Observable} &
\textbf{Constraint} &
\textbf{Scenario} & 
\textbf{Values} 
\\
\midrule
$\mathcal{B}(t \to c e^\pm \mu^\mp)$ & 
$< 8.9 \times 10^{-7}$ \cite{CMS:2022ztx} & 
$\mathcal{C}_{\chi \underset{23}{q(u)}}^{(6,2)}\,\mathcal{C}_{\chi \underset{12}{\ell(e)}}^{(6,2)}$ & 
$< 7.47 \times 10^{-9}$ \\
\midrule
$\mathcal{B}(t \to u e^\pm \mu^\mp)$ & 
$< 7.0 \times 10^{-8}$ \cite{CMS:2022ztx} & 
$\mathcal{C}_{\chi \underset{13}{q(u)}}^{(6,2)}\,\mathcal{C}_{\chi \underset{12}{\ell(e)}}^{(6,2)}$ & 
$< 2.10 \times 10^{-9}$ \\
\midrule
$\mathcal{B}(t \to c \tau^\pm \mu^\mp)$ & 
$< 8.7 \times 10^{-7}$ \cite{ATLAS:2024njy} & 
$\mathcal{C}_{\chi \underset{23}{q(u)}}^{(6,2)}\,\mathcal{C}_{\chi \underset{23}{\ell(e)}}^{(6,2)}$ & 
$< 7.39 \times 10^{-9}$ \\
\midrule
$\mathcal{B}(t \to u \tau^\pm \mu^\mp)$ & 
$< 8.7 \times 10^{-7}$ \cite{ATLAS:2024njy} & 
$\mathcal{C}_{\chi \underset{13}{q(u)}}^{(6,2)}\,\mathcal{C}_{\chi \underset{23}{\ell(e)}}^{(6,2)}$ & 
$< 7.39 \times 10^{-9}$ \\
\bottomrule
\end{tabular}
\caption{Constraints on the products of DMEFT WCs derived from top quark Lepton Flavour Violating decays for $m_{\chi}=800~\text{GeV}$.}
\label{tab:tLFV}
\end{table}
%--------------------------------
\subsection{Top FCNC decays}
%--------------------------------

\begin{itemize}
    \item \textbf{$t \to u_j Z$:}
As noted previously, intermediate $Z$-boson exchange does not mediate the three-body $t \to u_j \ell_i \ell_j$ transitions within our specific theoretical setup. However, the two-body flavour-changing decay $t \to u_j Z$ remains a highly potent, independent observable. The ATLAS collaboration has established stringent experimental upper limits on the branching fraction $\mathcal{B}(t \to u_j Z)$~\cite{ATLAS:2023qzr}. In our analysis, we leverage these direct experimental bounds to place rigorous constraints on the product of the underlying effective couplings, specifically $\mathcal{C}_{\chi q(u)}^{(6)} \mathcal{C}_{B\chi}^{(5)}$.
The general expression describing branching fraction $t \to u_j Z$ process can be written as \cite{Kala:2026xzo},
%------------------------------
\begin{align}\label{eq:BR_tqZ}
		\mathcal{B}\left(t \to u_j Z\right)&=\frac{g_W^2 m_t}{128 \pi c_W^2 \Gamma_t}\left(1-\frac{M_Z^2}{m_t^2}\right)^2 \Bigg[\left(2+\frac{m_t^2}{M_Z^2}\right)\left(\left|X_L^{u_jt}\right|^2+\left|X_R^{u_jt}\right|^2\right) \nonumber\,,\\
		&+4\left(2+\frac{M_Z^2}{m_t^2}\right)\left(\left|\kappa_L^{u_jt}\right|^2+\left|\kappa_R^{u_jt}\right|^2\right)\,,\\&+6\left(X_L^{u_jt} \kappa_R^{*\,u_jt}+X_L^{*\,u_jt} \kappa_R^{u_jt}+X_R^{u_jt} \kappa_L^{*\,u_jt}+X_R^{*\,u_jt} \kappa_L^{u_jt}\right)\Bigg]\,,\nonumber
	\end{align}
%------------------------------
Following the notation of Ref.~\cite{Kala:2026xzo}, the coefficients $X_{L,R}^{u_j t}$ denote the vector-type $tqZ$ couplings, while $\kappa_{L,R}^{u_j t}$ represent the tensor-type dipole interactions. Within our specific effective framework, the NP exclusively induces vector and axial-vector operators. As a result, the tensor contributions identically vanish ($\kappa_{L,R}^{u_j t} = 0$), and this decay rate simplifies significantly, governed strictly by the modified vector couplings $X_{L,R}^{u_j t}$. In Appendix~\ref{Append:Top_FCNC}, we provide the explicit matching relations for the effective vector couplings $X_{L,R}$ in terms of the  DMEFT WCs.

\item \textbf{$t \to u_j H$:} To complete our analysis of the top-quark FCNC sector, we evaluate the two-body decay of a top quark into an up-type quark and a SM Higgs boson, $t \to u_j H$ (where $u_j \in \{u,c\}$). Neglecting the mass of the light final-state quark, the general expression for the branching fraction is given by
\begin{align}\label{eq:BR_tqH}
    \mathcal{B}(t \to u_j H) &= \frac{m_t}{64 \pi \Gamma_t} \left(1-\frac{M_H^2}{m_t^2}\right)^2 \left( \left|\eta_L^{u_jt}\right|^2 + \left|\eta_R^{u_jt}\right|^2 \right) \,,
\end{align}
where $\eta_L^{u_jt}$ and $\eta_R^{u_jt}$ denote the effective left- and right-handed chiral couplings, respectively. Here, we adopt the notation of Ref.~\cite{Kala:2026xzo} for the effective parametrisetion of these $tu_jH$ interactions. In our analysis, we utilise the direct experimental bounds on $\mathcal{B}(t \to u_j H)$ to place constraints on the product of the underlying effective couplings, specifically $\mathcal{C}_{\chi q(u)}^{(6)} \mathcal{C}_{\chi D H}^{(5)}$. The explicit matching relations bridging these low-energy couplings to the underlying DMEFT WCs are provided in Appendix~\ref{Append:Top_FCNC}. By applying the current experimental upper limits on this branching fraction, we extract rigorous constraints on the corresponding coupling parameter space. 
\end{itemize}

%------------------------------
% top FCNC
\begin{table}[htbp]
\footnotesize
\centering
\renewcommand{\arraystretch}{1.5}
\begin{tabular}{l c c c}
\toprule
\textbf{Observable} &
\textbf{Constraint}  &
\textbf{Scenario} & 
\textbf{Values} 
\\
\midrule
\multirow{2}{*}{$\mathcal{B}(t \to c Z)$} & $<1.3 \times 10^{-4}$ (LH) \cite{ATLAS:2023qzr} & $\mathcal{C}^{(6,1)}_{\chi \underset{23}{q}}\mathcal{C}_{B\chi}^{(5,1)}$ & $<2.29 \times 10^{-7}$\\
 & $<1.2 \times 10^{-4}$ (RH)\ & $\mathcal{C}^{(6,1)}_{\chi \underset{23}{u}}\mathcal{C}_{B\chi}^{(5,1)}$ & $<2.20 \times 10^{-7}$ \\
\multirow{2}{*}{$\mathcal{B}(t \to u Z)$} & $<0.62 \times 10^{-4}$ (LH)\cite{ATLAS:2023qzr} & $\mathcal{C}^{(6,1)}_{\chi \underset{13}{q}}\mathcal{C}_{B\chi}^{(5,1)}$ & $< 1.58 \times 10^{-7}$\\
 & $<0.66 \times 10^{-4}$ (RH) & $\mathcal{C}^{(6,1)}_{\chi \underset{13}{u}}\mathcal{C}_{B\chi}^{(5,1)}$ & $< 1.63 \times 10^{-7}$\\
 \midrule
 $\mathcal{B}(t \to cH)$ & $<3.4 \times 10^{-4}$ \cite{ATLAS:2024mih} & $\mathcal{C}^{(6,2)}_{\chi \underset{23}{q(u)}}\mathcal{C}_{\chi DH}^{(5,2)}$ & $<3.32 \times 10^{-5}$\\
  $\mathcal{B}(t \to uH)$ & $<2.8\times 10^{-4}$ \cite{ATLAS:2024mih} & $\mathcal{C}^{(6,2)}_{\chi \underset{13}{q(u)}}\mathcal{C}_{\chi DH}^{(5,2)}$& $< 3.01 \times 10^{-5}$\\
\bottomrule
\end{tabular}
\caption{Constraints on the products of DMEFT WCs derived from top quark FCNC decays for $m_{\chi}=800~\text{GeV}$.}
\label{tab:top_FCNC}
\end{table}
%--------------------------------

The constraints obtained from the top-quark FCNC sector are summarised in Table~\ref{tab:top_FCNC}. As expected, the current experimental limits on the branching fractions of the rare decays $t\to qZ$ and $t\to qH$ ($q=u,c$) translate into bounds on products of DMEFT WCs, reflecting the loop-induced nature of the underlying amplitudes. Among the four channels considered, the strongest constraints arise from the $t\to qZ$ modes, where the products involving $\mathcal{C}_{B\chi}^{(5,1)}$ are constrained at the level of $\mathcal{O}(10^{-7})$. In particular, the bounds on $\mathcal{C}_{\chi \underset{13}{q}}^{(6,1)} \mathcal{C}_{B\chi}^{(5,1)}$ and $\mathcal{C}_{\chi \underset{13}{u}}^{(6,1)} \mathcal{C}_{B\chi}^{(5,1)}$ derived from $t\to uZ$ are slightly stronger than their counterparts involving the charm quark, reflecting the more stringent experimental limits currently available for the former decay mode. In contrast, the Higgs-mediated channels $t\to qH$ yield comparatively weaker bounds, constraining the combinations $\mathcal{C}_{\chi \underset{13}{q(u)}}^{(6,2)} \mathcal{C}_{\chi DH}^{(5,2)}$ and $\mathcal{C}_{\chi \underset{23}{q(u)}}^{(6,2)} \mathcal{C}_{\chi DH}^{(5,2)}$ at the level of $\mathcal{O}(10^{-5})$. Although these limits are less restrictive than those obtained from the $Z$-mediated channels, they provide an important and complementary probe of Higgs-current operators that are inaccessible in other FCNC top-quark observables. Overall, the top-FCNC sector probes a class of flavour-changing quark-DM interactions that is largely independent of the low-energy flavour observables considered previously. Consequently, these measurements furnish valuable complementary information on the DMEFT parameter space and play an important role in constraining operator combinations involving third-generation quarks, particularly those relevant for future precision top-quark studies at the HL-LHC and future collider facilities.

%------------------------------
\subsection{Magnetic Moment}
%------------------------------
%------------------------------
\begin{figure}[t!]
	\centering
%	\subfloat[]{
		\begin{tikzpicture}
			\begin{feynman}
				\vertex (a);
				\vertex[left=1.cm of a] (b) {\(\ell,q\)};
				\vertex[right=1.4cm of a] (c);
				\vertex[right=1.cm of c] (d) {\(\ell,q\)};
				\vertex[right=0.8cm of a] (a1);
				\vertex[thick, blob, above=1cm of a1] (e) {};
				\vertex[above=1.7cm of a1] {\(\chi\)};
				\vertex[right=0.7cm of a] (f);
				\vertex[below=0.8cm of f](g){\(\gamma\)};
				\diagram*{
					(b) -- [thick, fermion] (a) -- [thick, fermion] (c) -- [thick, fermion] (d),
					(a) -- [thick, boson, edge label=\(V_i\)] (e),
					(c) -- [thick, boson, edge label'=\(V_j\)] (e),
					(f) --[thick,boson](g),
				};
			\end{feynman}
	\end{tikzpicture}
	\caption{Leading Feynman topologies contributing to the lepton and quark EDMs generated by DMEFT operators.}
	\label{fig:EDM}
\end{figure}
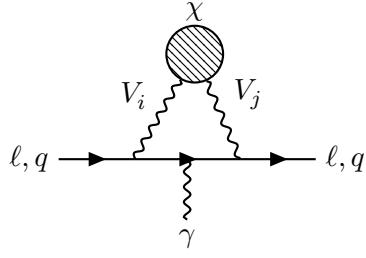
%-------------------------------

The anomalous magnetic moment is a CP-even observable, making it sensitive to CP-conserving new physics. Recently, the Muon $g-2$ collaboration at Fermilab reported its final measurement of the muon anomalous magnetic moment, yielding $a_{\mu} = 116\,592\,0710(162) \times 10^{-12}$ (139 ppb)~\cite{Muong-2:2025xyk}. This result establishes a new experimental world average of $a_{\mu}^{\rm exp} = 116\,592\,0715(145) \times 10^{-12}$. In contrast, the updated SM prediction provided by the recent Muon $g-2$ Theory Initiative White Paper (WP2025)~\cite{Aliberti:2025beg} stands at $a_{\mu}^{\rm SM} = 116\,592\,033(62) \times 10^{-11}$. The corresponding difference between the experimental measurement and the SM prediction is,
\begin{align}
    \Delta a_{\mu}=(38.5 \pm 63.7) \times 10^{-11}
\end{align}
is consistent with zero at the $1\sigma$ level, which was earlier $5.1 \sigma$ discrepancy. This substantially weakens the new-physics explanation of the $g-2$ anomaly. Within the DMEFT framework, the leading contributions to the muon $(g-2)_\mu$ arise from the exact same two-loop topologies depicted previously in fig.~\ref{fig:EDM}. The corresponding matching equation of $\Delta a_{\mu}$ in terms of DMEFT WCs are detailed in Appendix~\ref{Append:Magnetic_mom}. Using the updated world average, we extract the corresponding bounds of DMEFT WCs, summarised in Table~\ref{tab:del_a_mu}.

% Throughout this work, we take all WCs to be real and study the impact of the different observables on the resulting parameter space. For complex WCs, the imaginary parts correspond to CP-violating interactions and can be constrained by electric dipole moment (EDM) measurements. A brief discussion of these constraints is provided in Appendix~\ref{Append:EDM}.

%------------------------------
% Delta a mu
\begin{table}[t]
\footnotesize
\centering
\renewcommand{\arraystretch}{1.5}
\label{tab:NPconstraints}
\begin{tabular}{l c c}
\toprule
\textbf{Observable} &
\textbf{Scenario} & 
\textbf{Values} 
\\
\midrule
\multirow{3}{*}{$\Delta a_{\mu}$ ~\cite{Muong-2:2025xyk}} & $\left|\mathcal{C}_{B\chi}^{(5,1)}\right|^2$ &  $(1.39 \pm 2.30)\times 10^{-3}$ \\
& $\left|\mathcal{C}_{B\chi}^{(5,2)}\right|^2$ &  $(1.46 \pm 2.42)\times 10^{-4}$ \\
& $\left|\mathcal{C}_{B\chi}^{(5,1)} \mathcal{C}_{\chi D H}\right|$ &  $(1.78 \pm 2.95)\times 10^{-3}$ \\
\bottomrule
\end{tabular}
\caption{Constraints on the products of DMEFT WCs derived from $\Delta a_{\mu}$. The values in the table are presented for $m_{\chi} = 800 $ GeV and $\mu^{\rm ren} = 1$ TeV.}
\label{tab:del_a_mu}
\end{table}
%-----------------------------

%----------------------------------
\subsection{Constraints on individual WCs and their dominant probes}
%---------------------------------
This section presents the leading constraints on the DMEFT WCs obtained from the combined analysis of the observables considered in the preceding sections.
 As discussed, the precision measurements of EWPOs together with flavour observables allow us to place stringent constraints on a large subset of the DMEFT parameter space and, in particular, to extract information on several individual WCs. Nevertheless, additional complementary information can be obtained from processes that are highly suppressed or forbidden within the SM, such as LFV meson and charged-lepton decays. The current experimental upper limits on these channels provide an excellent opportunity to further constrain the lepton-DM interaction operators appearing in the DMEFT framework.

A common feature of the LFV observables considered in this work is that their amplitudes are generated through loop-induced processes and are therefore sensitive to products of two distinct WCs. Consequently, the corresponding experimental limits directly constrain combinations of couplings rather than individual coefficients. However, many of the WCs entering these products have already been tightly constrained by the EWPO, rare decay, and meson mixing analyses presented above. Exploiting these previously established limits, we can translate the experimental bounds on LFV observables into independent constraints on the remaining, otherwise poorly constrained, lepton-DM couplings. In this way, LFV processes provide an important complementary probe of the DMEFT parameter space and allow us to extend the coverage of the parameter regions accessible via current indirect searches.

%------------------------------------
\begin{figure}[t]
	\centering
	\includegraphics[width=1.\linewidth]{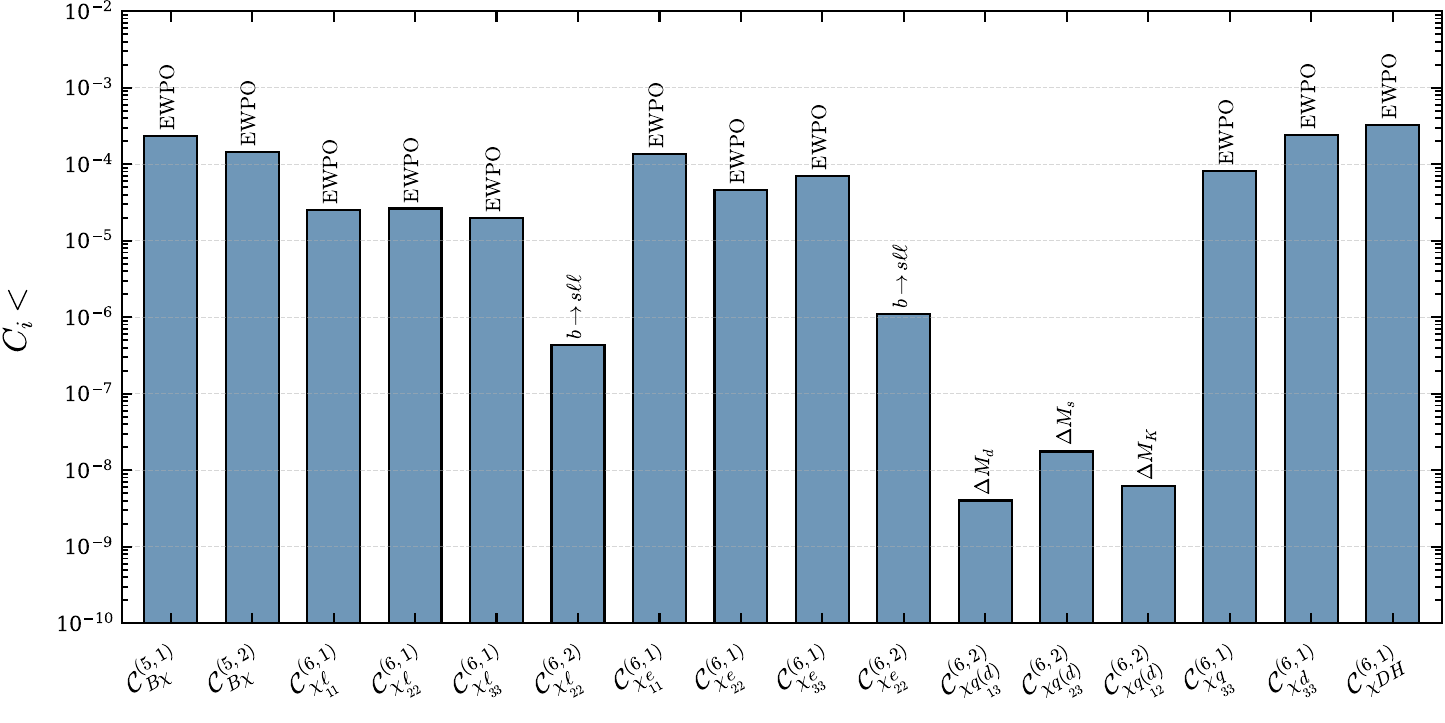}
	\caption{Upper bounds on the DMEFT WCs from the global
analysis for $m_\chi=800\,\mathrm{GeV}$. The dimension-five and
dimension-six Wilson coefficients are quoted in units of
$\mathrm{GeV}^{-1}$ and $\mathrm{GeV}^{-2}$, respectively.}
	\label{fig:exisiting_800}
\end{figure}
%-------------------------------------
%------------------------------------
\begin{figure}[h]
	\centering
	\includegraphics[width=1.\linewidth]{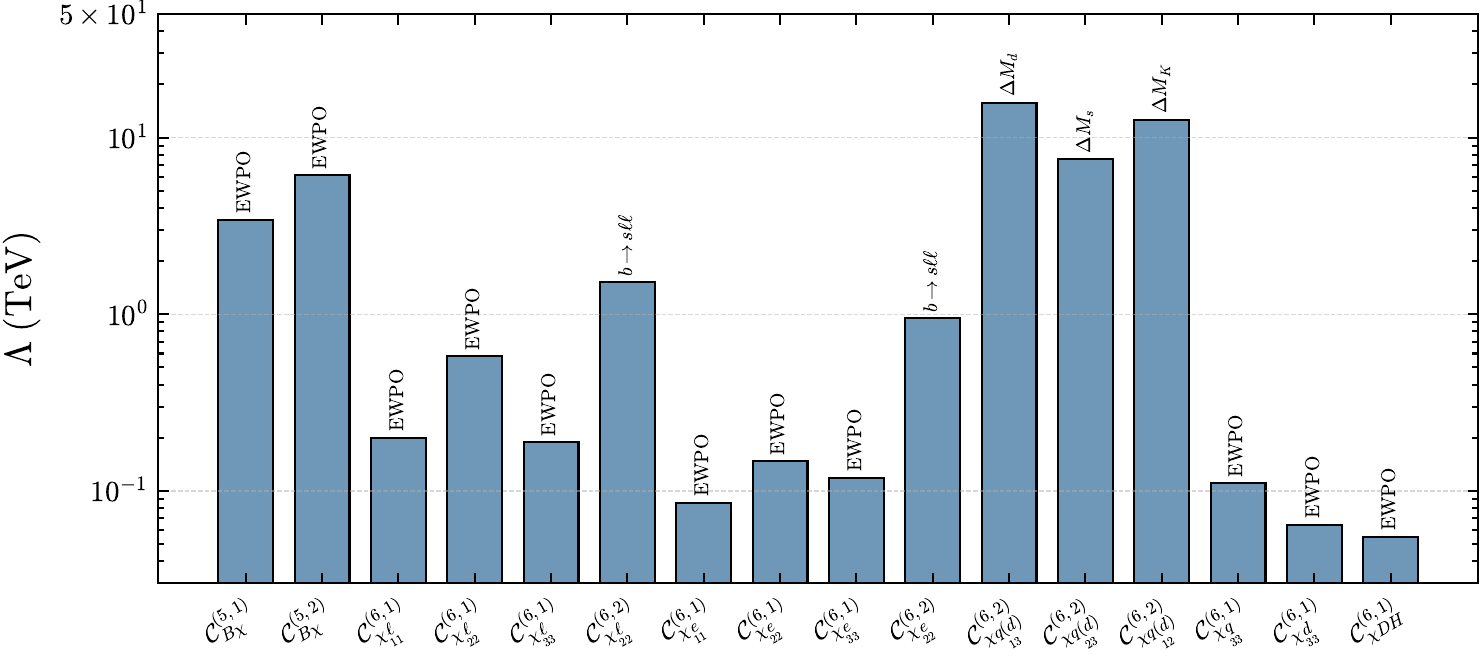}
	\caption{New Physics scale $\Lambda$ prove for individual DMEFT WCs at $m_\chi = 800\,\mathrm{GeV}$, categorized by their most sensitive observable.}
	\label{fig:exisitinglam_800}
\end{figure}
%-------------------------------------
%-------------------------------------
\begin{figure}[t]
	\centering
	\includegraphics[width=1.\linewidth]{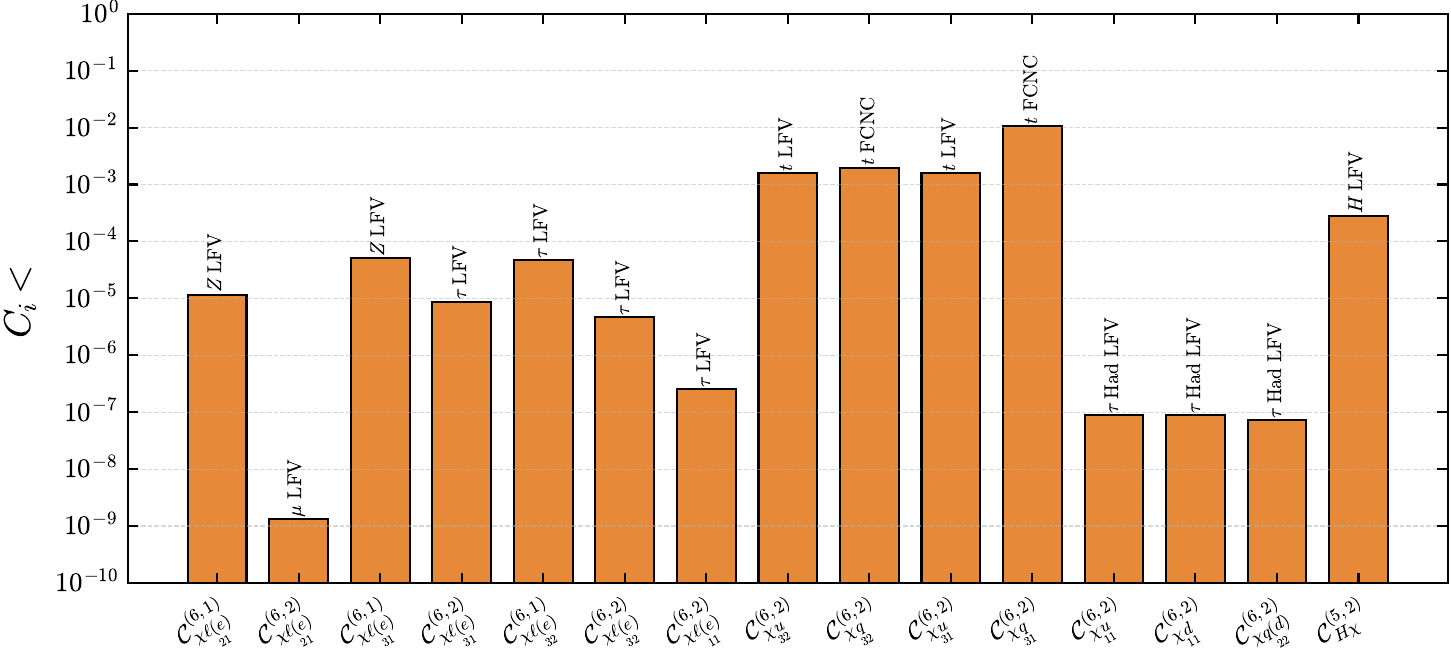}
	\caption{Upper bounds on the DMEFT WCs from the global
analysis for $m_\chi=800\,\mathrm{GeV}$, derived from product couplings.}
	\label{fig:derived_800}
\end{figure}
%-------------------------------------
%-------------------------------------
\begin{figure}[h]
	\centering
	\includegraphics[width=1.\linewidth]{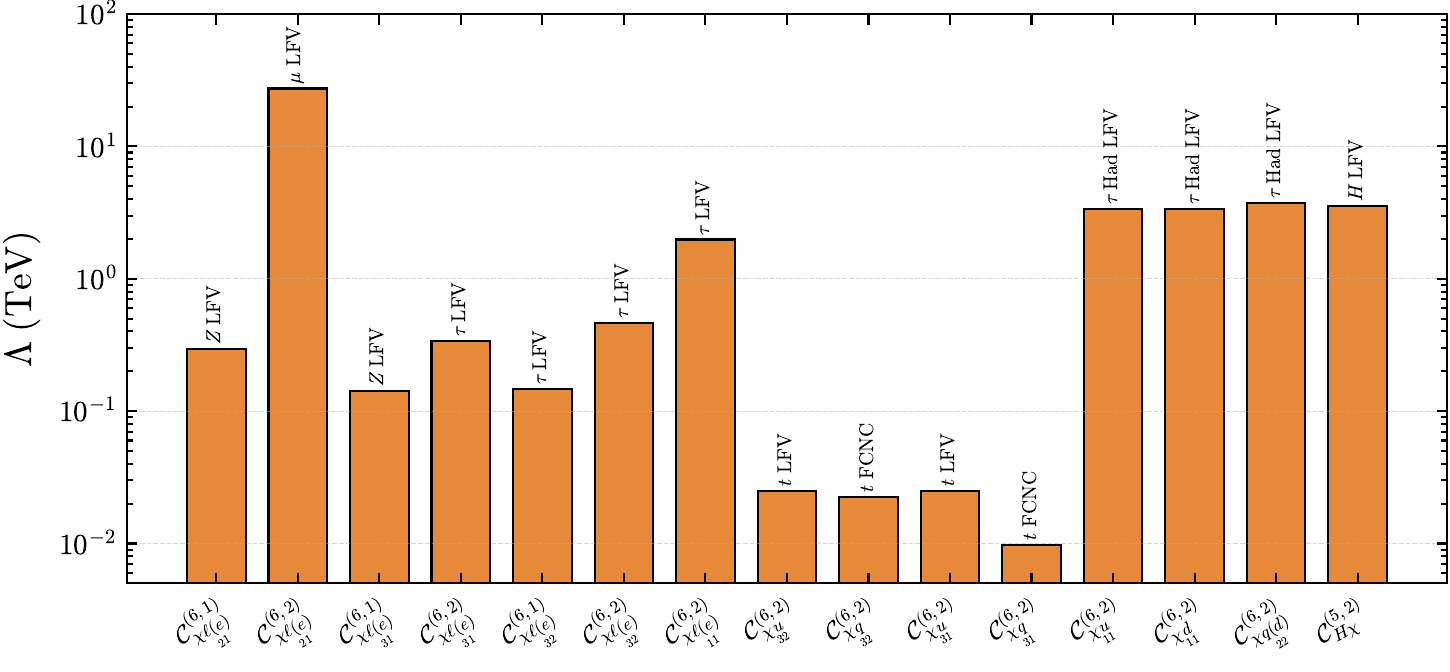}
	\caption{New Physics scale $\Lambda$ proves for individual DMEFT WCs derived from product couplings, at $m_\chi = 800\,\mathrm{GeV}$, categorised by their most sensitive observable.}
	\label{fig:derivedlam_800}
\end{figure}
%-------------------------------------

The resulting bounds on the relevant WCs are presented in figs.~\ref{fig:exisiting_800} and~\ref{fig:derived_800}, respectively. The former summarises the limits obtained directly from existing constraints on the relevant operator combinations, while the latter displays the derived bounds obtained after incorporating the constraints extracted from EWPOs and flavour observables. For all results presented in the bar plots, we employ the shorthand notation $\mathcal{C}^{(5)} \equiv \mathcal{C}^{(5)}/\Lambda$ and $\mathcal{C}^{(6)} \equiv \mathcal{C}^{(6)}/\Lambda^2$ for the dimension-5 and dimension-6 WCs, respectively. Furthermore, the dark matter mass is fixed to the benchmark value of $m_\chi = 800\text{ GeV}$. 

Furthermore, to quantify the reach of current indirect searches in a more model-independent manner, we consider the representative scenario in which the WCs are of order unity, $C_i \sim \mathcal{O}(1)$. Under this assumption, the experimental limits can be translated into lower bounds on the EFT cutoff scale $\Lambda$. The corresponding exclusion reaches are shown in figs.~\ref{fig:exisitinglam_800} and~\ref{fig:derivedlam_800}, which illustrate the scales up to which present indirect measurements are capable of probing or excluding new physics effects within the DMEFT framework.

% The resulting bounds on the individual WCs are summarized in figs.~\ref{fig:exisiting_800} and~\ref{fig:derived_800}. fig.~\ref{fig:exisiting_800} demonstrates the complementarity of the various low- and high-energy observables considered in this work.
The EWPOs provide the dominant constraints on the operators $\mathcal{C}_{B\chi}^{(5,1(2))}$, $\mathcal{C}_{\chi DH}^{(6,1)}$, and several lepton-DM operators, typically restricting their allowed magnitudes to the level of $\mathcal{O}(10^{-4})$ or stronger. On the other hand, flavour observables probe a different class of interactions and play a crucial role in constraining flavour-violating quark-DM operators. In particular, the neutral meson mixing observables $\Delta M_K$, $\Delta M_d$, and $\Delta M_s$ yield the strongest limits on the flavour-changing operators $\mathcal{C}_{\chi \underset{12}{q(d)}}^{(6,2)}, \mathcal{C}_{\chi \underset{13}{q(d)}}^{(6,2)}$, and $\mathcal{C}_{\chi \underset{23}{q(d)}}^{(6,2)}$, respectively, illustrating the exceptional sensitivity of meson mixing measurements to new sources of flavour violation. The figure therefore clearly demonstrates that no single observable class dominates the entire DMEFT parameter space and that a meaningful assessment of the model requires a simultaneous consideration of electroweak and flavour constraints.

The bounds derived using existing limits on LFV and FCNC processes are shown in fig.~\ref{fig:derived_800}. These results are obtained after combining the limits extracted above with the experimentally allowed ranges of the WCs entering the relevant operator products. Several noteworthy features emerge. The most stringent constraints are obtained for operators contributing to LFV processes involving the first and second lepton generations, with some coefficients constrained down to the $\mathcal{O}(10^{-10})$-$\mathcal{O}(10^{-9})$ level. The LFV $\tau$-decay observables also provide strong constraints, typically probing WCs in the range $\mathcal{O}(10^{-7})$-$\mathcal{O}(10^{-5})$. In contrast, the operators associated with top-quark FCNC and LFV transitions remain comparatively less constrained, allowing coefficients as large as $\mathcal{O}(10^{-3})$-$\mathcal{O}(10^{-2})$. This hierarchy reflects the vastly different experimental sensitivities currently available for rare charged-lepton, meson, and top-quark processes. Importantly, these constraints are not obtained directly from a single observable but result from the combined power of EWPO, flavour, and LFV data. Consequently, they illustrate how precision measurements in one sector can be used to derive powerful indirect limits on otherwise poorly constrained interactions in a different sector of the DMEFT framework.

Taken together, these figures highlight the broad reach of current indirect searches for dark-sector interactions. They show that present electroweak and flavour data already probe a substantial region of the DMEFT parameter space, with sensitivities extending from the percent level down to values below $10^{-9}$ for certain flavour-violating operators. This wide range of constraints underscores the importance of combining information from multiple sectors when assessing the viability of effective dark-sector scenarios. Moreover, the results indicate that future improvements in precision flavour and LFV measurements have the potential to significantly extend the sensitivity to DMEFT operators, particularly in regions of parameter space that remain only weakly constrained by current data.

To further illustrate the impact of the indirect constraints, we translate the bounds on the WCs into lower limits on the effective scale $\Lambda$ by assuming WCs of order unity, $C_i \sim \mathcal{O}(1)$. The resulting exclusion reaches are presented in figs.~\ref{fig:exisitinglam_800} and~\ref{fig:derivedlam_800}. These figures provide a complementary interpretation of the results, highlighting the regions of parameter space already disfavoured by current flavour and electroweak precision data. In particular, values of $\Lambda$ below the quoted limits are excluded by existing measurements, while the parameter region above these scales remains compatible with present observations and therefore constitutes the target range for future high-energy collider searches. The bounds derived from existing EWPO and flavour constraints span a wide range of scales, extending from the sub-TeV region up to $\mathcal{O}(10)$ TeV for several flavour-changing operators. Particularly strong limits are obtained for the flavour-violating quark-DM operators constrained by neutral meson mixing observables, where the current data probe scales approaching or even exceeding the multi-TeV regime. Similarly, rare semileptonic decays and electroweak precision measurements provide significant sensitivity to several lepton-DM and dipole-type operators, demonstrating the remarkable reach of indirect probes beyond the direct kinematic capabilities of present colliders.

An even more striking picture emerges from the derived limits shown in fig.~\ref{fig:derivedlam_800}, where the information obtained from the EWPO and flavour analyses is combined with the stringent experimental bounds on LFV and FCNC processes. In several cases, the resulting lower limits on $\Lambda$ extend to tens or even hundreds of TeV, substantially exceeding the scales directly accessible at current collider experiments. The strongest reaches are obtained from LFV observables, particularly those involving $\tau$ and $\mu$ flavour violation, reflecting the exceptional experimental sensitivity to such rare processes. These results clearly demonstrate the power of precision flavour measurements as indirect probes of new physics, capable of testing energy scales far beyond those accessible through direct production searches. Consequently, any future collider signal associated with the DMEFT operators considered in this work must originate from parameter regions lying above the exclusion scales shown in figs.~\ref{fig:exisitinglam_800} and~\ref{fig:derivedlam_800}. The comparison also illustrates the complementarity between indirect precision measurements and direct collider searches: while colliders probe the production of new states, flavour and EW precision observables already place stringent restrictions on the underlying effective interactions and significantly reduce the viable parameter space available for future discoveries.

%--------------------------------
\section{Low-Mass Dark Matter Probes from Invisible Decays} \label{sec:invisible}
%------------------------------
\begin {table}[t!]
\footnotesize
\centering
\renewcommand{\arraystretch}{1.5}
\begin{tabular}{c c c c c}
	\toprule
	\textbf{Observable} & 
	\textbf{Constraint} & 
	\textbf{Coupling combinations} & 
	\textbf{\shortstack{$m_{\chi}$ \\ (GeV)}} & 
	\textbf{\shortstack{Values \\ ($\rm GeV^{-2}$)}} 
	\\
	\midrule
	\multirow{4}{*}{$\mathcal{B}(B^{+} \to K^{+} \chi \bar \chi)$} & \multirow{4}{*}{$(2.30 \pm 0.67) \times 10^{-5}$ \cite{Belle-II:2023esi} } 
	& \multirow{2}{*}{$\big|\mathcal{C}^{(6,1)}_{\chi \underset{23}{d}}+\mathcal{C}^{(6,1)}_{\chi \underset{23}{q}}\big|$} & 1 & $< 7.20 \times 10^{-9}$ \\
	& & & 0.1 & $<6.31 \times 10^{-9}$ \\
	& & \multirow{2}{*}{$\big| \mathcal{C}^{(6,2)}_{\chi \underset{23}{q}}\big|$} & 1 & $<3.60 \times 10^{-9}$ \\
	& & & 0.1 & $<3.15 \times 10^{-9}$ \\
	\midrule
	%---------------------------------
	\multirow{4}{*}{$\mathcal{B}(B^{+} \to K^{*+} \chi \bar \chi)$} & \multirow{4}{*}{$< 4.0 \times 10^{-5}$ \cite{Belle:2013tnz}} 
	& \multirow{2}{*}{$\big| \mathcal{C}^{(6,1)}_{\chi \underset{23}{d}} +\mathcal{C}^{(6,1)}_{\chi \underset{23}{q}} \big| \,,\, \big| \mathcal{C}^{(6,2)}_{\chi \underset{23}{d}} +\mathcal{C}^{(6,2)}_{\chi \underset{23}{q}} \big|$} & 1 & $<1.36 \times 10^{-8}$ \\
	& & & 0.1 & $<1.31 \times 10^{-8}$ \\
	& & \multirow{2}{*}{$\big| \mathcal{C}^{(6,1)}_{\chi \underset{23}{d}} -\mathcal{C}^{(6,1)}_{\chi \underset{23}{q}} \big| \,,\, \big| \mathcal{C}^{(6,2)}_{\chi \underset{23}{d}} - \mathcal{C}^{(6,2)}_{\chi \underset{23}{q}} \big|$} & 1 & $<7.57 \times 10^{-9}$ \\
	& & & 0.1 & $<6.89 \times 10^{-9}$ \\
	\midrule
	%---------------------------------
	\multirow{4}{*}{$\mathcal{B}(B^{+} \to \pi^{+} \chi \bar \chi)$} & \multirow{4}{*}{$< 1.4 \times 10^{-5}$ \cite{Belle:2017oht}} 
	& \multirow{2}{*}{$\big|\mathcal{C}^{(6,1)}_{\chi \underset{13}{d}}+\mathcal{C}^{(6,1)}_{\chi \underset{13}{q}}\big|$} & 1 & $<1.69 \times 10^{-9}$ \\
	& & & 0.1 & $<1.44 \times 10^{-9}$ \\
	& & \multirow{2}{*}{$\big| \mathcal{C}^{(6,2)}_{\chi \underset{13}{q}}\big|$} & 1 & $<8.50 \times 10^{-10}$ \\
	& & & 0.1 & $<7.20 \times 10^{-10}$ \\
	\midrule
	%---------------------------------
	\multirow{2}{*}{$\mathcal{B}(K^{+} \to \pi^{+} \chi \bar \chi)$} & \multirow{2}{*}{$(13.0 \pm 3.15) \times 10^{-11}$ \cite{NA62:2024pjp} } 
	& $\big|\mathcal{C}^{(6,1)}_{\chi \underset{12}{d}}+\mathcal{C}^{(6,1)}_{\chi \underset{12}{q}}\big|$ & 0.1 & $<9.12 \times 10^{-11}$ \\
	& & $\big| \mathcal{C}^{(6,2)}_{\chi \underset{12}{q}}\big|$ & 0.1 & $<4.56 \times 10^{-11}$ \\
	\bottomrule
\end{tabular}
\caption{Bounds on the coupling combinations derived from invisible decay modes, considering the DM $\chi$ as the missing energy signature. For isospin-related transitions, we tabulate only the results from the specific process that imposes the strongest constraint on the given coupling combinations.}
\label{tab:invisible_P2MXX}
\end{table}
%--------------------------------

The analyses presented in the previous sections primarily probe the DMEFT parameter space through loop-induced electroweak and flavour observables and are therefore applicable over a broad range of dark matter masses, including the heavy dark matter regime. However, if the dark matter particle is sufficiently light, additional constraints can be obtained from rare invisible meson decays, where the dark matter is produced on-shell and escapes the detector, giving rise to a missing-energy signature. Such processes provide a complementary probe of the DMEFT framework, as they are sensitive to the same flavour-changing operators that enter the low-energy flavour observables but contribute at tree level rather than through quantum loops.

In the SM, invisible meson decays are mediated by final-state neutrinos, leading to processes of the form $P\to M\nu\bar{\nu}$. Since neutrinos are not directly observed in collider or flavour experiments, any light weakly interacting particle can generate an identical experimental signature. In our framework, if the dark matter mass satisfies the kinematic condition 
\begin{align} 
	m_\chi \leq \frac{m_P-m_M}{2},
 \end{align} 
The decay channel $P\to M\chi\bar{\chi}$ becomes accessible and contributes directly to the experimentally measured invisible decay rate. Consequently, the observed branching ratio receives contributions from both $P\to M\nu\bar{\nu}$ and $P\to M\chi\bar{\chi}$ final states. While the neutrino mode is generated through loop-induced FCNC transitions, the dark matter channel arises at tree level through the operators listed in Table~\ref{tab:dmeft_operators}. Therefore, in the low-mass region, the invisible decay observables become exceptionally sensitive probes of the corresponding WCs. Moreover, unlike the flavour observables discussed previously, which were sensitive primarily to products of WCs through loop effects, the invisible decay amplitudes depend directly on individual WCs or simple linear combinations of them. This allows robust constraints to be derived without requiring additional assumptions regarding other operator coefficients.

The resulting bounds are summarised in Table~\ref{tab:invisible_P2MXX} for two benchmark dark matter masses, $m_\chi=1~\mathrm{GeV}$ and $0.1~\mathrm{GeV}$. The constraints are found to be remarkably stringent, typically probing coupling combinations at the level of $10^{-9}$-$10^{-11}\,\mathrm{GeV}^{-2}$. Among the channels considered, the kaon decay $K^+\to \pi^+\chi\bar{\chi}$ provides the strongest sensitivity owing to the excellent experimental precision achieved in rare kaon measurements, constraining the relevant operator combinations at the level of $10^{-11}\,\mathrm{GeV}^{-2}$. The decays $B^+\to K^+\chi\bar{\chi}$, $B^+\to K^{*+}\chi\bar{\chi}$, and $B^+\to \pi^+\chi\bar{\chi}$ yield slightly weaker bounds but nevertheless probe the DMEFT parameter space very effectively. We also observe only a mild dependence on the dark matter mass within the considered kinematic range, indicating that the sensitivity is dominated by the experimental precision of the measured branching fractions rather than the precise value of $m_\chi$. Overall, these invisible decay observables provide a powerful and highly complementary probe of the low-mass dark matter regime, extending the coverage of the DMEFT parameter space beyond that accessible through the heavy-dark-matter analyses performed in the preceding sections.

%%%%%%%%%%%%%%%%%%%%%
\section{Summary and Conclusion}\label{sec:summary}
%--------------------
In this work, we have performed a comprehensive study of the fermionic DMEFT framework, beginning with the calculation of the anomalous dimension matrices for the complete set of relevant dimension-5 and dimension-6 operators involving Dirac fermionic dark matter. The resulting renormalization-group evolution allows us to consistently connect the WCs defined at the high scale to the low-energy scales relevant for phenomenological observables. Building on this framework, we have performed a comprehensive phenomenological study of a fermionic DMEFT framework by combining constraints from EWPOs, low-energy FCNC processes, neutral meson mixing observables, and existing bounds on lepton-flavour-violating (LFV) transitions. The analysis included the one-loop contributions of the relevant dimension-5 and dimension-6 operators to the effective $Zf\bar f$ couplings, the oblique parameters $(S,T,U)$, and the radiative parameter $\Delta r$, together with their impact on rare semileptonic decays $b\to s(d)\ell^+\ell^-$ and neutral meson mixing amplitudes. By performing dedicated $\chi^2$ analyses in each sector and subsequently combining the available information, we derived constraints on both products of WCs and individual operator couplings. Furthermore, exploiting existing experimental limits on LFV meson and lepton decays, we extracted complementary bounds on several lepton-DM operators that are otherwise difficult to constrain directly. Finally, assuming WCs of order unity, the obtained limits were translated into constraints on the effective new-physics scale $\Lambda$, thereby providing a direct measure of the energy reach of current indirect searches.

Our results demonstrate the remarkable complementarity of electroweak precision and flavour observables in probing the DMEFT parameter space. The EWPO fit constrains several WCs to values typically at or below the $\mathcal{O}(10^{-4})$ level, while neutral meson mixing observables provide some of the most stringent bounds on flavour-violating quark-DM interactions. Rare semileptonic decays further constrain non-trivial combinations of WCs and play an essential role in disentangling the flavour structure of the effective theory. The derived LFV limits yield even stronger restrictions on a number of lepton-DM operators, substantially extending the phenomenological reach of the analysis. When interpreted in terms of the EFT cutoff scale, the current data are found to probe energy scales ranging from the TeV regime to several tens or even hundreds of TeV, depending on the operator structure. These findings highlight the power of present precision measurements as indirect probes of dark-sector interactions and show that a significant portion of the DMEFT parameter space is already constrained by existing data, thereby providing important guidance for future flavour experiments and collider searches.

In addition to the heavy dark matter regime, we also investigated the complementary scenario in which the dark matter particle is sufficiently light to be produced on shell in rare invisible meson decays. In this case, tree-level processes such as $K^+\to\pi^+\chi\bar{\chi}$, $B^+\to K^{(*)+}\chi\bar{\chi}$, and $B^+\to\pi^+\chi\bar{\chi}$ provide direct constraints on individual WCs and their linear combinations. Owing to the tree-level nature of these transitions, the resulting bounds are particularly stringent, with the rare kaon modes providing the strongest sensitivity among the channels considered. These constraints offer an important complement to the heavy-dark-matter analysis and substantially extend the coverage of the DMEFT parameter space into the low-mass region.

%%%%%%%%%%%%%%%%%%%%%%%%%%%%%%%%%%%%%%%%%%%%%%%%%%%%
\acknowledgments
%%%%%%%%%%%%%%%%%%%%%%%%%%%%%%%%%%%%%%%%%%%%%%%%%%%%
LK acknowledges financial support from IIT Gandhinagar under project grant No. OTH/R\&D/13467. 
%----------------------------
\appendix

%-------------------------------
\section{Inputs related to ADM}\label{Append:List_of_diagrams}
The renormalisation group evolution equations and the resulting anomalous dimension matrices have been presented in section~\ref{sec:ADM}. In this section, we provide the relevant one-loop Feynman diagrams associated with the renormalisation of the effective operators listed in Table~\ref{tab:dmeft_operators}. These diagrams constitute the building blocks of the anomalous dimension matrix and clarify the origin of the operator mixing entering the RG evolution. 

\begin{figure}[t]
	\centering
	%%%%%%%%%%%%%%%%%%%%%%%%%%
	\subfloat[]{\label{fig:four_fermi_a}
		\begin{tikzpicture}
			\begin{feynman}
				\vertex[thick, crossed dot](a){};
				\vertex[above left=2.cm of a] (b) {\(\psi_1\)};
				\vertex[above left=1cm of a](g);
				\vertex[above right=2.cm of a] (c) {\(\psi_2\)};
				\vertex[above right=1.cm of a](h);
				\vertex[thick, crossed dot, below=.4cm of a](d){};
				\vertex[below left=2.cm of d] (e) {\(\psi_3\)};
				\vertex[below right=2.cm of d] (f) {\(\psi_4\)};
				
				\diagram* {
					(b) -- [thick, fermion] (a) -- [thick, fermion] (c),
					(e) -- [thick, fermion] (d) -- [thick, fermion] (f),
					(g) --[thick, boson, edge label=\(X_{\mu}\)](h),
				};
			\end{feynman}
		\end{tikzpicture}
	}
	%%%%%%%%%%%%%%%%%%%%%%%%
	\subfloat[]{\label{fig:four_fermi_b}
		\begin{tikzpicture}
			\begin{feynman}
				\vertex[thick,crossed dot](a){};
				\vertex[above left=2.cm of a] (b) {\(\psi_1\)};
				\vertex[above right=2.cm of a] (c) {\(\psi_2\)};
				\vertex[above right=1.cm of a] (g);
				\vertex[thick, crossed dot, below=.4cm of a](d){};
				\vertex[below left=2.cm of d] (e) {\(\psi_3\)};
				\vertex[below right=2.cm of d] (f) {\(\psi_4\)};
				\vertex[below right=1.cm of d] (h);
				\diagram* {
					(b) -- [thick, fermion] (a) -- [thick, fermion] (c),
					(e) -- [thick, fermion] (d) -- [thick, fermion] (f),
					(g) --[thick, boson, edge label=\(X_{\mu}\)](h),
				};
			\end{feynman}
		\end{tikzpicture}
	}
	%%%%%%%%%%%%%%%%%%%%%%%%
	\subfloat[]{\label{fig:four_fermi_c}
		\begin{tikzpicture}
			\begin{feynman}
				\vertex[thick, crossed dot](a){};
				\vertex[above left=2.cm of a] (b) {\(\psi_1\)};
				\vertex[above left=1.cm of a] (g);
				\vertex[above right=2.cm of a] (c) {\(\psi_2\)};
				\vertex[thick, crossed dot, below=.4cm of a](d){};
				\vertex[below left=2.cm of d] (e) {\(\psi_3\)};
				\vertex[below right=2.cm of d] (f) {\(\psi_4\)};
				\vertex[below right=1.cm of d] (h);
				\diagram* {
					(b) -- [thick, fermion] (a) -- [thick, fermion] (c),
					(e) -- [thick, fermion] (d) -- [thick, fermion] (f),
					(g) --[thick, boson,half left,looseness=1.5, edge label=\(X_{\mu}\)](h),
				};
			\end{feynman}
		\end{tikzpicture}
	}\\
	%%%%%%%%%%%%%%%%%%%%%%%%
	\subfloat[]{\label{fig:four_fermi_d}
		\begin{tikzpicture}
			\begin{feynman}
				\vertex[thick, crossed dot](a){};
				\vertex[above left=1.5cm of a] (b) {\(\psi_1\)};
				\vertex[above right=1.5cm of a] (c) {\(\psi_2\)};
				\vertex[thick, crossed dot, below=.4cm of a](d){};
				\vertex[below=0.7 cm of d](g);
				\vertex[below=1.1 cm of d](h);
				\vertex[below left=1.cm of h] (e) {\(\psi_3\)};
				\vertex[below right=1.cm of h] (f) {\(\psi_4\)};
				
				\diagram* {
					(b) -- [thick, fermion] (a) -- [thick, fermion] (c),
					(e) -- [thick, fermion] (h) -- [thick, fermion] (f),
					(g) --[thick, fermion, half left, looseness=1.2](d) --[thick, fermion, half left, looseness=1.2](g),
					(g) --[thick, boson, edge label=\(X_{\mu}\)](h),
				};
			\end{feynman}
		\end{tikzpicture}
	}
	%%%%%%%%%%%%%%%%%%%%%%%%
	\subfloat[]{\label{fig:four_fermi_e}
		\begin{tikzpicture}
			\begin{feynman}
				\vertex[thick, crossed dot](a){};
				\vertex[above left=1.5cm of a] (b) {\(\psi_1\)};
				\vertex[above right=1.5cm of a] (c) {\(\psi_2\)};
				\vertex[thick, crossed dot, below=.4cm of a](d){};
				\vertex[below=0.7 cm of d](g);
				\vertex[below=1.1 cm of d](h);
				\vertex[below left=1.cm of h] (e) {\(\psi_3\)};
				\vertex[below right=1.cm of h] (f) {\(\psi_4\)};
				
				\diagram* {
					(b) -- [thick, fermion] (a) -- [thick, fermion] (c),
					(e) -- [thick, fermion] (h) -- [thick, fermion] (f),
					(g) --[thick, scalar, half left, looseness=1.2](d) --[thick, scalar, half left, looseness=1.2](g),
					(g) --[thick, boson, edge label=\(X_{\mu}\)](h),
				};
			\end{feynman}
		\end{tikzpicture}
	}
    %------------------------
\subfloat[]{\begin{tikzpicture}
	\begin{feynman}
	\vertex (a1){\(\psi_1\)};
	\vertex[crossed dot,blue,right=1.3cm of a1](a2);
	\vertex[above=1.5cm of a2](a3);
	\vertex[left=1.cm of a3](a4){\(\psi_2\)};
	\vertex[crossed dot,thick,right=1.3cm of a2](a5){};
	\vertex[crossed dot,thick,right=1.3cm of a3](a6){};
	\vertex[right=1.3cm of a6](a7){\(\chi\)};
	\vertex[right=1.3cm of a5](a8){\(\chi\)};
	
	\diagram* { 
		(a1) --[ thick, fermion, arrow size=1pt](a2) --[thick, fermion, arrow size=1pt,edge label=\(\psi\)](a3) --[thick, fermion, arrow size=1pt](a4),
		(a2) --[ thick, boson,edge label'=\(B_{\mu}\)](a5),
		(a3) --[thick, boson,edge label={\(B_{\mu}\)}](a6),
		(a7) --[thick, fermion, arrow size=1pt](a6) --[ thick, fermion, arrow size=1pt,edge label=\(\chi\)](a5) --[thick, fermion, arrow size=1pt](a8),
	};	
	\end{feynman}
	\end{tikzpicture}}
    %------------------------
    \subfloat[]{\begin{tikzpicture}
	\begin{feynman}
	\vertex (a1){\(\psi_1\)};
	\vertex[crossed dot,blue,right=1.3cm of a1](a2);
	\vertex[above=1.5cm of a2](a3);
	\vertex[left=1.cm of a3](a4){\(\psi_2\)};
	\vertex[crossed dot,thick,right=1.3cm of a2](a5){};
	\vertex[crossed dot,thick,right=1.3cm of a3](a6){};
	\vertex[right=1.3cm of a6](a7){\(\chi\)};
	\vertex[right=1.3cm of a5](a8){\(\chi\)};
	
	\diagram* { 
		(a1) --[ thick, fermion, arrow size=1pt](a3) --[thick, fermion, arrow size=1pt,edge label=\(\psi\)](a2) --[thick, fermion, arrow size=1pt](a4),
		(a2) --[ thick, boson,edge label'=\(B_{\mu}\)](a5),
		(a3) --[thick, boson,edge label={\(B_{\mu}\)}](a6),
		(a7) --[thick, fermion, arrow size=1pt](a6) --[ thick, fermion, arrow size=1pt,edge label=\(\chi\)](a5) --[thick, fermion, arrow size=1pt](a8),
	};	
	\end{feynman}
	\end{tikzpicture}}
    %------------------------
	\caption{Possible topologies for one-loop electroweak corrections to the four-fermion operator. The generic gauge boson line $X_{\mu}$ represents $B_{\mu}$, $W_{\mu}$, or $G_{\mu}$. The fermion fields $\psi_i$ are $q,l,u,d,e\,,\mathrm{and}\,~ \chi$. Here, the dashed line represents scalar fields.}
	\label{fig:ADM_gauge_correction}
\end{figure}
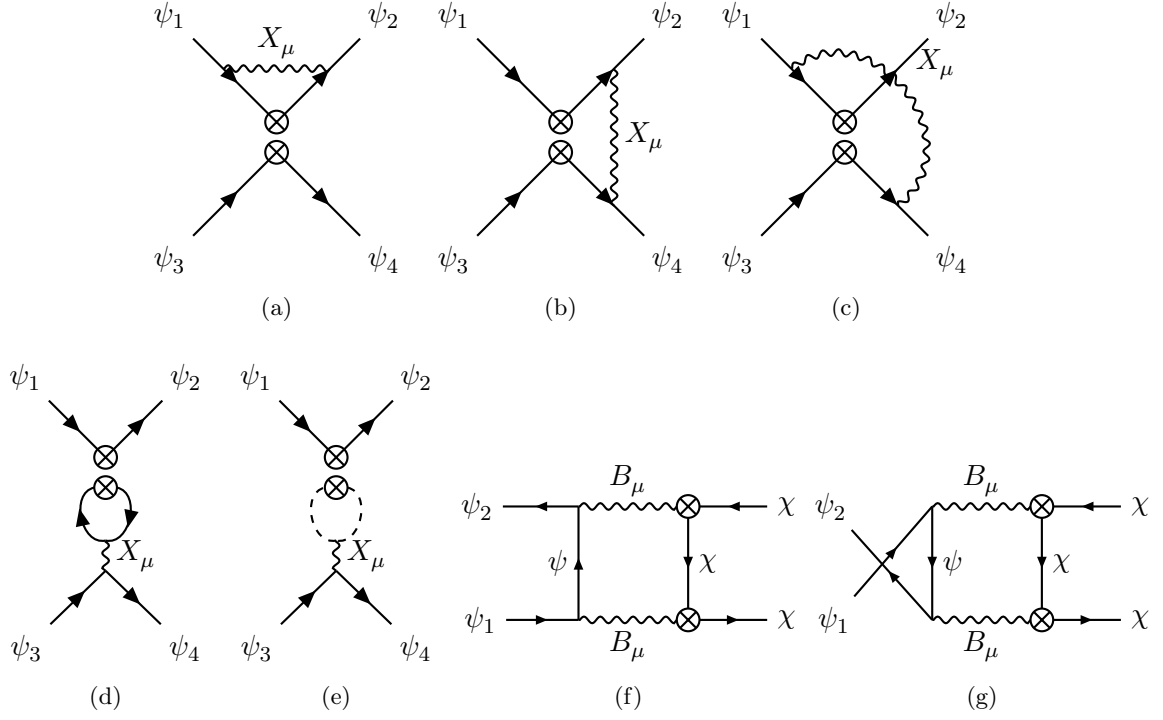

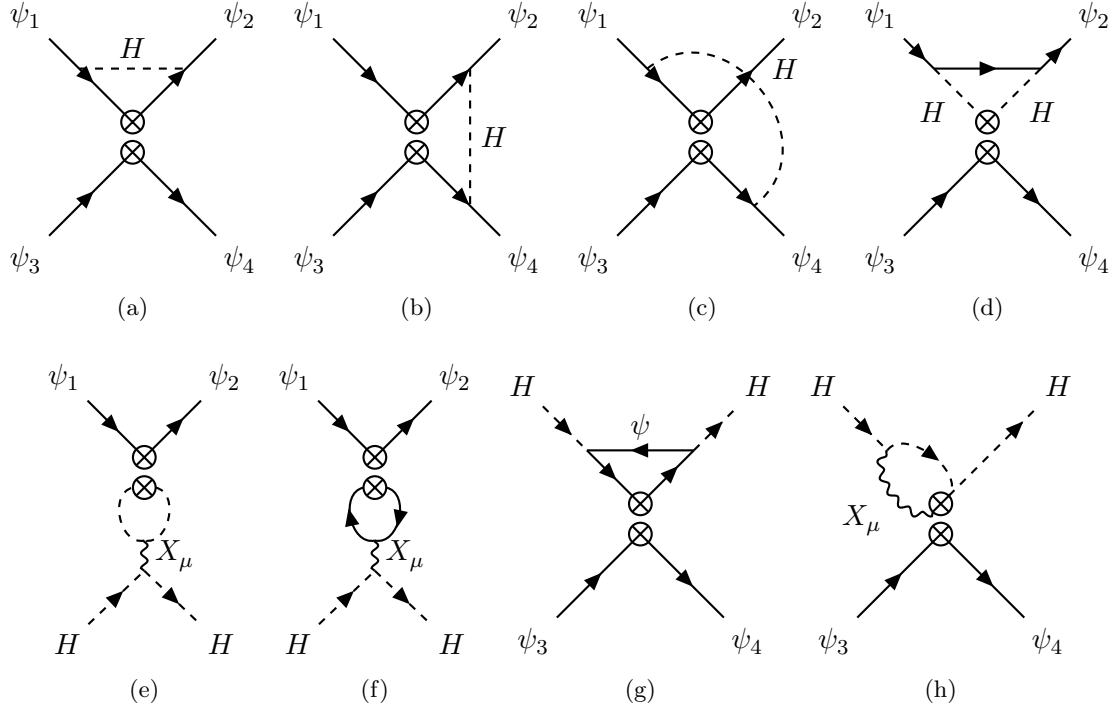
\begin{figure}[t]
	\centering
	%%%%%%%%%%%%%%%%%%%%%%%%%%%%%%
	\subfloat[]{\label{fig:four_fermi_f}
		\begin{tikzpicture}
			\begin{feynman}
				\vertex[thick, crossed dot](a){};
				\vertex[above left=2.cm of a] (b) {\(\psi_1\)};
				\vertex[above left=1cm of a](g);
				\vertex[above right=2.cm of a] (c) {\(\psi_2\)};
				\vertex[above right=1.cm of a](h);
				\vertex[thick, crossed dot, below=.4cm of a](d){};
				\vertex[below left=2.cm of d] (e) {\(\psi_3\)};
				\vertex[below right=2.cm of d] (f) {\(\psi_4\)};
				
				\diagram* {
					(b) -- [thick, fermion] (a) -- [thick, fermion] (c),
					(e) -- [thick, fermion] (d) -- [thick, fermion] (f),
					(g) --[thick, scalar, edge label=\(H\)](h),
				};
			\end{feynman}
		\end{tikzpicture}
	}
	%%%%%%%%%%%%%%%%%%%%%%%%
	\subfloat[]{\label{fig:four_fermi_g}
		\begin{tikzpicture}
			\begin{feynman}
				\vertex[thick, crossed dot](a){};
				\vertex[above left=2.cm of a] (b) {\(\psi_1\)};
				\vertex[above right=2.cm of a] (c) {\(\psi_2\)};
				\vertex[above right=1.cm of a] (g);
				\vertex[thick, crossed dot, below=.4cm of a](d){};
				\vertex[below left=2.cm of d] (e) {\(\psi_3\)};
				\vertex[below right=2.cm of d] (f) {\(\psi_4\)};
				\vertex[below right=1.cm of d] (h);
				\diagram* {
					(b) -- [thick, fermion] (a) -- [thick, fermion] (c),
					(e) -- [thick, fermion] (d) -- [thick, fermion] (f),
					(g) --[thick, scalar, edge label=\(H\)](h),
				};
			\end{feynman}
		\end{tikzpicture}
	}
	%%%%%%%%%%%%%%%%%%%%%%%%
	\subfloat[]{\label{fig:four_fermi_h}
		\begin{tikzpicture}
			\begin{feynman}
				\vertex[thick, crossed dot](a){};
				\vertex[above left=2.cm of a] (b) {\(\psi_1\)};
				\vertex[above left=1.cm of a] (g);
				\vertex[above right=2.cm of a] (c) {\(\psi_2\)};
				\vertex[thick, crossed dot, below=.4cm of a](d){};
				\vertex[below left=2.cm of d] (e) {\(\psi_3\)};
				\vertex[below right=2.cm of d] (f) {\(\psi_4\)};
				\vertex[below right=1.cm of d] (h);
				\diagram* {
					(b) -- [thick, fermion] (a) -- [thick, fermion] (c),
					(e) -- [thick, fermion] (d) -- [thick, fermion] (f),
					(g) --[thick, scalar,half left,looseness=1.5, edge label=\(H\)](h),
				};
			\end{feynman}
		\end{tikzpicture}
	}
	%%%%%%%%%%%%%%%%%%%%%%%%
	\subfloat[]{\label{fig:four_fermi_i}
		\begin{tikzpicture}
			\begin{feynman}
				\vertex[thick, crossed dot](a){};
				\vertex[above left=2.cm of a] (b) {\(\psi_1\)};
				\vertex[above left=1cm of a](g);
				\vertex[above right=2.cm of a] (c) {\(\psi_2\)};
				\vertex[above right=1.cm of a](h);
				\vertex[thick, crossed dot, below=.4cm of a](d){};
				\vertex[below left=2.cm of d] (e) {\(\psi_3\)};
				\vertex[below right=2.cm of d] (f) {\(\psi_4\)};
				
				\diagram* {
					(b) -- [thick, fermion] (g) -- [thick, fermion] (h) --[thick, fermion](c),
					(e) -- [thick, fermion] (d) -- [thick, fermion] (f),
					(g) --[thick, scalar, edge label'=\(H\)](a),
					(h) --[thick, scalar, edge label=\(H\)](a),
				};
			\end{feynman}
		\end{tikzpicture}
	}\\
	%%%%%%%%%%%%%%%%%%%%%%%%%%
	\subfloat[]{\label{fig:four_fermi_k}
		\begin{tikzpicture}
			\begin{feynman}
				\vertex[thick, crossed dot](a){};
				\vertex[above left=1.5cm of a] (b) {\(\psi_1\)};
				\vertex[above right=1.5cm of a] (c) {\(\psi_2\)};
				\vertex[thick, crossed dot, below=.4cm of a](d){};
				\vertex[below=0.7 cm of d](g);
				\vertex[below=1.1 cm of d](h);
				\vertex[below left=1.cm of h] (e) {\(H\)};
				\vertex[below right=1.cm of h] (f) {\(H\)};
				
				\diagram* {
					(b) -- [thick, fermion] (a) -- [thick, fermion] (c),
					(e) -- [thick, charged scalar] (h), (h) -- [thick, charged scalar] (f),
					(g) --[thick, scalar, half left, looseness=1.2](d) --[thick, scalar, half left, looseness=1.2](g),
					(g) --[thick, boson, edge label=\(X_{\mu}\)](h),
				};
			\end{feynman}
		\end{tikzpicture}
	}
	%%%%%%%%%%%%%%%%%%%%%%%%%%
	\subfloat[]{\label{fig:four_fermi_l}
		\begin{tikzpicture}
			\begin{feynman}
				\vertex[thick, crossed dot](a){};
				\vertex[above left=1.5cm of a] (b) {\(\psi_1\)};
				\vertex[above right=1.5cm of a] (c) {\(\psi_2\)};
				\vertex[thick, crossed dot, below=.4cm of a](d){};
				\vertex[below=0.7 cm of d](g);
				\vertex[below=1.1 cm of d](h);
				\vertex[below left=1.cm of h] (e) {\(H\)};
				\vertex[below right=1.cm of h] (f) {\(H\)};
				
				\diagram* {
					(b) -- [thick, fermion] (a) -- [thick, fermion] (c),
					(e) -- [thick, charged scalar] (h), (h) -- [thick, charged scalar] (f),
					(g) --[thick, fermion, half left, looseness=1.2](d) --[thick, fermion, half left, looseness=1.2](g),
					(g) --[thick, boson, edge label=\(X_{\mu}\)](h),
				};
			\end{feynman}
		\end{tikzpicture}
	}
	%%%%%%%%%%%%%%%%%%%%%%%%%%
	\subfloat[]{\label{fig:four_fermi_j}
		\begin{tikzpicture}
			\begin{feynman}
				\vertex[thick, crossed dot](a){};
				\vertex[above left=2.2cm of a] (b) {\(H\)};
				\vertex[above left=1cm of a](g);
				\vertex[above right=2.2cm of a] (c) {\(H\)};
				\vertex[above right=1.cm of a](h);
				\vertex[thick, crossed dot, below=.4cm of a](d){};
				\vertex[below left=2.cm of d] (e) {\(\psi_3\)};
				\vertex[below right=2.cm of d] (f) {\(\psi_4\)};
				
				\diagram* {
					(g) -- [thick, fermion] (a) -- [thick, fermion] (h),
					(e) -- [thick, fermion] (d) -- [thick, fermion] (f),
					(h) --[thick, fermion, edge label'=\(\psi\)](g),
					(b) --[thick, charged scalar](g),
					(h) --[thick, charged scalar](c),
				};
			\end{feynman}
		\end{tikzpicture}
	}
     %%%%%%%%%%%%%%%%%%%%%%%%%%
\subfloat[]{\label{fig:four_fermi_j}
		\begin{tikzpicture}
			\begin{feynman}
				\vertex[thick, crossed dot](a){};
				\vertex[above left=2.2cm of a] (b) {\(H\)};
				\vertex[above left=1.0cm of a](g);
				\vertex[above right=2.2cm of a] (c) {\(H\)};
				\vertex[above right=1.cm of a](h);
				\vertex[thick, crossed dot, below=.4cm of a](d){};
				\vertex[below left=2.cm of d] (e) {\(\psi_3\)};
				\vertex[below right=2.cm of d] (f) {\(\psi_4\)};
				
				\diagram* {
					% (g) -- [thick, fermion] (a) -- [thick, fermion] (h),
					(e) -- [thick, fermion] (d) -- [thick, fermion] (f),
					% (h) --[thick, fermion, edge label'=\(\psi\)](g),
					(b) --[thick, charged scalar](g),
					(h) --[thick, charged scalar](c),
                    (g) --[thick, half left, looseness=1., charged scalar](a),
                    (g) --[thick, half right, looseness=1., boson, edge label'=\(X_{\mu}\)](a),
                    (a) --[thick, scalar](h),
				};
			\end{feynman}
		\end{tikzpicture}
	}
    %%%%%%%%%%%%%%%%%%%%%%%%%%
	\caption{Possible topologies for the four-fermion and dimension-six Higgs current operators including Yukawa corrections. The dashed lines represent scalar fields, while $X_{\mu}$ denotes the SM gauge bosons.}
	\label{fig:ADM_Higgs_correction}
\end{figure}

Figure~\ref{fig:dim_five_ADM} illustrates the one-loop corrections to the dimension-five Higgs scalar operators, evaluated up to $\mathcal{O}(1/\Lambda^2)$. Furthermore, fig.~\ref{fig:ADM_gauge_correction} displays the one-loop corrections to the dimension-six operators arising from electroweak gauge boson exchanges, while fig.~\ref{fig:ADM_Higgs_correction} shows the corresponding corrections mediated by Yukawa interactions. 

\begin{figure}[htbp]
    \centering
    %%%%%%%%%%%%%%%%%%%%%%%%%%%%%%
    \subfloat[]{
    \begin{tikzpicture}
        \begin{feynman}
            \vertex (a) {\(\psi_i\)};
            \vertex [right=1 cm of a] (b);
            \vertex [right=1.5 cm of b] (c);
            \vertex [right=1 cm of c] (d){\(\psi_i\)};

            \diagram* {
                (a) -- [thick,fermion] (b) -- [thick, fermion, edge label'=\(\psi_j\)] (c) -- [thick,fermion] (d),
                (b) -- [thick,boson, half left, looseness=1.7, edge label=\(X_{\mu}\)] (c),
            };
        \end{feynman}
    \end{tikzpicture}
    }
    \hspace{1cm}
    %%%%%%%%%%%%%%%%%%%%%%%%%%%%%%
    \subfloat[]{
    \begin{tikzpicture}
        \begin{feynman}
            \vertex (a) {\(\psi_i\)};
            \vertex [right=1 cm of a] (b);
            \vertex [right=1.5 cm of b] (c);
            \vertex [right=1 cm of c] (d){\(\psi_i\)};

            \diagram* {
                (a) -- [thick,fermion] (b) -- [thick,fermion, edge label'=\(\psi_j\)] (c) -- [thick,fermion] (d),
                (b) -- [thick,scalar, half left, looseness=1.7, edge label=\(\phi\)] (c),
            };
        \end{feynman}
    \end{tikzpicture}
    }
    %%%%%%%%%%%%%%%%%%%%%%%%%%%%%%
    \caption{One-loop self-energy corrections to the fermion fields via gauge boson ($X_\mu$) and scalar ($\phi$) exchanges.}
    \label{fig:self_energy}
\end{figure}
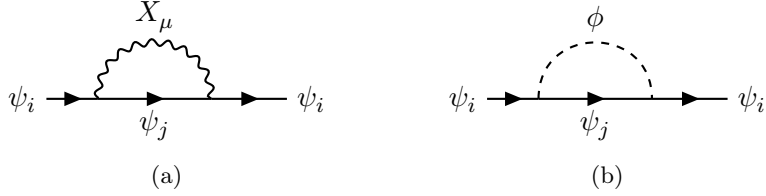

The determination of the anomalous dimension matrix also necessitates the evaluation of the wave-function renormalisation constants associated with the external fields appearing in the effective operators. These contributions arise from the one-loop self-energy corrections and enter the renormalisation of the operators through the corresponding field renormalisation factors. In this section, we display the relevant self-energy diagrams and summarise their contributions, which, when combined with the operator renormalisation effects, yield the complete RG equations for the WCs. The one-loop wavefunction renormalisation graphs for the SM fields are not depicted here; their explicit contributions will come from the diagrams \ref{fig:self_energy}. 
%---------------------------
\begin{figure}[t!]
	\centering
	%------------------
	\subfloat[]{\label{fig:dim_five_a}
		\begin{tikzpicture}
			\begin{feynman}
				\vertex[thick, crossed dot](a){};
				\vertex[above left=2.cm of a] (b) {\(\psi_1\)};
				\vertex[above right=2.cm of a] (c) {\(\psi_2\)};
				\vertex[thick, crossed dot, below=.4cm of a](d){};
				\vertex[below left=1cm of d](g);
				\vertex[below left=2.cm of d] (e) {\(H\)};
				\vertex[below right=1.cm of d](h);
				\vertex[below right=2.cm of d] (f) {\(H\)};
				
				\diagram* {
					(b) -- [thick, fermion] (a) -- [thick, fermion] (c),
					(e) -- [thick, charged scalar] (d) -- [thick, charged scalar] (f),
					(g) --[thick, boson, edge label'=\(X_{\mu}\)](h),
				};
			\end{feynman}
		\end{tikzpicture}
	}
	%------------------
	\subfloat[]{\label{fig:dim_five_b}
		\begin{tikzpicture}
			\begin{feynman}
				\vertex[thick, crossed dot](a){};
				\vertex[above left=1.5cm of a] (b) {\(\psi_1\)};
				\vertex[above right=1.5cm of a] (c) {\(\psi_2\)};
				\vertex[thick, crossed dot, below=.4cm of a](d){};
				\vertex[below=0.7 cm of d](g);
				\vertex[below=1. cm of d](h);
				\vertex[below left=1.cm of h] (e) {\(H\)};
				\vertex[below right=1.cm of h] (f) {\(H\)};
				
				\diagram* {
					(b) -- [thick, fermion] (a) -- [thick, fermion] (c),
					(e) -- [thick, charged scalar] (h), (h) -- [thick, charged scalar] (f),
					(h) --[thick, charged scalar, half left, looseness=1.4, edge label=\(H\)](d) --[thick, charged scalar, half left, looseness=1.4, edge label=\(H\)](h),
				};
			\end{feynman}
		\end{tikzpicture}
	}
	%-------------------------
	\subfloat[]{\label{fig:dim_five_c}
		\begin{tikzpicture}
			\begin{feynman}
				\vertex[thick,crossed dot](a){};
				\vertex[above left=2.cm of a] (b) {\(\psi_1\)};
				\vertex[above right=2.cm of a] (c) {\(\psi_2\)};
				\vertex[thick, crossed dot, above right=1.cm of a] (g){};
				\vertex[thick, crossed dot, below=.4cm of a](d){};
				\vertex[below left=2.cm of d] (e) {\(H\)};
				\vertex[below right=2.cm of d] (f) {\(H\)};
				\vertex[below right=1.cm of d] (h);
				\diagram* {
					(b) -- [thick, fermion] (a) -- [thick, fermion] (c),
					(e) -- [thick, charged scalar] (d) -- [thick, charged scalar] (f),
					(g) --[thick, boson, edge label=\(B_{\mu}\)](h),
				};
			\end{feynman}
		\end{tikzpicture}
	}\\
	%------------------
	\subfloat[]{\label{fig:dim_five_d}
		\begin{tikzpicture}
			\begin{feynman}
				\vertex[](a);
				\vertex[above left=2.cm of a] (b) {\(\psi_1\)};
				\vertex[thick, crossed dot, above left=1.cm of a](g){};
				\vertex[above right=2.cm of a] (c) {\(\psi_2\)};
				\vertex[thick, crossed dot, above right=1.cm of a](h){};
				\vertex[below left=1.5cm of a] (e) {\(H\)};
				\vertex[below right=1.5cm of a] (f) {\(H\)};
				
				\diagram* {
					(c) --[thick, fermion](h) --[thick, fermion](g) --[thick, fermion](b),
					(g) --[thick, boson,edge label'=\(B_{\mu}\)](a) --[thick, boson, edge label'=\(B_{\mu}\)](h),
					(f) --[thick, charged scalar](a) --[thick, charged scalar](e),
				};
			\end{feynman}
		\end{tikzpicture}
	}
	%------------------
	\subfloat[]{\label{fig:dim_five_e}
		\begin{tikzpicture}
			\begin{feynman}
				\vertex[thick, crossed dot](a){};
				\vertex[above left=1.4cm of a] (b) {\(\psi_1\)};
				\vertex[above right=1.4cm of a] (c) {\(H\)};
				\vertex[thick, crossed dot, below=.4cm of a](d);
				\vertex[thick, crossed dot, below=0.8 cm of d](h) {};
				\vertex[below left=1.4cm of h] (e) {\(\psi_3\)};
				\vertex[below right=1.4cm of h] (f) {\(H\)};
				
				\diagram* {
					(b) --[thick, fermion](a),
					(a) --[thick, fermion, half right, looseness=1.2, edge label'=\(\psi\)](h),
					(h) --[thick, fermion](e),
					(a) --[thick, charged scalar](c),
					(h) --[thick, charged scalar, half right, looseness=1.2,edge label'=\(H\)](a),
					(f) --[thick, charged scalar](h),
				};
			\end{feynman}
		\end{tikzpicture}
	}
	%----------------------
    \subfloat[]{\label{fig:dim_five_f}
		\begin{tikzpicture}
			\begin{feynman}
				\vertex[thick, crossed dot](a){};
				\vertex[above left=1.4cm of a] (b) {\(\psi_1\)};
				\vertex[above right=1.4cm of a] (c) {\(H\)};
				\vertex[thick, crossed dot, below=.4cm of a](d);
				\vertex[thick, crossed dot, below=0.8 cm of d](h) {};
				\vertex[below left=1.4cm of h] (e) {\(\psi_3\)};
				\vertex[below right=1.4cm of h] (f) {\(H\)};
				
				\diagram* {
					(b) --[thick, fermion](a),
					(a) --[thick, fermion, half right, looseness=1.2, edge label'=\(\psi\)](h),
					(h) --[thick, fermion](e),
					(c) --[thick, charged scalar](h),
					(a) --[thick, charged scalar](h),
					(f) --[thick, charged scalar](a),
				};
			\end{feynman}
		\end{tikzpicture}
	}
    %--------------------
	\caption{Possible topologies for the dimension-five Higgs scalar operator, including corrections up to $\mathcal{O}(1/\Lambda^2)$. Here $\psi_i$ is DM fermion.}
	\label{fig:dim_five_ADM}
\end{figure}
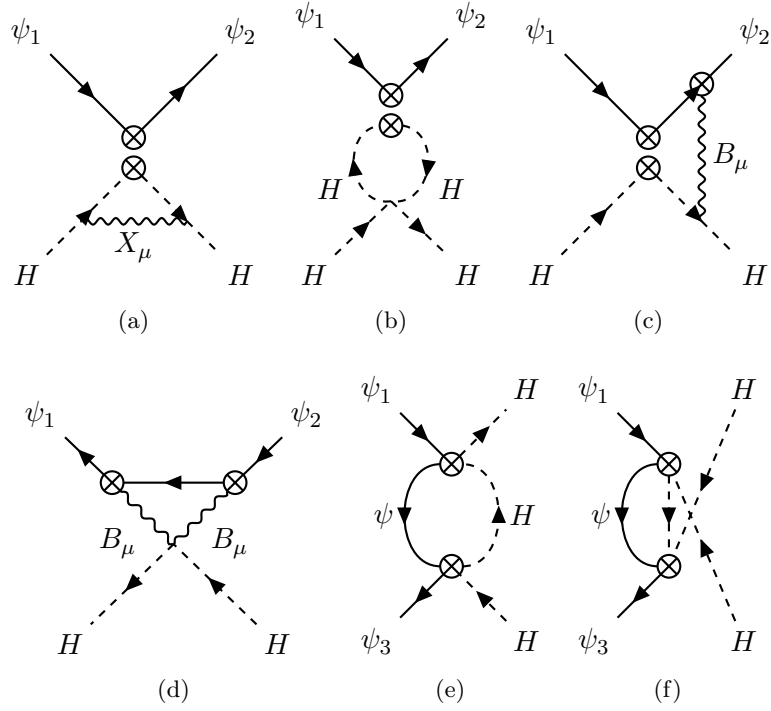
%----------------------------------

%----------------------------------
%-------------------------------

%------------------------------
%%%%%%%%%%%%%%%%%%%%%%%%%%%%%

%\subsection{Wave function renormalization}
%\label{Append:wavefunction_renorm}

As illustrated in fig.~\ref{fig:self_energy}, each fermion leg receives one-loop corrections via interactions with either the gauge bosons $X_{\mu}$ or the scalar sector (i.e., the Higgs boson $H$). A general expression for the wave function renormalisation of the fermionic fields $\psi_i \in \{l, e, q, u, d\}$ can be parameterized as
\begin{align}\label{eq:Wave_RG}
    \left(Z_{\rm WR}\right)_{\psi_i} = 1 + \frac{1}{\epsilon}\left(\frac{\alpha_n}{4\pi} b_{\psi_i}^n - \frac{\gamma^{(\Gamma)}_{rs}}{16\pi^2}\right) \,,
\end{align}
where $\alpha_n$ denotes the gauge couplings with $n=1,2,3$ for the $\mathrm{U}(1)_{\rm Y}$, $\mathrm{SU}(2)_{\rm L}$, and $\mathrm{SU}(3)_{\rm C}$ gauge groups, respectively. Expanding this for the SM fields yields the explicit divergent structures:

    \begin{align}
        \left(Z_{\rm WR}\right)_l &= 1 - \frac{1}{\epsilon}\left[ \frac{\alpha_1}{4\pi}(Y_l^2) + \frac{\alpha_2}{4\pi}\left(\frac{3}{4}\right) - \frac{1}{16\pi^2}\left(\frac{1}{2}\Gamma_e\Gamma_e^{\dagger}\right)_{rs} \right] \\
        \left(Z_{\rm WR}\right)_e &= 1 - \frac{1}{\epsilon}\left[ \frac{\alpha_1}{4\pi}(Y_e^2) - \frac{1}{16\pi^2}\left(2 \times \frac{1}{2}\Gamma_e^{\dagger}\Gamma_e\right)_{rs} \right] \\
        \left(Z_{\rm WR}\right)_q &= 1 - \frac{1}{\epsilon}\left[ \frac{\alpha_1}{4\pi}(Y_q^2) + \frac{\alpha_2}{4\pi}\left(\frac{3}{4}\right) + \frac{\alpha_3}{4\pi}\left(C_F\right) - \frac{1}{16\pi^2}\left(\frac{1}{2}\left(\Gamma_u\Gamma_u^{\dagger} + \Gamma_d \Gamma_d^{\dagger}\right)\right)_{rs} \right] \\
        \left(Z_{\rm WR}\right)_u &= 1 - \frac{1}{\epsilon}\left[ \frac{\alpha_1}{4\pi}(Y_u^2) + \frac{\alpha_3}{4\pi}\left(C_F\right) - \frac{1}{16\pi^2}\left(2 \times \frac{1}{2}\Gamma_u^{\dagger}\Gamma_u\right)_{rs} \right] \\
        \left(Z_{\rm WR}\right)_d &= 1 - \frac{1}{\epsilon}\left[ \frac{\alpha_1}{4\pi}(Y_d^2) + \frac{\alpha_3}{4\pi}\left(C_F\right) - \frac{1}{16\pi^2}\left(2 \times \frac{1}{2}\Gamma_d^{\dagger}\Gamma_d\right)_{rs} \right] \\
        \left(Z_{\rm WR}\right)_H &= 1 - \frac{1}{\epsilon}\left[ \frac{\alpha_1}{4\pi}\left(\frac{1}{2}\right) + \frac{\alpha_2}{4\pi}\left(\frac{3}{2}\right) - \frac{1}{16\pi^2}\mathrm{Tr}\left(N_c\Gamma_u\Gamma_u^{\dagger} + N_c\Gamma_d\Gamma_d^{\dagger} + \Gamma_e\Gamma_e^{\dagger}\right) \right]
    \end{align}

The QCD color factor is defined as $C_F = \frac{N_c^2 - 1}{2N_c}$, which evaluates to $C_F = \frac{4}{3}$ for $N_c = 3$.

%------------------------------
\section{List of Observables}
%-------------------------------
This appendix details the complete matching relations connecting the DMEFT WCs to the relevant low-energy observables. Because this connection is achieved via a two-step matching procedure onto the SMEFT, we also provide the tree-level matching expressions for the low-energy processes in terms of the standard SMEFT Warsaw basis \cite{Grzadkowski:2010es}. 
% Additionally, we provide the details of the experimental observables used in this analysis.
In our calculations, we utilise standard analytic one-loop functions, such as $\text{DiscB}$.
%-----------------------
\begin{align}
    \text{DiscB}[p^2,m_0,m_1]&=\frac{\sqrt{\lambda(p^2,m_0^2,m_1^2)}}{s}\log\Big[\frac{m_0^2+m_1^2-p^2+\sqrt{\lambda(p^2,m_0^2,m_1^2)}}{2 m_0 m_1}\Big]
\end{align}
where, $\lambda(x,y,z)$ is the standard K\"{a}ll\'{e}n kinematic function defined as:
\begin{align}
\lambda(x, y, z) &= x^2 + y^2 + z^2 - 2xy - 2xz - 2yz \,.
\end{align}
%-------------------------------------
\subsection{Electroweak Precision Observables}
%-------------------------------------
\subsubsection*{Matching Relations:} \label{Append:EWPO_match}
%-------------------------------------
The general $Zf\bar{f}$ effective Lagrangian introduced in Eq.~\eqref{eq:Z_eff_EWPO} receives NP contributions from two distinct sources: gauge boson self-energy corrections, parameterised by $s_{\rm eff}$, and direct vertex corrections, denoted by $\delta g_{L(R)}^{Zf}$. The matching relations for these new physics interactions in terms of the DMEFT WCs are given as follows:

\paragraph{Self-energy corrections of gauge bosons:} In Eq.~\eqref{eq:seff}, we established the relationship between $s_{\rm eff}^2$ and the various gauge boson self-energies. We now provide the corresponding one-loop matching relations. The general expression describing gauge boson self-energy corrections can be written as,
%-------------------------------------
\begin{align}
    \left(\Pi^{V_i V_j}_{\mu\nu}\right)(q^2)=-i g_{\mu\nu}\left(q^2-M_{V_i}^2\right)\delta_{V_i V_j}-i\left(g_{\mu\nu}-\frac{q_{\mu}q_{\nu}}{q^2}\right)\Pi_T^{V_i V_j}(q^2)-i\frac{q_{\mu}q_{\nu}}{q^2}\Pi_L^{V_i V_j}(q^2)
\end{align}
%-------------------------------------
where $V_{i,j}=\gamma,Z$ bosons respectively, and $M_{\gamma}=0$.
%-------------------------------------
\begin{subequations}
    \begin{align}\label{eq:Pi_gammagamma}
        % \Pi^{\gamma\gamma}_{\mu\nu}(q^2)&=\Pi^{\gamma\gamma}(q^2)\left(q^2 g^{\mu\nu}-q^{\mu}q^{\nu}\right)\approx\beta q^2 \left(q^2 g^{\mu\nu}-q^{\mu}q^{\nu}\right)\nonumber\\
        % \Pi^{\gamma\gamma}_T(q^2)&=\frac{ c_W^2 q^2}{\pi^2}\Bigg[ m_{\chi}^2 \left(-\left(|\mathcal{C}_{B\chi}^{5,1}|^2+|\mathcal{C}_{B\chi}^{5,2}|^2\right)+\left(|\mathcal{C}_{B\chi}^{5,1}|^2-|\mathcal{C}_{B\chi}^{5,2}|^2\right)\log\frac{\mu^2}{m_{\chi}^2}\right)\\
        % &+\frac{q^2}{3}\left(|\mathcal{C}_{B\chi}^{5,1}|^2+\frac{1}{2}\left(|\mathcal{C}_{B\chi}^{5,1}|^2-|\mathcal{C}_{B\chi}^{5,2}|^2\right)\log\frac{\mu^2}{m_{\chi}^2}\right)\bigg]\nonumber
        \Pi^{\gamma\gamma}_T(q^2)&=q^2\,\Pi^{\gamma\gamma}(q^2)=\frac{ c_W^2 q^2}{\pi^2}\Bigg[\frac{q^2}{3}\left(|\mathcal{C}_{B\chi}^{5,1}|^2+\frac{1}{2}\left(|\mathcal{C}_{B\chi}^{5,1}|^2+|\mathcal{C}_{B\chi}^{5,2}|^2\right)\log\frac{\mu^2}{m_{\chi}^2}\right)\bigg]
    \end{align}

    \begin{align}\label{eq:Pi_ZZ}
        % &\Pi_{\mu\nu}^{ZZ}(q^2)=\Pi^{ZZ}(q^2)\left(q^2 g^{\mu\nu}-q^{\mu}q^{\nu}\right)\nonumber\\
        &\Pi^{ZZ}_T(q^2)=q^2\,\Pi^{ZZ}(q^2)=\frac{ s_W^2 q^2}{\pi^2}\Bigg[m_{\chi}^2 \left(-\left(|\mathcal{C}_{B\chi}^{5,1}|^2+|\mathcal{C}_{B\chi}^{5,2}|^2\right)+\left(|\mathcal{C}_{B\chi}^{5,1}|^2-|\mathcal{C}_{B\chi}^{5,2}|^2\right)\log\frac{\mu^2}{m_{\chi}^2}\right)\\
       &+\frac{q^2}{3}\left(|\mathcal{C}_{B\chi}^{5,1}|^2+\frac{1}{2}\left(|\mathcal{C}_{B\chi}^{5,1}|^2+|\mathcal{C}_{B\chi}^{5,2}|^2\right)\log\frac{\mu^2}{m_{\chi}^2}\right)\bigg]-\frac{q^2}{48 \pi^2}(g_1^2+g_2^2)v^2|\mathcal{C}_{\chi D H}|^2\left(\log\frac{\mu^2}{m_{\chi}^2}\right)\nonumber
    \end{align}

    \begin{align}\label{eq:Pi_gammaZ}
        % &\Pi_{\mu\nu}^{\gamma Z}(q^2)=\Pi^{\gamma Z}(q^2)\left(q^2 g^{\mu\nu}-q^{\mu}q^{\nu}\right)\nonumber\\
        &\Pi^{\gamma Z}_T(q^2)=q^2\,\Pi^{\gamma Z}(q^2)= \frac{c_W s_W q^2}{\pi^2}\Bigg[m_{\chi}^2 \left(-\left(|\mathcal{C}_{B\chi}^{5,1}|^2+|\mathcal{C}_{B\chi}^{5,2}|^2\right)+\left(|\mathcal{C}_{B\chi}^{5,1}|^2-|\mathcal{C}_{B\chi}^{5,2}|^2\right)\log\frac{\mu^2}{m_{\chi}^2}\right)\\
        &+\frac{q^2}{3}\left(|\mathcal{C}_{B\chi}^{5,1}|^2+\frac{1}{2}\left(|\mathcal{C}_{B\chi}^{5,1}|^2+|\mathcal{C}_{B\chi}^{5,2}|^2\right)\log\frac{\mu^2}{m_{\chi}^2}\right)\bigg]+\sqrt{g_1^2+g_2^2}\, v\frac{c_W q^2}{4\pi^2} m_{\chi}\,\mathcal{C}_{B\chi}^{5,1}\mathcal{C}_{\chi D H}\nonumber
    \end{align}
\end{subequations}
%--------------------------------
% Fit results at different benchmark scenarios are, 
% \begin{table}[h]
% \footnotesize
% \centering
% \begin{tabular}{l c c c c}
% \toprule
% \makecell{\textbf{$\Lambda$}\\ (TeV)} &
% \makecell{\textbf{$m_{\chi}$}\\ (GeV)} &
% \makecell{\textbf{$\mathcal{C}_{B\chi}^{5,1}\times 10^3$}\\ $(\mathrm{GeV}^{-1})$} &
% \makecell{\textbf{$\mathcal{C}_{B\chi}^{5,2}\times 10^3$}\\ $(\mathrm{GeV}^{-1})$} &
% \makecell{\textbf{$\mathcal{C}_{\chi DH}^{6,1}\times 10^3$}\\ $(\mathrm{GeV}^{-2})$}
% \\
% \midrule
%  & $1$ & $(0.0\pm 0.22)$&  $(0.0\pm 0.23)$ & $(0.0\pm 0.37)$\\
%  $1$ & $250$ & $(0.0\pm 0.16)$ & $(-0.38 \pm 0.07)$ & $(0.0\pm 0.82)$ \\
%  & $800$ & $(0.23\pm 0.07)$ & $(0.14 \pm 0.04)$ & $(0.0\pm 2.04)$\\
% \bottomrule
% \end{tabular}
% \caption{Constraints on NP operators from EWPOs.}
% \label{tab:EWPO_fit}
% \end{table}

%----------------------------
\paragraph{Vertex corrections of the gauge boson:} The following DMEFT operators contribute direct vertex corrections $\delta g_{L(R)}$.
%-----------------------------
\begin{subequations}\label{eq:delta_gLR}
\begin{align}
    &\delta g_L^{Zf_i f_j}=\frac{c_W}{2g_Z\pi^2}m_{\chi}M_Z^2 \mathcal{C}_{B\chi}^{(5,1)}\mathcal{C}_{\overset{\chi\ell}{pqij}}^{(6,1)}\left(2+\mathrm{DiscB}[M_Z^2,m_{\chi},m_{\chi}]+\log\frac{\Lambda^2}{m_{\chi}^2}\right)  \,,\\
    &+\frac{v^2}{72\pi^2}\Bigg[\mathcal{C}_{\overset{\chi\ell}{pqij}}^{(6,1)}\mathcal{C}_{\chi D H}^{(6,1)}\left(5 M_Z^2+12 m_{\chi}^2+3\left(M_Z^2+2 m_{\chi}^2 \right)\mathrm{DiscB}[M_Z^2,m_{\chi},m_{\chi}]+3M_Z^2\log\frac{\Lambda^2}{m_{\chi}^2}\right)\nonumber\\
    &+\mathcal{C}_{\overset{\chi\ell}{pqij}}^{(6,2)}\mathcal{C}_{\chi D H}^{(6,2)}\left(5 M_Z^2-24 m_{\chi}^2+3 \left(M_Z^2-4m_{\chi}^2\right) \mathrm{DiscB}[M_Z^2,m_{\chi},m_{\chi}]+ 3(M_Z^2-6 m_{\chi}^2)\log\frac{\Lambda^2}{m_{\chi}^2}\right)\Bigg]\nonumber
\end{align}

\begin{align}
    &\delta g_R^{Zf_i f_j}=  \frac{c_W}{2g_Z\pi^2}m_{\chi}M_Z^2 \mathcal{C}_{B\chi}^{(5,1)}\mathcal{C}_{\overset{\chi e}{pqij}}^{(6,1)}\left(2+\mathrm{DiscB}[M_Z^2,m_{\chi},m_{\chi}]+\log\frac{\Lambda^2}{m_{\chi}^2}\right) \,\\
    &+\frac{v^2}{72\pi^2}\Bigg[\mathcal{C}_{\overset{\chi e}{pqij}}^{(6,1)}\mathcal{C}_{\chi D H}^{(6,1)}\left(5 M_Z^2+12 m_{\chi}^2+3\left(M_Z^2+2 m_{\chi}^2 \right)\mathrm{DiscB}[M_Z^2,m_{\chi},m_{\chi}]+3M_Z^2\log\frac{\Lambda^2}{m_{\chi}^2}\right)\nonumber\\
    &+\mathcal{C}_{\overset{\chi e}{pqij}}^{(6,2)}\mathcal{C}_{\chi D H}^{(6,2)}\left(5 M_Z^2-24 m_{\chi}^2+3 \left(M_Z^2-4m_{\chi}^2\right) \mathrm{DiscB}[M_Z^2,m_{\chi},m_{\chi}]+ 3(M_Z^2-6 m_{\chi}^2)\log\frac{\Lambda^2}{m_{\chi}^2}\right)\Bigg]\nonumber
\end{align}
\end{subequations}
%-----------------------------
Equation~\eqref{eq:delta_gLR} defines the general vertex modification $\delta g_{L(R)}^{Zf_i f_j}$, neglecting light lepton and quark masses. For standard EWPOs, these couplings are flavour-conserving ($i=j$ with $i=e,\mu,\tau$). However, our inclusion of $Z$-LFV observables activates the flavour-violating ($i \neq j$) components of these interactions. In the flavour-conserving case relevant for EWPOs, the specific combinations of couplings that enter the matching relations are shown in eq.~\eqref{eq:Z_coupling_shifts}:
{\footnotesize
\begin{align}\label{eq:Z_coupling_shifts}
    \delta g_L^{Zee}&\to \left\{\mathcal{C}_{B\chi}^{(5,1)}\mathcal{C}_{\chi \underset{11}{\ell}}^{(6,1)}\,,\,\mathcal{C}_{\chi D H}^{(6,1)}\mathcal{C}_{\chi \underset{11}{\ell}}^{(6,1)}\,,\,\mathcal{C}_{\chi D H}^{(6,2)}\mathcal{C}_{\chi \underset{11}{\ell}}^{(6,2)}\right\}\,,&\quad \delta g_R^{Zee}&\to \left\{\mathcal{C}_{B\chi}^{(5,1)}\mathcal{C}_{\chi \underset{11}{e}}^{(6,1)}\,,\,\mathcal{C}_{\chi D H}^{(6,1)}\mathcal{C}_{\chi \underset{11}{e}}^{(6,1)}\,,\,\mathcal{C}_{\chi D H}^{(6,2)}\mathcal{C}_{\chi \underset{11}{e}}^{(6,2)}\right\}\,,\nonumber\\
    \delta g_L^{Z\mu\mu}&\to \left\{\mathcal{C}_{B\chi}^{(5,1)}\mathcal{C}_{\chi \underset{22}{\ell}}^{(6,1)}\,,\,\mathcal{C}_{\chi D H}^{(6,1)}\mathcal{C}_{\chi \underset{22}{\ell}}^{(6,1)}\,,\,\mathcal{C}_{\chi D H}^{(6,2)}\mathcal{C}_{\chi \underset{22}{\ell}}^{(6,2)}\right\}\,,&\quad \delta g_R^{Z\mu\mu}&\to \left\{\mathcal{C}_{B\chi}^{(5,1)}\mathcal{C}_{\chi \underset{22}{e}}^{(6,1)}\,,\,\mathcal{C}_{\chi D H}^{(6,1)}\mathcal{C}_{\chi \underset{22}{e}}^{(6,1)}\,,\,\mathcal{C}_{\chi D H}^{(6,2)}\mathcal{C}_{\chi \underset{22}{e}}^{(6,2)}\right\}\,,\nonumber\\
    \delta g_L^{Z\tau\tau}&\to \left\{\mathcal{C}_{B\chi}^{(5,1)}\mathcal{C}_{\chi \underset{33}{\ell}}^{(6,1)}\,,\,\mathcal{C}_{\chi D H}^{(6,1)}\mathcal{C}_{\chi \underset{33}{\ell}}^{(6,1)}\,,\,\mathcal{C}_{\chi D H}^{(6,2)}\mathcal{C}_{\chi \underset{33}{\ell}}^{(6,2)}\right\}\,,&\quad \delta g_R^{Z\tau\tau}&\to \left\{\mathcal{C}_{B\chi}^{(5,1)}\mathcal{C}_{\chi \underset{33}{e}}^{(6,1)}\,,\,\mathcal{C}_{\chi D H}^{(6,1)}\mathcal{C}_{\chi \underset{33}{e}}^{(6,1)}\,,\,\mathcal{C}_{\chi D H}^{(6,2)}\mathcal{C}_{\chi \underset{33}{e}}^{(6,2)}\right\}\,,\nonumber\\
    % \delta g_L^{Zuu}&\to \left\{\mathcal{C}_{B\chi}^{(5,1)}\mathcal{C}_{\chi \underset{11}{q}}^{(6,1)}\,,\,\mathcal{C}_{\chi D H}^{(6,1)}\mathcal{C}_{\chi \underset{11}{q}}^{(6,1)}\,,\,\mathcal{C}_{\chi D H}^{(6,2)}\mathcal{C}_{\chi \underset{11}{q}}^{(6,2)}\right\}\,,&\quad \delta g_R^{Zuu}&\to \left\{\mathcal{C}_{B\chi}^{(5,1)}\mathcal{C}_{\chi \underset{11}{u}}^{(6,1)}\,,\,\mathcal{C}_{\chi D H}^{(6,1)}\mathcal{C}_{\chi \underset{11}{u}}^{(6,1)}\,,\,\mathcal{C}_{\chi D H}^{(6,2)}\mathcal{C}_{\chi \underset{11}{u}}^{(6,2)}\right\}\,,\nonumber\\
    % \delta g_L^{Zcc}&\to \left\{\mathcal{C}_{B\chi}^{(5,1)}\mathcal{C}_{\chi \underset{22}{q}}^{(6,1)}\,,\,\mathcal{C}_{\chi D H}^{(6,1)}\mathcal{C}_{\chi \underset{22}{q}}^{(6,1)}\,,\,\mathcal{C}_{\chi D H}^{(6,2)}\mathcal{C}_{\chi \underset{22}{q}}^{(6,2)}\right\}\,,&\quad \delta g_R^{Zcc}&\to \left\{\mathcal{C}_{B\chi}^{(5,1)}\mathcal{C}_{\chi \underset{22}{u}}^{(6,1)}\,,\,\mathcal{C}_{\chi D H}^{(6,1)}\mathcal{C}_{\chi \underset{22}{u}}^{(6,1)}\,,\,\mathcal{C}_{\chi D H}^{(6,2)}\mathcal{C}_{\chi \underset{22}{u}}^{(6,2)}\right\}\,,\nonumber\\
     \delta g_L^{Zbb}&\to \left\{\mathcal{C}_{B\chi}^{(5,1)}\mathcal{C}_{\chi \underset{33}{q}}^{(6,1)}\,,\,\mathcal{C}_{\chi D H}^{(6,1)}\mathcal{C}_{\chi \underset{33}{q}}^{(6,1)}\,,\,\mathcal{C}_{\chi D H}^{(6,2)}\mathcal{C}_{\chi \underset{33}{q}}^{(6,2)}\right\}\,,&\quad \delta g_R^{Zbb}&\to \left\{\mathcal{C}_{B\chi}^{(5,1)}\mathcal{C}_{\chi \underset{33}{d}}^{(6,1)}\,,\,\mathcal{C}_{\chi D H}^{(6,1)}\mathcal{C}_{\chi \underset{33}{d}}^{(6,1)}\,,\,\mathcal{C}_{\chi D H}^{(6,2)}\mathcal{C}_{\chi \underset{33}{d}}^{(6,2)}\right\}\,.
\end{align}
}

 Again these modified couplings $\delta g_{L(R)}^{Z f_i f_j}$ can be written in terms of SMEFT WCs as,
 %---------------------------
\begin{align}
    \delta g_{L}^{Z \ell_i \ell_j}&=-\frac{v^2}{2} \left(\mathcal{C}_{\overset{\phi l}{ij}}^{(1)}+\mathcal{C}_{\overset{\phi l}{ij}}^{(3)}\right) \,,\quad &
    \delta g_{R}^{Z \ell_i \ell_j}&=-\frac{v^2}{2} \mathcal{C}_{\overset{\phi e}{ij}}^{(1)}\,,\\
    \delta g_{L}^{Z d}&=-\frac{v^2}{2} \left(\mathcal{C}_{\overset{\phi q}{ii}}^{(1)}+\mathcal{C}_{\overset{\phi q}{ii}}^{(3)}\right) \,,\quad &
    \delta g_{R}^{Z d}&=-\frac{v^2}{2} \mathcal{C}_{\overset{\phi d}{ii}}^{(1)}\,.
\end{align}
 %---------------------------

 %-------------------------------------
\subsubsection*{Observables:}\label{Append:EWPO_obs}
%-------------------------------------
The $Z$-pole observables considered in our analysis are summarised in Table~\ref{tab:Z_pole}. These consist of the width ratios ($R$), the various asymmetries ($A, A^{\rm FB}$), and the partial decay widths ($\Gamma$). Furthermore, we incorporate the $\Delta r$ parameter to account for associated electroweak corrections to the $W$-boson mass. 
\begin{align} \label{eq:deltar_exp}
    M_W^2 \left( 1 - \frac{M_W^2}{M_Z^2} \right) = \frac{\pi \alpha_{em}}{\sqrt{2} G_F} \frac{1}{1 - \Delta r}.
\end{align}

The parameter $\Delta r$ can also be written in terms of oblique parameters,

\begin{align} \label{eq:deltar_theo}
    \Delta r=\Delta r_{\rm SM}+\frac{\alpha_{em}}{s_W^2}\left( \frac{S}{2}-c_W^2 T-\frac{c_W^2-s_W^2}{4s_W^2}U\right)
\end{align}
where the oblique parameters are,
\begin{align}\label{eq:oblique_param}
    S&=\left(\frac{4 s_W^2 c_W^2}{\alpha_{em}}\right) \left(\Bigg[\frac{ \Pi^{ZZ}_T(M_Z^2)- \Pi_T^{ZZ}(0)}{M_Z^2}\Bigg]-\frac{c_W^2-s_W^2}{c_W s_W}\frac{\Pi_T^{\gamma Z}(M_Z^2)}{M_Z^2}-\frac{\Pi_T^{\gamma\gamma}(M_Z^2)}{M_Z^2} \right)\,, \nonumber \\
    T&=\frac{1}{\alpha_{em}}\left(\frac{\Pi_T^{WW}(0)}{M_W^2} -\frac{\Pi^T_{ZZ}(0)}{M_Z^2}\right)\,,\\
    U&=\frac{4 s_W^2}{\alpha_{em}} \left(\Bigg[\frac{\Pi_T^{WW}(M_W^2)- \Pi_T^{WW}(0)}{M_W^2}\Bigg]-\frac{c_W}{s_W}\frac{\Pi_T^{\gamma Z}(M_Z^2)}{M_Z^2}-\frac{ \Pi_T^{\gamma\gamma}(M_Z^2)}{M_Z^2}\right)-S\,. \nonumber
\end{align}

We derive the $\Delta r$ parameter for different experimental measurements of the $W$-boson mass via eq.~\eqref{eq:deltar_exp}, using $M_Z = 91.1884 \pm 0.0019~\mathrm{GeV}$~ \cite{ParticleDataGroup:2024cfk}, and summarize the results in Table~\ref{tab:W_mass_deltar}. We then relate these empirical values to the underlying new physics; specifically, the DMEFT operators that modify the gauge boson self-energies, which in turn modify the theoretical expectation for $\Delta r$ via eq.~\eqref{eq:deltar_theo}.

%--------------------------------------

%--------------------------------
\begin{table}[h]
\centering
\footnotesize
\renewcommand{\arraystretch}{1.25}
\begin{tabular}{c c c}
\toprule
\textbf{$M_W\,\mathrm{(GeV)}$} & \textbf{Reference} & $\Delta r$ \\
\midrule
$80.356 \pm 0.005$   & SM\,\cite{ParticleDataGroup:2024cfk} & $-0.032000 \pm 0.001605$ \\
$80.3602 \pm 0.0099$ & CMS\,\cite{CMS:2024lrd}              & $-0.032266 \pm 0.001696$ \\
$80.4335 \pm 0.0094$ & CDF\,\cite{CDF:2022hxs}              & $-0.036966 \pm 0.001706$ \\
$80.354 \pm 0.030$   & LHCb\,\cite{LHCb:2021bjt}            & $-0.031872 \pm 0.002471$ \\
$80.3665 \pm 0.0159$ & ATLAS\,\cite{ATLAS:2024erm}          & $-0.032668 \pm 0.001873$ \\
$80.367 \pm 0.023$   & D0\,\cite{D0:2012kms}                & $-0.032699 \pm 0.002152$ \\
\bottomrule
\end{tabular}
\caption{Measured values of the $W$-boson mass from different experimental collaborations and the corresponding extracted values of the $\Delta r$ parameter using eq.~\eqref{eq:deltar_exp}.}
\label{tab:W_mass_deltar}
\end{table}
%--------------------------------
%-------------------------------------
\subsection{Rare \texorpdfstring{$B$}{B} decays}\label{Append:Rare_B_decay}
%-------------------------------------

\subsubsection*{Matching Relations:}
The general effective Hamiltonian describing the $b \to d_i \ell^+ \ell^-$ transition is given in Eq.~\eqref{eq:Heff_b2di}. As previously discussed, within the WET basis, our new physics framework generates contributions exclusively to the vector and axial-vector operators. The matching relations in terms of the DMEFT basis,

\begin{subequations}
  \begin{align}
    C_9 &= \lambda_{c} \, \frac{m_{\chi}^2}{\pi^2} \, \log\frac{\mu^2}{m_{\chi}^2}\, \left[ \mathcal{C}_{\chi \underset{i3}{q}}^{(6,2)} \left( \mathcal{C}_{\chi \underset{22}{e}}^{(6,2)} + \mathcal{C}_{\chi \underset{22}{\ell}}^{(6,2)}\right) + \frac{4 c_W \sqrt{\pi \alpha }}{m_{\chi}} \mathcal{C}_{\chi \underset{i3}{q}}^{(6,1)} \mathcal{C}_{B \chi}^{(5,1)}\right],\\
    C_9^{\prime}&=  \lambda_{c} \, \frac{m_{\chi}^2}{\pi^2} \, \log\frac{\mu^2}{m_{\chi}^2}\, \left[  \mathcal{C}_{\chi \underset{i3}{d}}^{(6,2)} \left( \mathcal{C}_{\chi \underset{22}{e}}^{(6,2)} + \mathcal{C}_{\chi \underset{22}{\ell}}^{(6,2)}\right) + \frac{4 c_W \sqrt{\pi \alpha }}{m_{\chi}} \mathcal{C}_{\chi \underset{i3}{d}}^{(6,1)} \mathcal{C}_{B \chi}^{(5,1)}\right]\,,\\
    C_{10}&= \lambda_{c} \, \frac{m_{\chi}^2}{\pi^2} \, \log\frac{\mu^2}{m_{\chi}^2}\,  \mathcal{C}_{\chi \underset{i3}{q}}^{(6,2)} \left( \mathcal{C}_{\chi \underset{22}{e}}^{(6,2)} - \mathcal{C}_{\chi \underset{22}{\ell}}^{(6,2)} \right)  \,,\\
    C_{10}^{\prime}&= \lambda_{c} \, \frac{m_{\chi}^2}{\pi^2} \, \log\frac{\mu^2}{m_{\chi}^2}\,  \mathcal{C}_{\chi \underset{i3}{d}}^{(6,2)} \left( \mathcal{C}_{\chi \underset{22}{e}}^{(6,2)} - \mathcal{C}_{\chi \underset{22}{\ell}}^{(6,2)} \right)  \,,
\end{align}
with the normalisation constant $\lambda_c = \sqrt{2} \pi / \alpha G_{F} V_{tb} V_{td_j}^*$. 
In terms of the SMEFT basis, matching relations are, 
\begin{align}\label{eq:b2dill_SMEFT}
   \lambda~ C_9 &= \left( \mathcal{C}^{(1)}_{\overset{lq}{\ell \ell d_i b}}+\mathcal{C}^{(3)}_{\overset{lq}{\ell \ell d_i b}}+\mathcal{C}_{\overset{qe}{d_i b\ell\ell}}\right)-\frac{g_Z^2 v^2}{2 M_Z^2}\left(\frac{1}{2}+2 ~q_e s_W^2\right)\left(\mathcal{C}_{\overset{\phi q}{d_i b}}^{(1)}+\mathcal{C}_{\overset{\phi q}{d_i b}}^{(3)}\right)\,,\\
    \lambda~C_9^{\prime}&= \left( \mathcal{C}_{\overset{ld}{\ell\ell d_i b}}+\mathcal{C}_{\overset{ed}{\ell\ell d_i b} }\right)-\frac{g_Z^2 v^2}{2 M_Z^2}\left(\frac{1}{2}+2 ~q_e s_W^2\right) \mathcal{C}_{\overset{\phi d}{d_i b}}\,,\\
    \lambda~C_{10}&=\left(- \mathcal{C}^{(1)}_{\overset{lq}{\ell \ell d_i b}}-\mathcal{C}^{(3)}_{\overset{lq}{\ell \ell d_i b}}+\mathcal{C}_{\overset{qe}{d_i b\ell\ell}}\right)+\frac{g_Z^2 v^2}{4 M_Z^2}\left(\mathcal{C}_{\overset{\phi q}{d_i b}}^{(1)}+\mathcal{C}_{\overset{\phi q}{d_i b}}^{(3)}\right) \,,\\
    \lambda~ C_{10}^{\prime}&=  \left( -\mathcal{C}_{\overset{ld}{\ell\ell d_i b}}+\mathcal{C}_{\overset{ed}{\ell\ell d_i b} }\right)+\frac{g_Z^2 v^2}{4 M_Z^2}\mathcal{C}_{\overset{\phi d}{d_i b}} \,.
\end{align}
where the prefactor $\lambda$ is,
\begin{align}
    \lambda \equiv -\frac{8 G_F}{\sqrt{2}}\frac{e^2}{16\pi^2} V_{tb} V_{t d_j}^*\,
\end{align}
\end{subequations}
In Eq.~\eqref{eq:b2dill_SMEFT}, the terms in the first set of parentheses represent the direct contributions from the four-fermion operators, while the terms in the second set correspond to the $Z$-boson mediated contributions to the $b \to d_i \ell^+ \ell^-$ process.
%-----------------------------------
\subsubsection*{Observables:} \label{Append:rare_B_obs}
%----------------------------------
Our analysis begins with the observables driven by $b \to s \ell^+ \ell^-$ transitions. A major fraction of these constraints originates from the semileptonic decays $B \to K^{(*)}\mu^+\mu^-$ and $B_s \to \phi \mu^+\mu^-$. The relevant datasets include binned differential decay rates, optimised angular observables, and both CP-averaged and CP-asymmetric quantities. The theoretical frameworks for these decays are well-established and have been discussed extensively in the literature. The expressions for the differential decay rates as functions of the dilepton invariant mass squared, $q^2$, for the decays $B\to K\mu^+\mu^-$ and $B\to K^{*}\mu^+\mu^-$, taken from refs.~\cite{Bobeth:2007dw} and \cite{Altmannshofer:2008dz}, respectively.

The theoretical expression for the branching fraction of the rare dileptonic decay $B_q \to \mu^+ \mu^-$ can be written as 
\begin{equation}
	\begin{split} \label{eq:BR_Bqll}
		\mathcal{B}(B_q \to \mu^+ \mu^-) = & \tau_{B_q} f_{B_q}^2 m_{B_q}^3 \frac{G_F^2 \alpha^2}{64 \pi^3} |V^*_{tq}V_{tb}|^2 \beta_{\mu}(m_{B_q}^2) \left[ \frac{m_{B_q}^2}{m_b^2} |C_S - C'_S|^2 \left(1-\frac{4m_{\mu}^2}{m_{B_q}^2}\right) \right.\\ 
		& \left. + \bigg|\frac{m_{B_q}}{m_b}(C_P - C'_P) + 2\frac{m_{\mu}}{m_{B_q}} (C_{10} - C'_{10})\bigg|^2 \right]\,,
	\end{split}
\end{equation}
%----------------------------------------
where $\beta_{\mu}(q^2) = \sqrt{1 - \frac{4 m_{\mu}^2}{q^2}}$, and the $B_q$ meson decay constant is defined via the matrix element 
\begin{equation}
	\langle 0| \bar{q} \gamma_{\mu} P_L b | B_q(p) \rangle = \frac{i}{2} f_{B_q} p_\mu \,,
\end{equation}
with $q \in \{d, s\}$. In our numerical analysis, we adopt the decay constant values $f_{B} = (190.0 \pm 1.3)\,\text{MeV}$ and $f_{B_s} = (230.3 \pm 1.3)\,\text{MeV}$~\cite{FlavourLatticeAveragingGroupFLAG:2024oxs}.

The resulting constraints are summarised in Table~\ref{tab:Rare_B_decays}, highlighting the sensitivity of low-energy flavour observables to various combinations of DMEFT WCs. In particular, the semileptonic channels are sensitive to the products $\mathcal{C}_{\overset{\chi q(d)}{pqi3}}^{(6,2)}\mathcal{C}_{\overset{\chi \ell}{pq22}}^{(6,2)}$ and $\mathcal{C}_{\overset{\chi q(d)}{pqi3}}^{(6,2)}\mathcal{C}_{\overset{\chi e}{pq22}}^{(6,2)}$ (with $i=1,2$) through the topology shown in fig.~\ref{fig:b2dill}(b). The corresponding bounds are found to be approximately four orders of magnitude stronger than those obtained for the combination $\mathcal{C}_{\overset{\chi q(d)}{pqi3}}^{(6,2)}\mathcal{C}_{B\chi}^{(5,1)}$, which contributes through the diagram in fig.~\ref{fig:b2dill}(a). This hierarchy can be understood from the structure of the underlying amplitudes. The $Z$-mediated contributions to the $b\to d_i \ell^+\ell^-$ transitions are suppressed by a factor of $m_b^2/M_Z^2$ and therefore provide only relatively weak constraints. Consequently, the dominant sensitivity to the combination $\mathcal{C}_{\overset{\chi q(d)}{pqi3}}^{(6,2)}\mathcal{C}_{B\chi}^{(5,1)}$ originates from the photon-mediated penguin diagrams. A similar suppression is observed for DM-mediated loop contributions involving vector-current operators, $\mathcal{C}_{\overset{\chi f}{pqij}}^{(6,1)}$, inserted at both vertices, as these amplitudes are also proportional to light-quark masses. Finally, the stringent bounds obtained on $\mathcal{C}_{B\chi}^{(5,1)}$ from the EWPO analysis can be translated into indirect constraints on the flavour-changing coefficients $\mathcal{C}_{\overset{\chi q(d)}{pqi3}}^{(6,2)}$. These derived limits are presented at the end of the results section.

%-------------------------------------
\begin{table}[t!]
	\centering
	\footnotesize
	\renewcommand{\arraystretch}{1.5}
	\begin{tabular}{c c c}
		\toprule
		\textbf{Scenario} & \textbf{$m_\chi = 800$ GeV} & \textbf{$m_\chi = 250$ GeV} \\
		\midrule
		$\mathcal{C}_{\chi \underset{23}{q}}^{(6,2)} \, \mathcal{C}_{\chi \underset{22}{\ell}}^{(6,2)}$ & $(-5.32 \pm 1.89) \times 10^{-15}$ & $(1.20 \pm 0.45) \times 10^{-14}$ \\
		
		$\mathcal{C}_{\chi \underset{23}{q}}^{(6,2)} \, \mathcal{C}_{\chi \underset{22}{e}}^{(6,2)}$ & $(-1.00 \pm 0.62) \times 10^{-14}$ & $(2.26 \pm 1.26) \times 10^{-14}$ \\
		
		$\mathcal{C}_{\chi \underset{23}{d}}^{(6,2)} \, \mathcal{C}_{\chi \underset{22}{\ell}}^{(6,2)}$ & $(-2.63 \pm 1.92) \times 10^{-15}$ & $(5.95 \pm 4.42) \times 10^{-15}$ \\
		
		$\mathcal{C}_{\chi \underset{23}{q}}^{(6,2)} \, \mathcal{C}_{\chi \underset{22}{\ell}}^{(6,2)}$ & $(-4.67 \pm 7.65) \times 10^{-15}$ & $(1.05 \pm 1.74) \times 10^{-14}$ \\
		
		$\mathcal{C}_{\chi \underset{23}{q}}^{(6,2)} \, \mathcal{C}_{B \chi}^{(5,1)}$ & $(-2.13 \pm 0.59 ) \times 10^{-11}$ & $(1.50 \pm 0.42 ) \times 10^{-11}$  \\
		
		$\mathcal{C}_{\chi \underset{23}{d}}^{(6,2)} \, \mathcal{C}_{B \chi}^{(5,1)}$ & $( -9.10 \pm 5.46) \times 10^{-12}$ & $( 6.45 \pm 3.86 ) \times 10^{-12}$ \\
		
		\midrule
		
		$\mathcal{C}_{\chi \underset{13}{q}}^{(6,2)} \, \mathcal{C}_{\chi \underset{22}{\ell}}^{(6,2)}$ & $(-2.28 \pm 1.98) \times 10^{-14}$ & $(5.17 \pm 4.49) \times 10^{-14}$ \\
		
		$\mathcal{C}_{\chi \underset{13}{q}}^{(6,2)} \, \mathcal{C}_{\chi \underset{22}{e}}^{(6,2)}$ & $(-0.15 \pm 1.98) \times 10^{-14}$ & $ (0.32 \pm 4.49) \times 10^{-14} $ \\
		
		$\mathcal{C}_{\chi \underset{13}{d}}^{(6,2)} \, \mathcal{C}_{\chi \underset{22}{\ell}}^{(6,2)}$ & $(0.00 \pm 1.97)\times 10^{-14}$ & $(0.02 \pm 4.48) \times 10^{-14}$ \\
		
		$\mathcal{C}_{\chi \underset{13}{d}}^{(6,2)} \, \mathcal{C}_{\chi \underset{22}{e}}^{(6,2)}$ & $ (-0.02 \pm 1.97) \times 10^{-14} $ & $(0.04 \pm 4.48) \times 10^{-14}$  \\
		
		$\mathcal{C}_{\chi \underset{13}{q}}^{(6,2)} \, \mathcal{C}_{B \chi}^{(5,1)}$ & $(-3.14 \pm 3.88) \times 10^{-11}$ & $(2.22 \pm 2.75) \times 10^{-11}$ \\
		
		$\mathcal{C}_{\chi \underset{13}{d}}^{(6,2)} \, \mathcal{C}_{B \chi}^{(5,1)}$ & $(-0.03 \pm 3.86) \times 10^{-11}$ & $(0.02 \pm 2.74) \times 10^{-11}$ \\
		
		\bottomrule
	\end{tabular}
	\caption{Constraints on DMEFT WCs from semileptonic decays of $B$ mesons.}
	\label{tab:Rare_B_decays}
\end{table}
\subsection{Meson mixings (\texorpdfstring{$\Delta F=2$}{F2} observables):}\label{Append:Meson_mixing}
%-------------------------------------
\subsubsection*{Matching Relations}\label{Append:meson_mixing_match}
%-------------------------------------
The general effective Hamiltonian governing $\Delta F=2$ transitions is given in Eq.~\eqref{eq:mixing_delF_2}. As detailed in the main text, our DMEFT framework exclusively generates loop-level contributions to the vector and axial-vector operators $Q_1$ and $\tilde{Q}_1$. For the $B_q^0 -\bar{B}_q^0$ mixing systems, the explicit matching relations for the corresponding coefficients are evaluated as
%--------------------------
\begin{align}
    \Delta C_1^{qb} &= \frac{m_{\chi}^2}{2\pi^2}\log\left(\frac{\mu^2}{m_{\chi}^2} \right)\mathcal{C}^{(6,2)}_{\chi \underset{3q}{q}}\mathcal{C}^{(6,2)\,*}_{\chi \underset{3q}{q}}\,, & 
    \Delta \tilde{C}_1^{qb} &= \frac{m_{\chi}^2}{2\pi^2}\log\left(\frac{\mu^2}{m_{\chi}^2}\right)\mathcal{C}^{(6,2)}_{\chi \underset{3q}{d}}\mathcal{C}^{(6,2)\,*}_{\chi \underset{3q}{d}}\,,
\end{align}
%--------------------------
where $q \in \{d,s\}$. Similarly, the matching relations for the $K^0 -\bar{K}^0$ system, $ \Delta C_1^{sd}$ and $ \Delta \tilde{C}_1^{sd} $ can be found by appropriate replacement of $\mathcal{C}^{(6,2)}_{\chi \underset{3q}{q(d)}} \to \mathcal{C}^{(6,2)}_{\chi \underset{21}{q(d)}}$. 
%------------------------
% \begin{align}
%     \Delta C_1^{sd}&= \frac{m_{\chi}^2}{2\pi^2}\log\left(\frac{\mu^2}{m_{\chi}^2} \right)\mathcal{C}^{(6,2)}_{\overset{\chi q}{pr21}}\mathcal{C}^{(6,2)\,*}_{\overset{\chi q}{pr21}}\,, & \quad \Delta \tilde{C}_1^{sd}&=\frac{m_{\chi}^2}{2\pi^2}\log\left(\frac{\mu^2}{m_{\chi}^2}\right)\mathcal{C}^{(6,2)}_{\overset{\chi d}{pr21}}\mathcal{C}^{(6,2)\,*}_{\overset{\chi d}{pr21}}\,.
% \end{align}
%------------------------
The matching relations in terms of SMEFT WCs can be given as:
%------------------------
\begin{subequations}
    \begin{align}
    \Delta C_1^{sb}&= -\left(\mathcal{C}_{\overset{qq}{2323}}^{(1)}+\mathcal{C}_{\overset{qq}{2323}}^{(3)}\right)\,, &  
    \Delta \tilde{C}_1^{sb}&= -\mathcal{C}_{\overset{dd}{2323}}\,,\\
    \Delta C_1^{db}&= -\left(\mathcal{C}_{\overset{qq}{1313}}^{(1)}+\mathcal{C}_{\overset{qq}{1313}}^{(3)}\right)\,, & 
    \Delta \tilde{C}_1^{db}&= -\mathcal{C}_{\overset{dd}{1313}}\,,\\
    \Delta C_1^{sd}&= -\left(\mathcal{C}_{\overset{qq}{1212}}^{(1)}+\mathcal{C}_{\overset{qq}{1212}}^{(3)}\right)\,, & 
    \Delta \tilde{C}_1^{sd}&= -\mathcal{C}_{\overset{dd}{1212}}\,.
\end{align}\end{subequations}
%------------------------
%------------------------------
\subsubsection*{Observables:}\label{Append:Meson_mixing_obs}
As discussed in sec.~\ref{subsec:Meson_Mixing}, the mass difference can be expressed as a mixing observable, given by $\Delta M = \left| \mathcal{M} \right| / m_P$. For specific neutral $B$-meson systems ($q = d,s$), the new-physics contribution to this mass difference takes the form
\begin{align}
    \Delta M_{q}^{\rm NP} = \frac{2}{3}\, m_{B_q} f_{B_q}^2 B_{B_q}(\mu)  \left( \Delta C_1^{qb} +\Delta \tilde{C}_1^{qb} \right) .
\end{align}
Since parity is conserved in QCD, the pseudoscalar-to-pseudoscalar matrix elements of the parity-conjugate operators are identical, yielding $\langle Q_1^q\rangle = \langle \tilde{Q}_1^q\rangle = \frac{2}{3} m_{B_q}^2 f_{B_q}^2 B_{B_q} (\mu)$. Here, $f_{B_q}$ is the meson decay constant and $B_{B_q}(\mu)$ is the bag parameter evaluated at the scale $\mu_b$. Similarly, for $K^0-\bar{K}^0$ mixing the mass difference takes the form,
\begin{align}
    \Delta M_{q}^{\rm NP} = \frac{2}{3}\, m_{K} f_{K}^2 B_{K}(\mu)  \left( \Delta C_1^{sd} +\Delta \tilde{C}_1^{sd} \right) .
\end{align}
We have used the following values of the constants for $B$-meson: $B_{B_s} = 1.232 (53), \, B_{B_d} = 1.222(61), \, f_{B_s} = 230.3 (1.3) {~\rm MeV}, \, f_{B_d} = 190.0(1.3) ~{\rm MeV}$ \cite{FlavourLatticeAveragingGroupFLAG:2024oxs}. For $K$-meson, we have used: $B_{B_K} = 0.717 (24),f_{K^{\pm}} = 155.7 (3)  \rm{\, MeV},  f_K = \frac{f_{K^{\pm}}}{\sqrt{1+\delta_{\rm SU(2)}}} \,, M_{K} = 495  \rm{~MeV}, \delta_{\rm SU(2)} = -0.0052(9) $ \cite{FlavourLatticeAveragingGroupFLAG:2024oxs, ParticleDataGroup:2024cfk, Bazavov:2017lyh}. 
\subsection{Lepton Flavour Violating decays:}
%------------------------------------
\subsubsection*{LFV \texorpdfstring{$B$}{B} decay modes} \label{Append:B_LFV}
%------------------------------------
The corresponding matching relations are (for the quark level transition: $q_{i} \bar q_{j} \to \ell_{\alpha} \ell_{\beta}$),

\begin{subequations}
\begin{align}
    C_{V_{LL}}&= \frac{m_{\chi}^2}{ 8 \pi^2} \log \frac{\mu^2}{m_{\chi}^2} \mathcal{C}_{\chi \underset{ij}{q}}^{(6,2)} \mathcal{C}_{\chi \underset{\alpha \beta}{\ell}}^{(6,2)}   \,, \quad \quad   C_{V_{LR}} = \frac{m_{\chi}^2}{ 8 \pi^2} \log \frac{\mu^2}{m_{\chi}^2} \mathcal{C}_{\chi \underset{ij}{q}}^{(6,2)} \mathcal{C}_{\chi \underset{\alpha \beta}{e}}^{(6,2)} \,, \\
    C_{V_{RL}}&= \frac{m_{\chi}^2}{ 8 \pi^2} \log \frac{\mu^2}{m_{\chi}^2} \mathcal{C}_{\chi \underset{ij}{e}}^{(6,2)} \mathcal{C}_{\chi \underset{\alpha \beta}{\ell}}^{(6,2)}\,, \quad \quad   C_{V_{RR}} = \frac{m_{\chi}^2}{ 8 \pi^2} \log \frac{\mu^2}{m_{\chi}^2} \mathcal{C}_{\chi \underset{ij}{d}}^{(6,2)} \mathcal{C}_{\chi \underset{\alpha \beta}{e}}^{(6,2)} \,.
\end{align}
\end{subequations}

\subsubsection*{LFV \texorpdfstring{$\mu,\tau$}{mu, tau} decays}\label{Append:lepton_cLFV}
%-------------------------------------
\paragraph{Leptonic Modes:} 
The expressions for the $\ell_i\to \ell_j\ell_k\bar{\ell}_l$ can be split into three subclasses, depending on the composition of the final state leptons, and the corresponding matching relations are as follows
%-------------------------------
\begin{itemize}
    \item \textbf{Decay \,$\ell_i\to\ell_j\ell_j\bar{\ell}_j:$} $\mu^{\pm}\to e^{\pm}e^+e^-$, $\tau^{\pm}\to e^{\pm}e^+e^-$ and $\tau^{\pm}\to \mu^{\pm}\mu^+\mu^-$.
%-----------------------------    
\begin{align}
    \mathcal{C}^V_{LL}&= \frac{m_{\chi}^2}{2\pi^2}\log\left(\frac{\mu^2}{m_{\chi}^2}\right)\Big[\mathcal{C}_{\chi \underset{ji}{\ell}}^{(6,2)}\mathcal{C}_{\chi \underset{jj}{\ell}}^{(6,2)}+\frac{e c_W}{m_{\chi}} \mathcal{C}_{\chi \underset{ji}{\ell}}^{(6,1)} \mathcal{C}_{B\chi}^{(5,1)}\Big]+\frac{m_i^2 m_{\chi}}{2\pi^2 M_Z^2} s_W g_Z g_Z^L\log\left(\frac{\mu^2}{m_{\chi}^2}\right) \mathcal{C}_{\chi \underset{ji}{\ell}}^{(6,1)} \mathcal{C}_{b\chi}^{(5,1)}\,, \nonumber\\
    \mathcal{C}^V_{LR}&=\frac{m_{\chi}^2}{2\pi^2}\log\left(\frac{\mu^2}{m_{\chi}^2}\right)\Big[\mathcal{C}_{\chi \underset{ji}{\ell}}^{(6,2)}\mathcal{C}_{\chi \underset{jj}{e}}^{(6,2)}  +\frac{e c_W}{m_{\chi}} \mathcal{C}_{\chi \underset{ji}{\ell}}^{(6,1)} \mathcal{C}_{B\chi}^{(5,1)}\Big] +\frac{m_i^2 m_{\chi}}{2\pi^2 M_Z^2} s_W g_Z g_Z^R\log\left(\frac{\mu^2}{m_{\chi}^2}\right) \mathcal{C}_{\chi \underset{ji}{\ell}}^{(6,1)} \mathcal{C}_{b\chi}^{(5,1)}\,,\nonumber\\
    \mathcal{C}^V_{RL}&= \frac{m_{\chi}^2}{2\pi^2}\log\left(\frac{\mu^2}{m_{\chi}^2}\right)\Big[\mathcal{C}_{\chi \underset{ji}{e}}^{(6,2)}\mathcal{C}_{\chi \underset{jj}{\ell}}^{(6,2)}+ \frac{e c_W}{m_{\chi}} \mathcal{C}_{\chi \underset{ji}{e}}^{(6,1)} \mathcal{C}_{B\chi}^{(5,1)}\Big] +\frac{m_i^2 m_{\chi}}{2\pi^2 M_Z^2} s_W g_Z g_Z^L\log\left(\frac{\mu^2}{m_{\chi}^2}\right) \mathcal{C}_{\chi \underset{ji}{e}}^{(6,1)} \mathcal{C}_{b\chi}^{(5,1)}\,,\nonumber\\
    \mathcal{C}^V_{RR}&= \frac{m_{\chi}^2}{2\pi^2}\log\left(\frac{\mu^2}{m_{\chi}^2}\right)\Big[\mathcal{C}_{\chi \underset{ji}{e}}^{(6,2)}\mathcal{C}_{\chi \underset{jj}{e}}^{(6,2)}+  \frac{e c_W}{m_{\chi}} \mathcal{C}_{\chi \underset{ji}{e}}^{(6,1)} \mathcal{C}_{B\chi}^{(5,1)}\Big]+\frac{m_i^2 m_{\chi}}{2\pi^2 M_Z^2} s_W g_Z g_Z^R\log\left(\frac{\mu^2}{m_{\chi}^2}\right) \mathcal{C}_{\chi \underset{ji}{e}}^{(6,1)} \mathcal{C}_{b\chi}^{(5,1)}\,.
\end{align}
%-------------------------------
    \item \textbf{Decay \,$\ell_i\to\ell_j\ell_k\bar{\ell}_k:$} $\tau^{\pm}\to e^{\pm}\mu^+\mu^-$ and $\tau^{\pm}\to\mu^{\pm} e^+ e^-$.
%-----------------------------    
% \begin{align}
%     \mathcal{C}^V_{LL}&= \frac{m_{\chi}^2}{2\pi^2}\log\left(\frac{\mu^2}{m_{\chi}^2}\right)\mathcal{C}_{\overset{\chi\ell}{pqji}}^{(6,2)}\mathcal{C}_{\overset{\chi\ell}{pq kk}}^{(6,2)}\,,&\quad
%     \mathcal{C}^V_{LR}&=\frac{m_{\chi}^2}{2\pi^2}\log\left(\frac{\mu^2}{m_{\chi}^2}\right)\mathcal{C}_{\overset{\chi\ell}{pqji}}^{(6,2)}\mathcal{C}_{\overset{\chi e}{pq kk}}^{(6,2)}  \,,\nonumber\\
%     &+\frac{m_i^2 m_{\chi}}{2\pi^2 M_Z^2} s_W g_Z g_Z^L\log\left(\frac{\mu^2}{m_{\chi}^2}\right) \mathcal{C}_{\overset{\chi\ell}{pqji}}^{(6,1)} \mathcal{C}_{b\chi}^{(5,1)}\,, &\quad  & +\frac{m_i^2 m_{\chi}}{2\pi^2 M_Z^2} s_W g_Z g_Z^R\log\left(\frac{\mu^2}{m_{\chi}^2}\right) \mathcal{C}_{\overset{\chi\ell}{pqji}}^{(6,1)} \mathcal{C}_{b\chi}^{(5,1)} \,,\\
%     \mathcal{C}^V_{RL}&= \frac{m_{\chi}^2}{2\pi^2}\log\left(\frac{\mu^2}{m_{\chi}^2}\right)\mathcal{C}_{\overset{\chi e}{pqji}}^{(6,2)}\mathcal{C}_{\overset{\chi\ell}{pq kk}}^{(6,2)} \,,& \quad
%     \mathcal{C}^V_{RR}&= \frac{m_{\chi}^2}{2\pi^2}\log\left(\frac{\mu^2}{m_{\chi}^2}\right)\mathcal{C}_{\overset{\chi e}{pqji}}^{(6,2)}\mathcal{C}_{\overset{\chi e}{pq kk}}^{(6,2)} \,,\nonumber\\
%     &+\frac{m_i^2 m_{\chi}}{2\pi^2 M_Z^2} s_W g_Z g_Z^L\log\left(\frac{\mu^2}{m_{\chi}^2}\right) \mathcal{C}_{\overset{\chi e}{pqji}}^{(6,1)} \mathcal{C}_{b\chi}^{(5,1)}\,, & & +\frac{m_i^2 m_{\chi}}{2\pi^2 M_Z^2} s_W g_Z g_Z^R\log\left(\frac{\mu^2}{m_{\chi}^2}\right) \mathcal{C}_{\overset{\chi e}{pqji}}^{(6,1)} \mathcal{C}_{b\chi}^{(5,1)} ,
% \end{align}

\begin{align}
    \mathcal{C}^V_{LL}&= \frac{m_{\chi}^2}{2\pi^2}\log\left(\frac{\mu^2}{m_{\chi}^2}\right)\Big[\mathcal{C}_{\chi \underset{ji}{\ell}}^{(6,2)}\mathcal{C}_{\chi \underset{kk}{\ell}}^{(6,2)}+\frac{e c_W}{m_{\chi}} \mathcal{C}_{\chi \underset{ji}{\ell}}^{(6,1)} \mathcal{C}_{B\chi}^{(5,1)}\Big]+\frac{m_i^2 m_{\chi}}{2\pi^2 M_Z^2} s_W g_Z g_Z^L\log\left(\frac{\mu^2}{m_{\chi}^2}\right) \mathcal{C}_{\chi \underset{ji}{\ell}}^{(6,1)} \mathcal{C}_{b\chi}^{(5,1)}\,, \nonumber\\
    \mathcal{C}^V_{LR}&=\frac{m_{\chi}^2}{2\pi^2}\log\left(\frac{\mu^2}{m_{\chi}^2}\right)\Big[\mathcal{C}_{\chi \underset{ji}{\ell}}^{(6,2)}\mathcal{C}_{\chi \underset{kk}{e}}^{(6,2)}  +\frac{e c_W}{m_{\chi}} \mathcal{C}_{\chi \underset{ji}{\ell}}^{(6,1)} \mathcal{C}_{B\chi}^{(5,1)}\Big] +\frac{m_i^2 m_{\chi}}{2\pi^2 M_Z^2} s_W g_Z g_Z^R\log\left(\frac{\mu^2}{m_{\chi}^2}\right) \mathcal{C}_{\chi \underset{ji}{\ell}}^{(6,1)} \mathcal{C}_{b\chi}^{(5,1)}\,,\nonumber\\
    \mathcal{C}^V_{RL}&= \frac{m_{\chi}^2}{2\pi^2}\log\left(\frac{\mu^2}{m_{\chi}^2}\right)\Big[\mathcal{C}_{\chi \underset{ji}{e}}^{(6,2)}\mathcal{C}_{\chi \underset{kk}{\ell}}^{(6,2)}+ \frac{e c_W}{m_{\chi}} \mathcal{C}_{\chi \underset{ji}{e}}^{(6,1)} \mathcal{C}_{B\chi}^{(5,1)}\Big] +\frac{m_i^2 m_{\chi}}{2\pi^2 M_Z^2} s_W g_Z g_Z^L\log\left(\frac{\mu^2}{m_{\chi}^2}\right) \mathcal{C}_{\chi \underset{ji}{e}}^{(6,1)} \mathcal{C}_{b\chi}^{(5,1)}\,,\nonumber\\
    \mathcal{C}^V_{RR}&= \frac{m_{\chi}^2}{2\pi^2}\log\left(\frac{\mu^2}{m_{\chi}^2}\right)\Big[\mathcal{C}_{\chi \underset{ji}{e}}^{(6,2)}\mathcal{C}_{\chi \underset{kk}{e}}^{(6,2)}+  \frac{e c_W}{m_{\chi}} \mathcal{C}_{\chi \underset{ji}{e}}^{(6,1)} \mathcal{C}_{B\chi}^{(5,1)}\Big]+\frac{m_i^2 m_{\chi}}{2\pi^2 M_Z^2} s_W g_Z g_Z^R\log\left(\frac{\mu^2}{m_{\chi}^2}\right) \mathcal{C}_{\chi \underset{ji}{e}}^{(6,1)} \mathcal{C}_{b\chi}^{(5,1)}\,.
\end{align}
%-------------------------------
    \item \textbf{Decay \,$\ell_i^{\pm}\to \bar{\ell}_j^{\mp}\ell_k^{\pm}\ell_k^{\pm}:$} $\tau^{\pm}\to e^{\mp}\mu^{\pm}\mu^{\pm}$ and $\tau^{\pm}\to \mu^{\mp}e^{\pm}e^{\pm}$
%-----------------------------    
\begin{align}
    \mathcal{C}^V_{LL}&= \frac{m_{\chi}^2}{2\pi^2}\log\left(\frac{\mu^2}{m_{\chi}^2}\right)\mathcal{C}_{\chi \underset{ki}{\ell}}^{(6,2)}\mathcal{C}_{\chi \underset{jk}{\ell}}^{(6,2)}\,,&\quad
    \mathcal{C}^V_{LR}&=\frac{m_{\chi}^2}{2\pi^2}\log\left(\frac{\mu^2}{m_{\chi}^2}\right)\mathcal{C}_{\chi \underset{ki}{\ell}}^{(6,2)}\mathcal{C}_{\chi \underset{jk}{e}}^{(6,2)}  \,,\\
    \mathcal{C}^V_{RL}&= \frac{m_{\chi}^2}{2\pi^2}\log\left(\frac{\mu^2}{m_{\chi}^2}\right)\mathcal{C}_{\chi \underset{ki}{e}}^{(6,2)}\mathcal{C}_{\chi \underset{jk}{\ell}}^{(6,2)} \,,& \quad
    \mathcal{C}^V_{RR}&= \frac{m_{\chi}^2}{2\pi^2}\log\left(\frac{\mu^2}{m_{\chi}^2}\right)\mathcal{C}_{\chi \underset{ki}{e}}^{(6,2)}\mathcal{C}_{\chi \underset{jk}{e}}^{(6,2)} \,,
\end{align}
%-------------------------------
\end{itemize}
Here, $c_W \equiv \cos\theta_W$ and $s_W \equiv \sin\theta_W$ denote the cosine and sine of the weak mixing angle, respectively. The left- and right-handed chiral couplings of the $Z$ boson to leptons are parameterized as $g_Z^L =  (I_W^{(3)} - q_e s_W^2)$ and $g_Z^R = - q_e s_W^2$, where $g_Z = g/c_W$, $I_W^{(3)}$ is the third component of the weak isospin, and $q_e$ represents the electric charge of the lepton.

The matching relations mapped from the SMEFT framework are given by
% \begin{align}
% \mathcal{C}^{V}_{LL} &= -\mathcal{C}_{\overset{ll}{prst}} \,, &
% \mathcal{C}^{V}_{LR} &= -\mathcal{C}_{\overset{le}{prst}} \,, &
% \mathcal{C}^{V}_{RR} &= -\mathcal{C}_{\overset{ee}{prst}} \,,
% \end{align}

%--------------------------------
\begin{itemize}
    \item \textbf{Decay \,$\ell_i\to\ell_j\ell_j\bar{\ell}_j:$}
    \begin{align}
        \mathcal{C}^V_{LL}&= 2\left( \mathcal{C}_{\overset{ll}{jijj}}-2 g_Z^L\left(\mathcal{C}_{\overset{\phi l}{ji}}^{(1)}+\mathcal{C}_{\overset{\phi l}{ji}}^{(3)}\right)\right)\,, &\quad \mathcal{C}^V_{LR}&= \left(\mathcal{C}_{\overset{\ell e}{jijj}}-2 g_Z^R\left(\mathcal{C}_{\overset{\phi l}{ji}}^{(1)}+\mathcal{C}_{\overset{\phi l}{ji}}^{(3)} \right) \right)\,\\
         \mathcal{C}^V_{RL}&=\left( \mathcal{C}_{\overset{le}{jjji}}-2 g_Z^L\mathcal{C}_{\overset{\phi e}{ji}}^{(1)}\right) \,, &\quad \mathcal{C}^V_{RR}&= 2\left( \mathcal{C}_{\overset{ee}{jijj}}-2 g_Z^R\mathcal{C}_{\overset{\phi e}{ji}}\right)\,.
    \end{align}

    \item \textbf{Decay \,$\ell_i\to\ell_j\ell_k\bar{\ell}_k:$}
\begin{align}
        \mathcal{C}^V_{LL}&= \left( \mathcal{C}_{\overset{\ell \ell}{jikk}}-2 g_Z^L\left(\mathcal{C}_{\overset{\phi \ell}{ji}}^{(1)}+\mathcal{C}_{\overset{\phi \ell}{ji}}^{(3)}\right)\right)\,, &\quad \mathcal{C}^V_{LR}&= \left(\mathcal{C}_{\overset{\ell e}{jikk}}-2 g_Z^R\left(\mathcal{C}_{\overset{\phi \ell}{ji}}^{(1)}+\mathcal{C}_{\overset{\phi \ell}{ji}}^{(3)} \right) \right)\,\\
         \mathcal{C}^V_{RL}&=\left( \mathcal{C}_{\overset{\ell e}{kkji}}-2 g_Z^L\mathcal{C}_{\overset{\phi e}{ji}}^{(1)}\right) \,, &\quad \mathcal{C}^V_{RR}&= \left( \mathcal{C}_{\overset{ee}{jikk}}-2 g_Z^R\mathcal{C}_{\overset{\phi e}{ji}}\right)\,.
    \end{align}
    
    \item \textbf{Decay \,$\ell_i^{\pm} \to \bar{\ell}_j^{\mp}\ell_k^{\pm} \ell_k^{\pm}:$}

    \begin{align}
        \mathcal{C}^V_{LL}&= 2~\mathcal{C}_{\overset{\ell \ell}{kikj}}\,, &\quad \mathcal{C}^V_{LR}&= \mathcal{C}_{\overset{\ell e}{kikj}} \,,\\
         \mathcal{C}^V_{RL}&=\mathcal{C}_{\overset{\ell e}{kjki}} \,, &\quad \mathcal{C}^V_{RR}&= 2~\mathcal{C}_{\overset{ee}{kikj}}\,.
    \end{align}
    
\end{itemize}

%--------------------------------
where the generation indices $i, j ~ \text{and}~ k$ map directly onto the specific initial- and final-state leptons defining each individual LFV process.

\paragraph*{Hadronic Modes}  We evaluate two-body decays featuring a charged lepton produced in association with a neutral vector ($V \in \{\rho^0, \phi\}$) or pseudoscalar ($P \in \{\pi^0, K^0\}$) meson. The explicit matching relations for these respective channels are given by:
%--------------------------------
\begin{align}
    C_{VLL}^{\tau\ell \,ii}&=\frac{m_{\chi}^2}{2\pi^2}\log\frac{\Lambda^2}{m_{\chi}^2}\mathcal{C}^{(6,2)}_{\chi \underset{3j}{\ell}}\mathcal{C}^{(6,2)}_{\chi \underset{ii}{q}}\,,&\quad C_{VRR}^{\tau\ell \,ii}&=\frac{m_{\chi}^2}{2\pi^2}\log\frac{\Lambda^2}{m_{\chi}^2}\mathcal{C}^{(6,2)}_{\chi \underset{3j}{e}}\mathcal{C}^{(6,2)}_{\chi \underset{ii}{u(d)}}\,,\\
    C_{VLR}^{\tau\ell \,ii}&=\frac{m_{\chi}^2}{2\pi^2}\log\frac{\Lambda^2}{m_{\chi}^2}\mathcal{C}^{(6,2)}_{\chi \underset{3j}{\ell}}\mathcal{C}^{(6,2)}_{\chi \underset{ii}{u(d)}}\,,&\quad C_{VLR}^{ii\,\tau\ell}&=\frac{m_{\chi}^2}{2\pi^2}\log\frac{\Lambda^2}{m_{\chi}^2}\mathcal{C}^{(6,2)}_{\chi \underset{3j}{e}}\mathcal{C}^{(6,2)}_{\chi \underset{ii}{q}}\,,
\end{align}
Here $i,j$ represent flavour indices of quark and lepton, respectively. Again, in terms of SMEFT WCs matching relations are
%-------------------------------
\begin{align}
    C_{VLL}^{\tau\ell \,ii}&=\mathcal{C}_{\overset{\ell q}{3jii}}\,,&\quad C_{VRR}^{\tau\ell \,ii}&=\mathcal{C}_{\overset{eu(d)}{3jii}}\,,\\
    C_{VLR}^{\tau\ell \,ii}&=\mathcal{C}_{\overset{\ell u(d)}{3j ii}}\,,&\quad C_{VLR}^{ii\,\tau\ell}&=\mathcal{C}_{\overset{qe}{ii3j}}\,.
\end{align}
%--------------------------------
\begin{itemize}
    \item \textbf{$\tau \to \ell ~V(\rho^0,\phi)$ :}
    
For the $V=\phi$ meson,
\begin{align}
    g_L^{\tau \ell \phi}&= \frac{m_{\chi}^2}{2 \pi^2} \log\frac{\Lambda^2}{m_{\chi}^2}~\left(\mathcal{C}^{(6,2)}_{\chi \underset{3j}{\ell}}\mathcal{C}^{(6,2)}_{\chi \underset{22}{q}}+ \mathcal{C}^{(6,2)}_{\chi \underset{3j}{\ell}}\mathcal{C}^{(6,2)}_{\chi \underset{22}{d}} \right)\,,\\
     g_R^{\tau \ell \phi}&=\frac{m_{\chi}^2}{2 \pi^2} \log\frac{\Lambda^2}{m_{\chi}^2}~\left(\mathcal{C}^{(6,2)}_{\chi \underset{3j}{e}}\mathcal{C}^{(6,2)}_{\chi \underset{22}{d}}+\mathcal{C}^{(6,2)}_{\chi \underset{3j}{e}}\mathcal{C}^{(6,2)}_{\chi \underset{22}{q}}\right)\,.
\end{align}
For the $V=\rho$ meson,

\begin{align}
    g_L^{\tau \ell\rho}&=\frac{m_{\chi}^2}{2\sqrt{2} \pi^2} \log\frac{\Lambda^2}{m_{\chi}^2}~\left( \mathcal{C}^{(6,2)}_{\chi \underset{3j}{\ell}}\mathcal{C}^{(6,2)}_{\chi \underset{11}{u}}-\mathcal{C}^{(6,2)}_{\chi \underset{3j}{\ell}}\mathcal{C}^{(6,2)}_{\chi \underset{11}{d}}\right)\,, \\
    g_R^{\tau \ell\rho}&=\frac{m_{\chi}^2}{2\sqrt{2} \pi^2} \log\frac{\Lambda^2}{m_{\chi}^2}~ \left( \mathcal{C}^{(6,2)}_{\chi \underset{3j}{e}}\mathcal{C}^{(6,2)}_{\chi \underset{11}{u}}-\mathcal{C}^{(6,2)}_{\chi \underset{3j}{e}}\mathcal{C}^{(6,2)}_{\chi \underset{11}{d}}\right)\,.
\end{align}

\item \textbf{$\tau\to P(\pi^0)\ell$ :}
For the $P=\pi^0$ meson,
%-----------------------------
\begin{align}
    g_L^{\tau\ell\pi^0}&=-m_{\ell}\frac{m_{\chi}^2}{4 \pi^2} \log\frac{\Lambda^2}{m_{\chi}^2} \left(\mathcal{C}^{(6,2)}_{\chi \underset{3j}{\ell}}\mathcal{C}^{(6,2)}_{\chi \underset{11}{u}}-\mathcal{C}^{(6,2)}_{\chi \underset{3j}{\ell}}\mathcal{C}^{(6,2)}_{\chi \underset{11}{d}} \right)\,,\nonumber\\
    &+m_{\tau}\frac{m_{\chi}^2}{4 \pi^2} \log\frac{\Lambda^2}{m_{\chi}^2} \left( \mathcal{C}^{(6,2)}_{\chi \underset{3j}{e}}\mathcal{C}^{(6,2)}_{\chi \underset{11}{u}}-\mathcal{C}^{(6,2)}_{\chi \underset{3j}{e}}\mathcal{C}^{(6,2)}_{\chi \underset{11}{d}}\right)  \\
    g_R^{\tau\ell\pi^0}&=-m_{\ell}\frac{m_{\chi}^2}{4 \pi^2} \log\frac{\Lambda^2}{m_{\chi}^2} \left( \mathcal{C}^{(6,2)}_{\chi \underset{3j}{e}}\mathcal{C}^{(6,2)}_{\chi \underset{11}{u}}-\mathcal{C}^{(6,2)}_{\chi \underset{3j}{e}}\mathcal{C}^{(6,2)}_{\chi \underset{11}{d}}\right)\,,\nonumber\\
    &+ m_{\tau}\frac{m_{\chi}^2}{4 \pi^2} \log\frac{\Lambda^2}{m_{\chi}^2} \left(\mathcal{C}^{(6,2)}_{\chi \underset{3j}{\ell}}\mathcal{C}^{(6,2)}_{\chi \underset{11}{u}}-\mathcal{C}^{(6,2)}_{\chi \underset{3j}{\ell}}\mathcal{C}^{(6,2)}_{\chi \underset{11}{d}} \right)\,.
\end{align}
%-----------------------------

% \begin{align}
%     C_{9}^{sd\tau\ell}&=\frac{m_{\chi}^2}{4\pi^2}\log\frac{\Lambda^2}{m_{\chi}^2}\left(\mathcal{C}^{(6,2)}_{\overset{\chi \ell}{pq3j}}\mathcal{C}^{(6,2)}_{\overset{\chi q}{pq21}}+\mathcal{C}^{(6,2)}_{\overset{\chi e}{pq3j}}\mathcal{C}^{(6,2)}_{\overset{\chi q}{pq21}} \right)\,,& C_{10}^{sd\tau\ell}&=\frac{m_{\chi}^2}{4\pi^2}\log\frac{\Lambda^2}{m_{\chi}^2}\left(\mathcal{C}^{(6,2)}_{\overset{\chi e}{pq3j}}\mathcal{C}^{(6,2)}_{\overset{\chi q}{pq21}}-\mathcal{C}^{(6,2)}_{\overset{\chi \ell}{pq3j}}\mathcal{C}^{(6,2)}_{\overset{\chi q}{pq21}} \right)\,,\notag\\
%     C_{9}^{\prime\,sd\tau\ell}&=\frac{m_{\chi}^2}{4\pi^2}\log\frac{\Lambda^2}{m_{\chi}^2}\left(\mathcal{C}^{(6,2)}_{\overset{\chi e}{pq3j}}\mathcal{C}^{(6,2)}_{\overset{\chi d}{pq21}}+\mathcal{C}^{(6,2)}_{\overset{\chi \ell}{pq3j}}\mathcal{C}^{(6,2)}_{\overset{\chi d}{pq21}} \right)\,,& C_{10}^{\prime\,sd\tau\ell}&=\frac{m_{\chi}^2}{4\pi^2}\log\frac{\Lambda^2}{m_{\chi}^2}\left(\mathcal{C}^{(6,2)}_{\overset{\chi e}{pq3j}}\mathcal{C}^{(6,2)}_{\overset{\chi d}{pq21}}-\mathcal{C}^{(6,2)}_{\overset{\chi \ell}{pq3j}}\mathcal{C}^{(6,2)}_{\overset{\chi d}{pq21}} \right)\,.\notag
% \end{align}

\end{itemize}
\subsubsection*{LFV \texorpdfstring{$t\to q^{\prime} \ell_i\ell_j$}{tql1l2} decay modes:} \label{Append:top_LFV}
%-----------------------------------

Here the form factors can be written in terms of,
%------------------------------
%-----------------------------    
\begin{align}
    \mathcal{C}^V_{LL}&= \frac{m_{\chi}^2}{2\pi^2}\log\left(\frac{\mu^2}{m_{\chi}^2}\right)\mathcal{C}_{\chi \underset{j3}{q}}^{(6,2)}\mathcal{C}_{\chi \underset{kl}{\ell}}^{(6,2)}\,,&\quad
    \mathcal{C}^V_{LR}&=\frac{m_{\chi}^2}{2\pi^2}\log\left(\frac{\mu^2}{m_{\chi}^2}\right)\mathcal{C}_{\chi \underset{j3}{q}}^{(6,2)}\mathcal{C}_{\chi \underset{kl}{e}}^{(6,2)}  \,,\nonumber\\
    \mathcal{C}^V_{RL}&= \frac{m_{\chi}^2}{2\pi^2}\log\left(\frac{\mu^2}{m_{\chi}^2}\right)\mathcal{C}_{\chi \underset{j3}{u}}^{(6,2)}\mathcal{C}_{\chi \underset{kl}{\ell}}^{(6,2)} \,,& \quad
    \mathcal{C}^V_{RR}&= \frac{m_{\chi}^2}{2\pi^2}\log\left(\frac{\mu^2}{m_{\chi}^2}\right)\mathcal{C}_{\chi \underset{j3}{u}}^{(6,2)}\mathcal{C}_{\chi \underset{kl}{e}}^{(6,2)} \,.
\end{align}
%-------------------------------
In terms of SMEFT WCs, these form factors are written as,
%-------------------------------
\begin{align}
    \mathcal{C}^V_{LL}&= \left(\mathcal{C}_{\overset{\ell q}{ij3q}}^{(1)}-\mathcal{C}_{\overset{\ell q}{ij3q}}^{(3)}\right) \,,&\quad \mathcal{C}^V_{LR}&= \mathcal{C}_{\overset{qe}{3qij}}\,,\nonumber\\
    \mathcal{C}^V_{RL}&= \mathcal{C}_{\overset{\ell u}{ij3q}} \,,&\quad \mathcal{C}^V_{RR}&= \mathcal{C}_{\overset{eu}{ij3q}}\,.
\end{align}
%-------------------------------
\subsection{Top FCNC decays:}\label{Append:Top_FCNC}
%-------------------------------
For $t \to q Z$ process, matching relations are,
%--------------------------------
\begin{align}
    X_L^{u_j t}&=-\frac{M_Z^2 m_{\chi}}{2\pi^2}s_W \left(2 +\text{DiscB}[M_Z^2,m_{\chi},m_{\chi}]+ \log\left(\frac{\mu^2}{m_{\chi}^2} \right)\right)\mathcal{C}^{(6,1)}_{\chi \underset{j3}{q}} \mathcal{C}^{(5,1)}_{B\chi} \,,\\
    X_R^{u_j t}&=-\frac{M_Z^2 m_{\chi}}{2\pi^2}s_W \left(2 +\text{DiscB}[M_Z^2,m_{\chi},m_{\chi}]+ \log\left(\frac{\mu^2}{m_{\chi}^2} \right)\right) \mathcal{C}^{(6,1)}_{\chi \underset{j3}{u}} \mathcal{C}^{(5,1)}_{B\chi}\,.
\end{align}
%--------------------------------
For $ t \to q H$ process, matching relations are
%------------------------------
\begin{align}
    \eta_L^{u_j t}&=\frac{m_t m_{\chi}}{2 \pi^2}s_W \left(2 +\text{DiscB}[M_H^2,m_{\chi},m_{\chi}]+ \log\left( \frac{\mu^2}{m_{\chi}^2}\right)\right)\mathcal{C}^{(6,2)}_{\chi \underset{j3}{u}}\mathcal{C}_{\chi D H}^{(5,2)} \,,\\
    \eta_R^{u_j t}&=\frac{m_t m_{\chi}}{2 \pi^2}s_W \left(2 +\text{DiscB}[M_H^2,m_{\chi},m_{\chi}]+ \log\left( \frac{\mu^2}{m_{\chi}^2}\right)\right)\mathcal{C}^{(6,2)}_{\chi \underset{j3}{q}}\mathcal{C}_{\chi D H}^{(5,2)} \,.
\end{align}
%------------------------------
In terms of SMEFT WCs, the matching relations are
%------------------------------
\begin{align}
    X_L^{u_j t}&= v^2 \left(\mathcal{C}_{\overset{\phi q}{j3}}^{(1)}-\mathcal{C}_{\overset{\phi q}{j3}}^{(3)}\right)\,, &\quad X_R^{u_j t}&= v^2 \mathcal{C}_{\overset{\phi u}{j3}}\,,\\
    \eta_L^{u_j t}&=\frac{3}{2}v^2 \left( \mathcal{C}_{\overset{u \phi}{j3}}\right)^* \,, &\quad \eta_R^{u_j t}&= \frac{3}{2}v^2 \mathcal{C}_{\overset{u \phi}{j 3}}\,.
\end{align}
%-----------------------------
%-------------------------------------
\subsection{Magnetic Moment}\label{Append:Magnetic_mom}
%-------------------------------------

%-------------------------------
As shown in fig.~\ref{fig:EDM}(b), gauge boson mixing provides further contributions to the anomalous magnetic moment. The distinct effect of each sector is indicated in the corresponding expressions for $\Delta a_{\ell}$.

\begin{subequations}
\begin{align}
    \big[\Delta a_{\ell}\big]^{(b)}_{\gamma\gamma}&=\frac{e^2 q_{\ell}}{8\pi^2} \left(1-\frac{2}{3}\beta_0 m_{\ell}^2\right)
\end{align}

\begin{align}
    \big[\Delta a_{\ell}\big]^{(b)}_{\gamma Z}&= \frac{e q_{\ell} g_Z g_V}{8\pi^2}\left(\frac{m_{\ell}^2}{3M_Z^2}(\beta_1-2\beta_2 M_Z^2)\right)
\end{align}

\begin{align}
    \big[\Delta a_{\ell}\big]^{(b)}_{ZZ}&= \frac{g_Z^2(g_A^2+g_V^2)}{48\pi^2}m_{\ell}^2 \beta_3
\end{align}
\end{subequations}
%-------------------------------
where,
%------------------------------
\begin{subequations}
\begin{align}
\beta_0&=\frac{ c_W^2}{3\pi^2}\Bigg[\left(|\mathcal{C}_{B\chi}^{5,1}|^2+\frac{1}{2}\left(|\mathcal{C}_{B\chi}^{5,1}|^2-|\mathcal{C}_{B\chi}^{5,2}|^2\right)\log\frac{\mu^2}{m_{\chi}^2}\right)\bigg]\\
    \beta_1&=\frac{c_W s_W}{\pi^2}m_{\chi}^2 \left(\left(|\mathcal{C}_{B\chi}^{5,1}|^2+|\mathcal{C}_{B\chi}^{5,2}|^2\right)-\left(|\mathcal{C}_{B\chi}^{5,1}|^2-|\mathcal{C}_{B\chi}^{5,2}|^2\right)\log\frac{\mu^2}{m_{\chi}^2}\right)+g_Z v\frac{c_W}{4\pi^2} m_{\chi}\,\mathcal{C}_{B\chi}^{5,1}\mathcal{C}_{\chi D H}
\end{align}

\begin{align}
    \beta_2&=\frac{c_W s_W}{3\pi^2}\left(|\mathcal{C}_{B\chi}^{5,1}|^2+\frac{1}{2}\left(|\mathcal{C}_{B\chi}^{5,1}|^2+|\mathcal{C}_{B\chi}^{5,2}|^2\right)\log\frac{\mu^2}{m_{\chi}^2}\right)\\
    \beta_3&=\frac{s_W^2}{3\pi^2}\left(|\mathcal{C}_{B\chi}^{5,1}|^2+\frac{1}{2}\left(|\mathcal{C}_{B\chi}^{5,1}|^2+|\mathcal{C}_{B\chi}^{5,2}|^2\right)\log\frac{\mu^2}{m_{\chi}^2}\right)
\end{align}
\end{subequations}
%----------------------------
Again, in terms of SMEFT WCs, the matching relation is,
\begin{align}
    a_{\ell}&=\frac{e q_{\ell}}{2m_{\ell}} v\sqrt{2} \left(s_W~ \text{Re}(\mathcal{C}_{\overset{eW}{ii}})-c_W ~\text{Re}(\mathcal{C}_{\overset{eB}{ii}}) \right)\,.
\end{align}

\label{Bibliography}
\bibliographystyle{JHEP}
\bibliography{biblio}
%%%%%%%%%%%%%%%%%%%%%%%%%%%%%%
\end{document}